\documentclass[aip,pof,preprint,amsmath,amssymb]{revtex4-1}

\usepackage{graphicx}

\begin{document}

\newcommand\etal{\mbox{\textit{et al.}}}
\newcommand\Real{\mbox{Re}} 
\newcommand\Imag{\mbox{Im}} 
\newcommand\Rey{\mbox{\textit{Re}}} 
\newcommand\Pra{\mbox{\textit{Pr}}} 
\newcommand\Pec{\mbox{\textit{Pe}}} 

\title{Roughness characteristics of an ice surface grown in the presence of a supercooled water film driven by gravity and wind drag}

\author{Kazuto Ueno}

\email{k.ueno@kyudai.jp}


\author{Masoud Farzaneh}

\affiliation
{NSERC/Hydro-Quebec/UQAC Industrial Chair on Atmospheric Icing of Power Network Equipment (CIGELE)
and Canada Research Chair on Atmospheric Icing Engineering of Power Networks (INGIVRE), 
Universit$\acute{e}$ du Qu$\acute{e}$bec $\grave{a}$ Chicoutimi, 
Chicoutimi, Qu$\acute{e}$bec G7H 2B1, Canada}


\begin{abstract}
A theoretical model is proposed to explain the roughness characteristics of an ice surface grown from a gravity and wind-driven supercooled water film flowing over an inclined plane.
The effects of the water supply rate, plane slope and air stream velocity on the spacing and height of ice surface roughness are investigated from a new type of morphological instability of the ice-water interface. The proposed macro-scale morphological instability under a supercooled water film is quite different from the micro-scale one which results in dendritic growth.
It was found that ice surface roughness spacing depends mainly on water layer thickness, and that surface roughness height is very sensitive to the convective heat transfer rate at the water-air interface. The present model takes into account the interaction between air and water flows through the boundary conditions at the water-air interface. This leads us to a major finding that tangential and normal shear stress disturbances due to airflow at the water-air interface play a crucial role not only on the convective heat transfer rate at the disturbed water-air interface but also on the height of the ice surface roughness. This is confirmed by comparison of the amplification rate of the ice-water interface disturbance predicted by the model with the roughness height observed experimentally.  
\end{abstract}

\keywords{Supercooled water film, Air shear stress, Morphological instability, Linear stability analysis}

\maketitle

\section{Introduction}
\begin{figure}
\begin{center}
\includegraphics[width=10cm,height=10cm,keepaspectratio,clip]{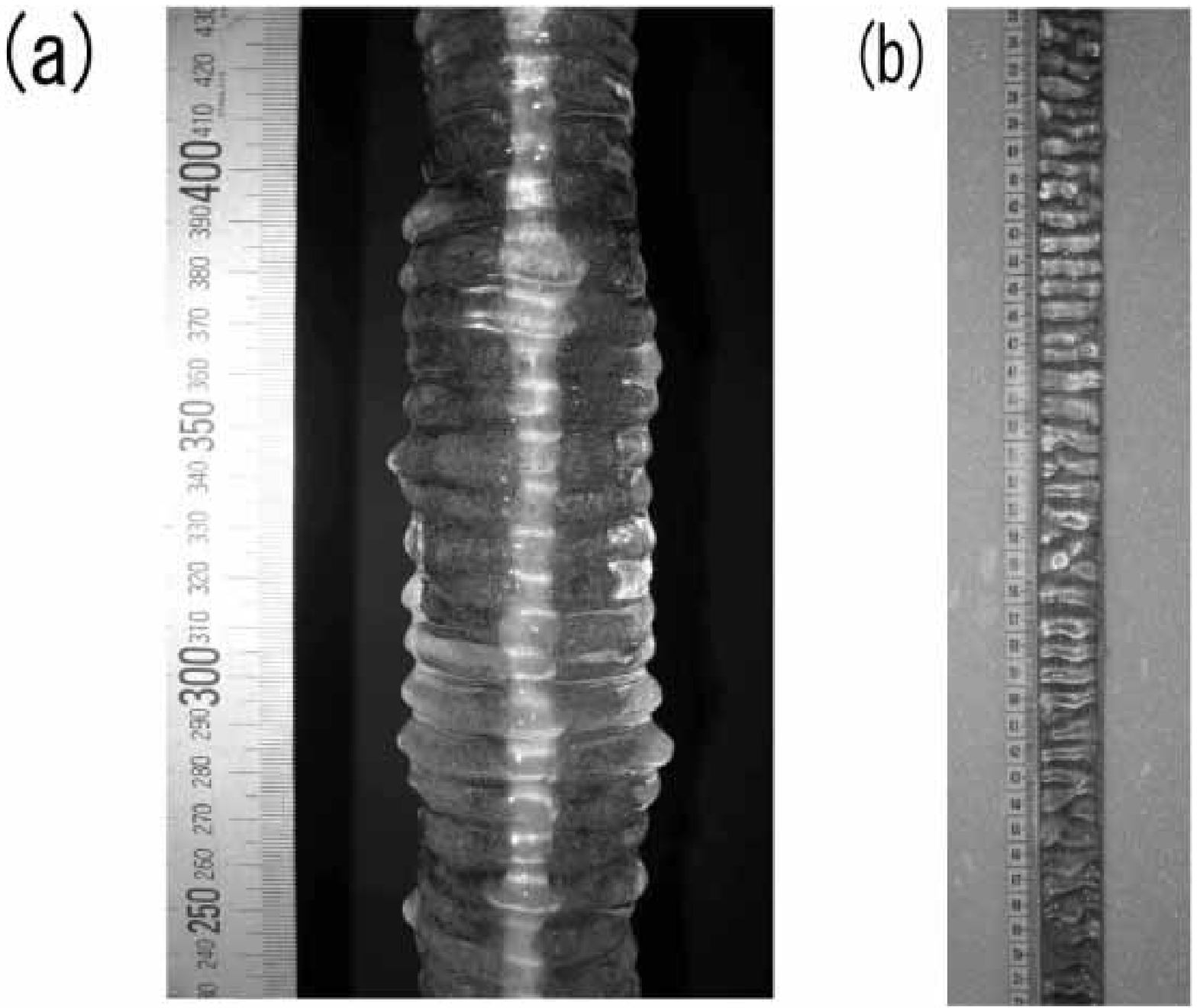}\\[5mm]
\includegraphics[width=15cm,height=15cm,keepaspectratio,clip]{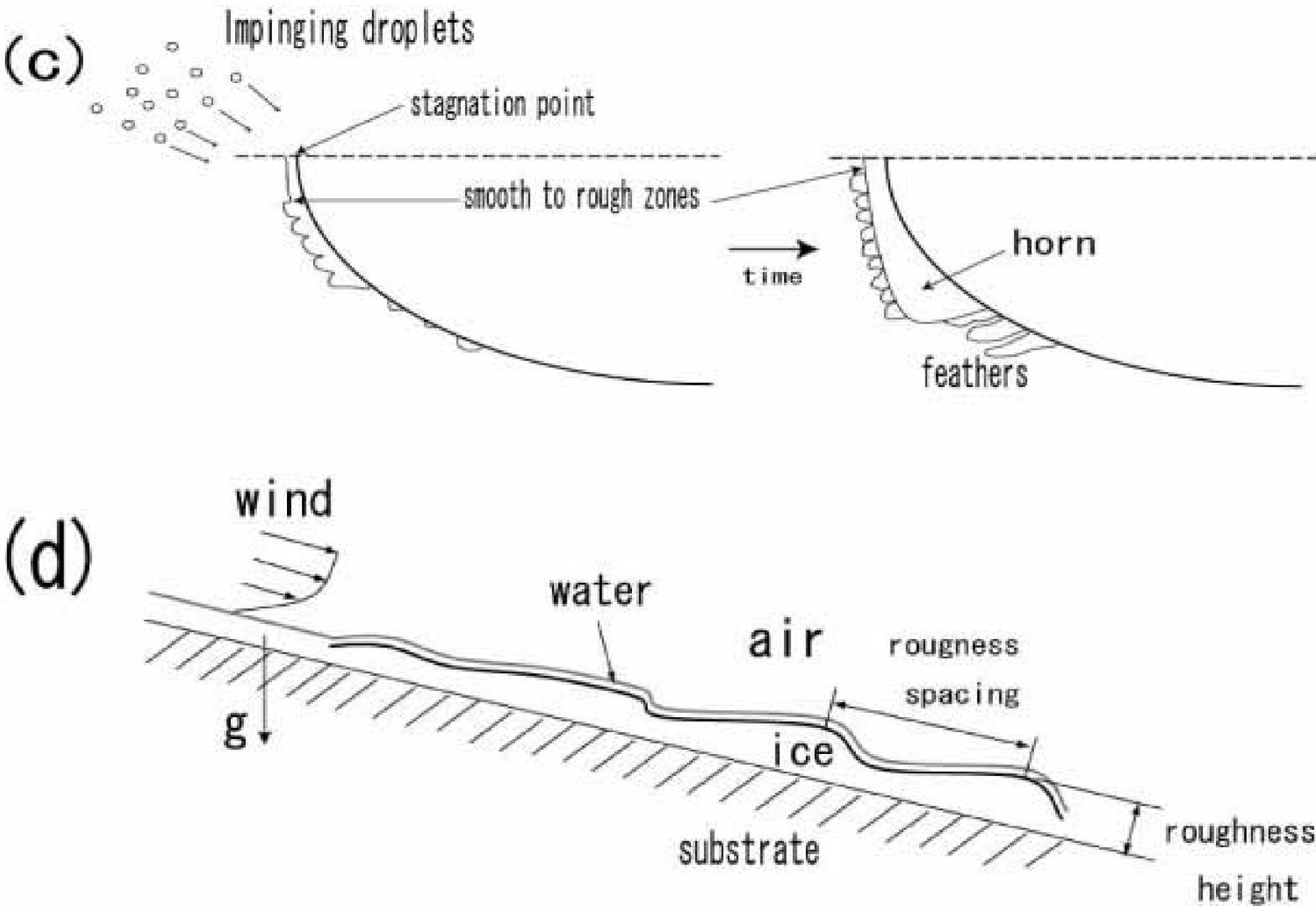}
\end{center}
\caption{
(a) Ice ripples on (a) a stick and (b) a plane. \cite{UFYT10} 
(c) Schematic of ice roughness formation on the parabolic leading-edge of a NACA 0012 airfoil in glaze icing conditions. \cite{Shin96}
(d) Schematic of an initial aufeis formation on a sloped surface. \cite{Streitz02}}
\label{fig:ripples}
\end{figure}

A variety of surface features of growing crystal under a thin layer of moving fluid, which separates the developing solid from the surrounding air, are observed in natural phenomena. \cite{Oron97, Meakin10} 
A first example is the ring-like ripples on the surface of icicles. \cite{Maeno94} A pattern similar to icicle ripples can be experimentally produced on the surface of a wooden round stick and that on a gutter on an inclined plane in a cold room, as shown in figures~\ref{fig:ripples} (a) and (b). \cite{Matsuda97, UFYT10} These ripples appear clearly when water dripping from the top of the stick and gutter spreads effectively and covers the entire ice surface uniformly. The latent heat from the ice-water interface to the environment through the water film must be released during the freezing process.
Consequently, a negative temperature gradient develops ahead of the growing ice beneath the water film and the water is a supercooled. \cite{Makkonen00} The spacing between ice ripples formed on the vertical stick and gutter was nearly 1-cm long, like natural icicle ripples. The wavelength of the ice ripples on the gutter decreases as the slope of the inclined plane increases. It increases only gradually as water supply rate increases, and the ripples move upwards very slightly with time. \cite{UFYT10, Chen11} 

As the second example, figure~\ref{fig:ripples} (c) is a schematic view of ice roughness formation on the parabolic leading edge of a NACA 0012 airfoil under glaze icing conditions (i.e. air temperature close to freezing and high liquid water content (LWC)), observed by Shin. \cite{Shin96} LWC is the mass of water contained in a unit volume of air. In glaze icing, the portion of the impinging water droplets that cannot be frozen runs off the surface due to gravity or wind drag. The latent heat released in the freezing process must be transferred from the ice-water interface through the unfrozen water film to the air. \cite{Makkonen00, Gent00} Shin defines roughness as surface irregularity growing on the top of macro-ice shape with horns and feathers. Smooth to rough zones in figure~\ref{fig:ripples} (c) is defined as the region where surface condition changes from a smooth to a rough one. Roughness size was measured for various airspeeds, air temperatures and LWCs. Roughness height increases with increasing air temperature and LWC, whereas airspeed has little effect on roughness height. Roughness spacing is of the order of a millimeter, decreases with increasing airspeed, and increases with increasing air temperature and LWC. The boundary between smooth and rough zones moves upstream towards stagnation region with time, as shown in figure~\ref{fig:ripples} (c).

For the third example, figure~\ref{fig:ripples} (d) is a schematic view of an initial aufeis (also referred to as icings) formation in frigid air when a shallow sheet of water, introduced at the upstream end of the wind tunnel, flows or trickles over a sloped frigid surface. \cite{Streitz02} Both gravity and wind drag drive the spreading of shallow flows of freezing water. The initial aufeis morphologies, characterized by wavelike or terraced forms, are shown in figure~\ref{fig:ripples} (d). Their roughness spacing and height was found to vary with slope and wind speed. As the slope increases, their roughness spacing decreases and roughness height increases. Moreover, their roughness spacing and height decrease as wind speed increases. \cite{Streitz02} 

Finally, travertine terracing is among the most spectacular geological phenomena on earth, not only in limestone caves and around hot springs, but also in streams and rivers in limestone terrain. The interactions between hydrodynamics, water chemistry, calcium carbonate precipitation and carbon dioxide degassing constitute a complex pattern formation of travertine terracing. \cite{Hammer10} The relationships between slope, discharge, terrace wavelength, depth and height are discussed by Pentecost: \cite{Pentecost05} Inter-dam distances increase with large discharge, where dam is defined as terraces that are filled with water, forming pools and lakes. Pools are shorter on steep slopes, instead height is larger. Interestingly, variations of the inter-dam distances and ice ripple wavelength with slope show the same trend (see Fig. 5 in Ref. \onlinecite{Hammer10} and Fig. 8 (a) in Ref. \onlinecite{UFYT10}), regardless of different crystals.

We can see common features for the surface roughness characteristics such as roughness spacing and height among phenomena mentioned above.
For the first example, a theoretical model of the origin of ripples on icicles was proposed, \cite{Ueno03, Ueno04, Ueno07, UF10} and the results were in good agreements with the experimental results. \cite{UFYT10, Chen11}
For the second example, in order to explain experimental results on glaze ice roughness diameters, accreted on NACA 0012 airfoil leading edges reported by Shin\cite{Shin96}, Tsao and Rothmayer developed a high Reynolds number triple-deck theory \cite{Schlichting99} to describe the interaction between the air boundary layer, water film and glaze ice sheet. \cite {Tsao98} A novel broad-band ice instability mode was found in regimes with simultaneous air and wall cooling, but there was no well-defined maximum amplification rate wavenumber (or equivalently, wavelength). To overcome this issue, the Gibbs-Thomson effect was introduced to stabilize the smallest scale icing disturbances. However, the length scale predicted by their theory was much smaller than the roughness spacing of the order of millimeters observed in the experiments of Shin. \cite{Tsao00} 
For the third and final examples, the quantitative morphology and size distribution of aufeis and travertine terraces as a function of parameters such as slope, water flux and airspeed have not yet been studied in detail.
Therefore, it is necessary to construct a comprehensive model to elucidate common features for the roughness characteristics described above. 

In this paper, in \ref{sec:model} we propose a theoretical model to explain the effects of water supply rate, plane slope and wind speed on the roughness spacing and height of an initial aufeis (icings), from the morphological instability of growing crystal under gravity and air shear stress driven water film. 
In a previous ice growth model, water is driven by gravity alone. \cite{UFYT10}
In another model, water is driven by air shear stress alone, \cite{UF11} but the model is valid on a horizontal surface. 
However, by combining the two driving forces, the resulting model becomes more complex than the previous ones because air and water flows and temperature fields are highly coupled with the water film thickness.   
Therefore, a numerical method is proposed to solve the governing equations for an air-water-ice multi-phase system.
In \ref{sec:RandD}, the experimental results concerning the roughness characteristics of the initial aufeis observed by Streitz and Ettema \cite{Streitz02} will be explained theoretically.
It will also be shown that the growth conditions of the ice-water interface disturbances are strongly affected by variable air stresses exerted on the water-air interface by the airflow. 
Crucial evidence of the importance of such air shear stress disturbances will be shown by comparing theoretically calculated amplification rates of the ice-water interface disturbance with the ice surface roughness heights observed by Streitz and Ettema.
Concluding remarks are made in \ref{sec:concl}.

\section{\label{sec:model}Model}

\begin{figure}
\begin{center}
\includegraphics[width=15cm,height=15cm,keepaspectratio,clip]{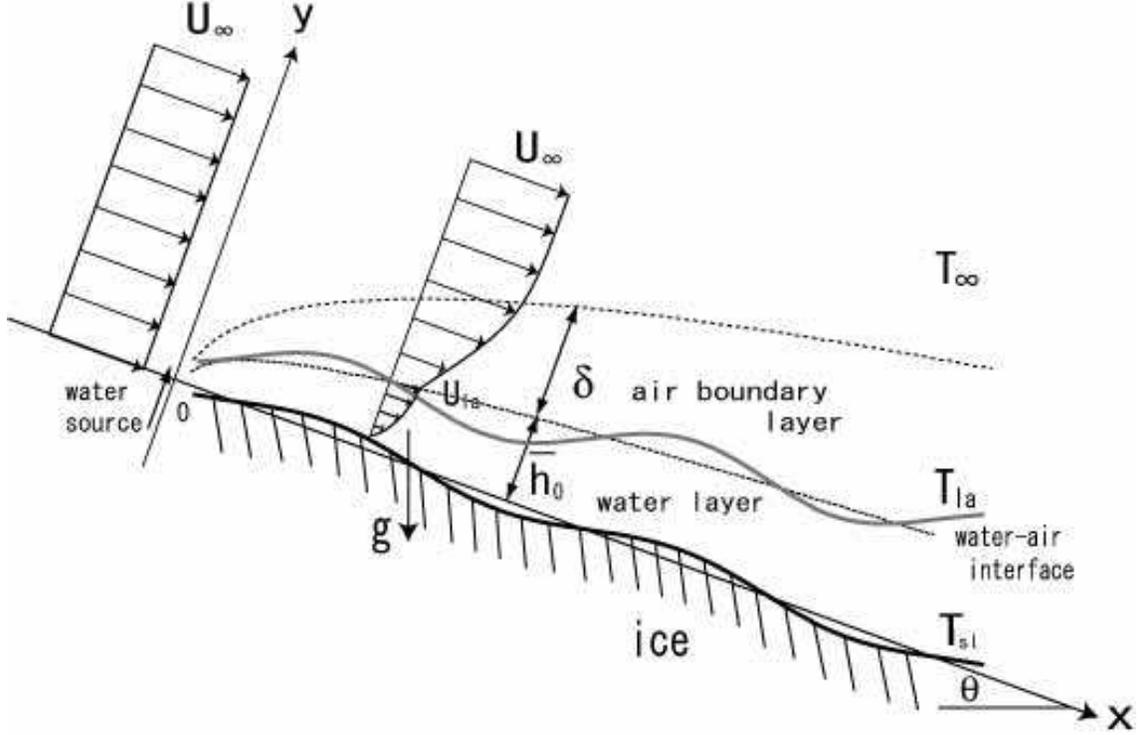}
\end{center}
\caption{Diagram of physical model and coordinate system (vertical height is not to scale).
The solid curves represent the locations of the disturbed ice-water and water-air interfaces.
The dashed curves represent the locations of the undisturbed water-air interface and air boundary layer thickness.}
\label{fig:ice-water-air}
\end{figure}
 
In the experiments of Streitz and Ettema \cite{Streitz02} for an initial formation of aufeis on an inclined plane shown in figure~\ref{fig:ripples} (d), a wind tunnel set up in a refrigerated laboratory was designed to tilt only downwards, allowing the water to flow as a thin sheet driven by gravity and wind drag. The water originates from a row of holes located at the top of the plane. The experiments was conducted with a water flow rate of about 1692 [(ml/h)/cm] at an initial water temperature of 0.1 $^{\circ}$C, in air at a temperature of -5 $^{\circ}$C. The initial formation of aufeis was defined as the initial layer of aufeis formed when shallow water first flows as a laminar sheet over a frigid surface and freezes onto it. The initial morphologies appeared essentially wavelike or terraced, and their spacing and height indicated in figure~\ref{fig:ripples} (c) were measured for plane slopes up to $15^{\circ}$ and wind speeds up to 48 km/h.

The current model configuration and coordinate system is shown in figure~\ref{fig:ice-water-air}, which is based on the laboratory experiments of Streitz and Ettema mentioned above. $x$ is the position along the inclined plane measured from the location of the water source, and $y$ is the position measured from a flat ice-water interface. $\theta$ is the angle of the inclined plane with respect to the horizontal. Air temperature is at $T_{\infty}$, which is lower than the temperature $T_{sl}$ at a flat ice-water interface. We assume a steady laminar airflow parallel to the $x$ axis with free stream velocity $u_{\infty}$. By releasing the latent heat to the air through the water-air interface at temperature $T_{la}$, ice grows from the portion of the supercooled water film driven by gravity and wind drag force exerted by the airflow. $\delta$ and $\bar{h}_{0}$ are the air boundary layer and water film thickness, respectively, and  $u_{la}$ is the water surface velocity. As shown in figure~\ref{fig:ice-water-air}, the water-air interface and flow and temperature in the air are disturbed due to a change in ice shape. As a result, the air shear stress exerted on the water-air interface and the heat transfer rate from the water-air interface to the air are variable. These in turn affect the flow and temperature distributions in the water layer. In response to this change, further ice-water interface development is determined. In this sense, the air boundary layer, the unfrozen water and the ice substrate become a complex air-water-ice multi-phase system.

The following assumptions are used in the present model:
(1) Water is discharged from only the top of the plane and there is no airborne water droplets impinging on the plane surface. 
(2) The free stream velocity $u_{\infty}$ is constant in space. 
(3) Density remains constant through the phase change. 
(4) Due to the long time scale of the ice-water interface motion, a quasi-stationary approximation is used for the disturbed fields, and unsteadiness only enters through the Stephan condition.
(5) Even in the presence of the undisturbed temperature gradient in ice, the morphological instability occurs when the ice thickness exceeds a critical thickness. As far as the ice thickness is large, it is a good approximation to neglect heat conduction into a substrate beneath the ice sheet. \cite{UF11}
(6) The presence of waves on the water film is ignored because the waves did not interact with the forming ice in any observable manner in the experiments, except for enhancing the spreading of the water over the aufeis surface. \cite{Streitz02} 
(7) Freshwater icing sponginess containing non-negligible amount of liquid water was observed in aufeis. \cite{Streitz02} The spongy ice formation, in which a portion of the surface liquid is incorporated into the ice matrix, is not considered.

\subsection{\label{sec:GovEq}Governing equations}

The following basic equations and boundary conditions governing the air-water-ice multi-phase system are based on previous papers \cite{UFYT10, UF11} and are reviewed here to ensure a relatively self-contained treatment.
The velocity components in the $x$ and $y$ directions in the air, $u_{a}$ and $v_{a}$, are governed by  
\begin{equation}
\frac{\partial u_{a}}{\partial t}
+u_{a}\frac{\partial u_{a}}{\partial x}
+v_{a}\frac{\partial u_{a}}{\partial y} 
=-\frac{1}{\rho_{a}}\frac{\partial p_{a}}{\partial x}
+\nu_{a}\left(\frac{\partial^{2}u_{a}}{\partial x^{2}}
+\frac{\partial^{2}u_{a}}{\partial y^{2}}\right),
\label{eq:geq-ua} 
\end{equation}

\begin{equation}
\frac{\partial v_{a}}{\partial t}
+u_{a}\frac{\partial v_{a}}{\partial x}
+v_{a}\frac{\partial v_{a}}{\partial y}
=-\frac{1}{\rho_{a}}\frac{\partial p_{a}}{\partial y}
+\nu_{a}\left(\frac{\partial^{2}v_{a}}{\partial x^{2}}
+\frac{\partial^{2}v_{a}}{\partial y^{2}}\right), 
\label{eq:geq-va}
\end{equation}

\begin{equation}
\frac{\partial u_{a}}{\partial x}+\frac{\partial v_{a}}{\partial y}=0,
\label{eq:continuity-air}
\end{equation}
where $p_{a}$ is the air pressure, $\rho_{a}=1.3$ ${\rm kg/m^{3}}$, the density of air, and $\nu_{a}=1.3 \times 10^{-5}$ ${\rm m^{2}/s}$, the kinematic viscosity of air. 
The velocity components in the $x$ and $y$ directions in the water layer, $u_{l}$ and $v_{l}$, are governed by 
\begin{equation}
\frac{\partial u_{l}}{\partial t}
+u_{l}\frac{\partial u_{l}}{\partial x}
+v_{l}\frac{\partial u_{l}}{\partial y} 
=-\frac{1}{\rho_{l}}\frac{\partial p_{l}}{\partial x}
+\nu_{l}\left(\frac{\partial^{2}u_{l}}{\partial x^{2}}
+\frac{\partial^{2}u_{l}}{\partial y^{2}}\right)
+g\sin\theta,
\label{eq:geq-ul} 
\end{equation}

\begin{equation}
\frac{\partial v_{l}}{\partial t}
+u_{l}\frac{\partial v_{l}}{\partial x}
+v_{l}\frac{\partial v_{l}}{\partial y}
=-\frac{1}{\rho_{l}}\frac{\partial p_{l}}{\partial y}
+\nu_{l}\left(\frac{\partial^{2}v_{l}}{\partial x^{2}}
+\frac{\partial^{2}v_{l}}{\partial y^{2}}\right)
-g\cos\theta, 
\label{eq:geq-vl}
\end{equation}

\begin{equation}
\frac{\partial u_{l}}{\partial x}+\frac{\partial v_{l}}{\partial y}=0,
\label{eq:continuity-water}
\end{equation}
where $\nu_{l}=1.8 \times 10^{-6}$ ${\rm m^{2}/s}$ and $\rho_{l}=1.0 \times 10^{3}$ ${\rm kg/m^{3}}$ are the kinematic viscosity and density of water, respectively, $p_{l}$ is the water pressure, $g$ is the gravitational acceleration, and $\theta$ is the plane angle.
The continuity equations (\ref{eq:continuity-air}) and (\ref{eq:continuity-water}) can be satisfied by introducing the stream functions $\psi_{a}$ and $\psi_{l}$ such that
$u_{a}=\partial \psi_{a}/\partial y$, 
$v_{a}=-\partial \psi_{a}/\partial x$,
$u_{l}=\partial \psi_{l}/\partial y$, and  
$v_{l}=-\partial \psi_{l}/\partial x$.

Both velocity components $u_{l}$ and $v_{l}$ at a disturbed ice-water interface, $y=\zeta(t,x)$, must satisfy the no-slip condition:
\begin{equation}
u_{l}|_{y=\zeta}=0,
\hspace{1cm}
v_{l}|_{y=\zeta}=0.
\label{eq:bc-ul-vl-zeta}
\end{equation}
Since there is no impingement of supercooled water droplets on the water film,
the kinematic condition at a disturbed water-air interface, $y=\xi(t,x)$, is 
\begin{equation}
\frac{\partial \xi}{\partial t}+u_{l}|_{y=\xi}\frac{\partial \xi}{\partial x}=v_{l}|_{y=\xi}.
\label{eq:bc-kinematic-xi}
\end{equation}
The continuity of velocities of water film flow and airflow at the water-air interface is 
\begin{equation}
u_{l}|_{y=\xi}=u_{a}|_{y=\xi},
\qquad
v_{l}|_{y=\xi}=v_{a}|_{y=\xi}.
\label{eq:bc-ul-vl-ua-va-xi}
\end{equation}
The continuity of tangential and normal stresses at the water-air interface is 
\begin{equation}
\mu_{l}\left(\frac{\partial u_{l}}{\partial y}\Big|_{y=\xi}
+\frac{\partial v_{l}}{\partial x}\Big|_{y=\xi}\right)
=\mu_{a}\left(\frac{\partial u_{a}}{\partial y}\Big|_{y=\xi}
+\frac{\partial v_{a}}{\partial x}\Big|_{y=\xi}\right),
\label{eq:bc-shear-stress-xi}
\end{equation}
\begin{equation}
-p_{a}|_{y=\xi}+2\mu_{a}\frac{\partial v_{a}}{\partial y}\Big|_{y=\xi}
-\left(-p_{l}|_{y=\xi}+2\mu_{l}\frac{\partial v_{l}}{\partial y}\Big|_{y=\xi}\right)
=-\gamma\frac{\partial^{2}\xi}{\partial x^{2}}\left[1+\left(\frac{\partial \xi}{\partial x}\right)^{2}\right]^{-3/2},
\label{eq:bc-normal-stress-xi}
\end{equation}
where $\mu_{l}=\rho_{l}\nu_{l}=1.8\times 10^{-3}$ $\rm N\,s/m^{2}$ and 
$\mu_{a}=\rho_{a}\nu_{a}=1.69\times 10^{-5}$ $\rm N\,s/m^{2}$ are the viscosities of water and air, respectively, and $\gamma=7.6 \times 10^{-2}$ N/m is the water-air surface tension. 

The equations for the temperatures in the air $T_{a}$, water $T_{l}$ and ice $T_{s}$ are 
\begin{equation}
\frac{\partial T_{a}}{\partial t}
+u_{a}\frac{\partial T_{a}}{\partial x}
+v_{a}\frac{\partial T_{a}}{\partial y}
=\kappa_{a}\left(\frac{\partial^{2} T_{a}}{\partial x^{2}}
+\frac{\partial^{2} T_{a}}{\partial y^{2}}\right),
\label{eq:geq-Ta}
\end{equation}

\begin{equation}
\frac{\partial T_{l}}{\partial t}
+u_{l}\frac{\partial T_{l}}{\partial x}
+v_{l}\frac{\partial T_{l}}{\partial y}
=\kappa_{l}\left(\frac{\partial^{2} T_{l}}{\partial x^{2}}+\frac{\partial^{2} T_{l}}{\partial y^{2}}\right),
\label{eq:geq-Tl}
\end{equation}

\begin{equation}
\frac{\partial T_{s}}{\partial t}
=\kappa_{s}\left(\frac{\partial^{2} T_{s}}{\partial x^{2}}
+\frac{\partial^{2} T_{s}}{\partial y^{2}}\right),
\label{eq:geq-Ts}
\end{equation}
where $\kappa_{a}=1.87 \times 10^{-5}$ ${\rm m^{2}/s}$, $\kappa_{l}=1.33 \times 10^{-7}$ ${\rm m^{2}/s}$ and  $\kappa_{s}=1.15 \times 10^{-6}$ ${\rm m^{2}/s}$ are the thermal diffusivities of air, water and ice, respectively. 

The continuity condition of temperature at the ice-water interface is 
\begin{equation}
T_{l}|_{y=\zeta}=T_{s}|_{y=\zeta}=T_{i},
\label{eq:Tsl}
\end{equation}
in which the interfacial temperature $T_{i}$ is an unknown to be determined.
The conventional Stefan problem cannot describe the pattern formation observed in nature. \cite{Langer80}
Likewise, all linear stability analyses based on the assumption that the temperature at the disturbed ice-water interface remains at the equilibrium freezing temperature, $T_{l}|_{y=\zeta}=T_{s}|_{y=\zeta}=T_{sl}$ ($T_{sl}=$0 $^{\circ}$C for pure water), showed that the ice-water interface disturbance becomes unstable for all wavenumbers, and that there is no dominant amplification rate to select a preferred wavelength. \cite{UFYT10, Tsao98, UF11} 
The Stephan condition is 
\begin{equation}
L\left(\bar{V}+\frac{\partial \zeta}{\partial t} \right)
=K_{s}\frac{\partial T_{s}}{\partial y}\Big|_{y=\zeta}
      -K_{l}\frac{\partial T_{l}}{\partial y}\Big|_{y=\zeta},
\label{eq:heatflux-zeta}
\end{equation}
where $L=3.3 \times 10^{8}$ ${\rm J/m^{3}}$ is the latent heat per unit volume, $\bar{V}$ is the undisturbed ice growth rate, and $K_{s}=2.22$ ${\rm J/(m\,K\,s)}$ and $K_{l}=0.56$ ${\rm J/(m\,K\,s)}$ are thermal conductivities of ice and water, respectively. 

The continuity condition of temperature at the water-air interface is
\begin{equation}
T_{l}|_{y=\xi}=T_{a}|_{y=\xi}=T_{la},
\label{eq:bc-Tla}
\end{equation}
where $T_{la}$ is a temperature at the water-air interface and will be determined later.
The continuity of heat flux at the water-air interface is
\begin{equation}
-K_{l}\frac{\partial T_{l}}{\partial y}\Big|_{y=\xi}
=-K_{a}\frac{\partial T_{a}}{\partial y}\Big|_{y=\xi},
\label{eq:heatflux-xi}
\end{equation}
where $K_{a}=0.024$ ${\rm J/(m\,K\,s)}$ is the thermal conductivity of air. 
Far away from the air boundary layer, the velocities and temperature asymptote to their far-field values:
\begin{equation}
u_{a}|_{y=\infty}=u_{\infty}, \qquad
v_{a}|_{y=\infty}=0, \qquad
T_{a}|_{y=\infty}=T_{\infty}.
\label{eq:bc-infinity}
\end{equation}

\subsection{\label{sec:LSA}Linear stability analysis}

Since most of the derivations of the stability analysis follow the same procedure given in Refs. \onlinecite{UFYT10}
and \onlinecite{UF11}, except for the modified undisturbed velocity profile in the water film and the boundary conditions of (\ref{eq:bc-basic_air}), (\ref{eq:bc-fa}), (\ref{eq:bc-shearstress}), (\ref{eq:bc-normalstress}), (\ref{eq:shearstress_a}), (\ref{eq:normalstress_a}) herein, full details will not be given. 

The field variables in ~\S\,\ref{sec:GovEq} are assumed to be decomposed into undisturbed and disturbed parts, as follows:
\begin{equation}
\left(
\begin{array}{c}
\zeta \\
\xi \\ 
\psi_{a} \\ 
\psi_{l} \\ 
p_{a} \\ 
p_{l} \\ 
T_{a} \\ 
T_{l} \\ 
T_{s}  
\end{array}
\right)
=
\left(
\begin{array}{c}
0 \\
\bar{h}_{0} \\ 
\bar{\psi}_{a} \\ 
\bar{\psi}_{l} \\ 
\bar{p}_{a} \\ 
\bar{p}_{l} \\ 
\bar{T}_{a} \\ 
\bar{T}_{l} \\ 
\bar{T}_{s}  
\end{array}
\right)
+
\left(
\begin{array}{c}
\zeta_{k} \\
\xi_{k} \\ 
u_{\infty}f_{a}(\eta)\xi_{k} \\ 
u_{la}f_{l}(y_{*})\zeta_{k} \\
(\rho_{a}u_{\infty}^{2}/\delta_{0})g_{a}(\eta)\xi_{k} \\ 
(\rho_{l}u_{la}^{2}/\bar{h}_{0})g_{l}(y_{*})\zeta_{k} \\ 
H_{a}(\eta)\bar{G}_{a}\xi_{k} \\ 
H_{l}(y_{*})\bar{G}_{l}\zeta_{k} \\ 
H_{s}(y_{*})\bar{G}_{l}\zeta_{k} 
\end{array}
\right)
\exp[\sigma t+i kx].
\label{eq:pertset}
\end{equation}
A simple normal-mode analysis is applied to the ice-water interface disturbance $\zeta$ and the corresponding fields variables 
$(\xi', \psi'_{a}, \psi'_{l}, p'_{a}, p'_{l}, T'_{a}, T'_{l}, T'_{s})$, which are the disturbed part in (\ref{eq:pertset}).
Here 
$\bar{h}_{0}$ is the undisturbed water film thickness, 
$u_{la}$ is the surface velocity of the water film driven by gravity and air shear stress,
$\delta_{0}=(2\nu_{a}x/u_{\infty})^{1/2}$ is a scaled measure in the air, \cite{Schlichting99}
$u_{\infty}$ is the free stream velocity, 
$x$ is the distance from the leading edge where water is supplied,
$\eta=(y-\bar{h}_{0})/\delta_{0}$,
$y_{*}=y/\bar{h}_{0}$,
$\bar{G}_{a} \equiv -\partial \bar{T}_{a}/\partial y|_{y=\bar{h}_{0}}$,
$\bar{G}_{l} \equiv -\partial \bar{T}_{l}/\partial y|_{y=0}$,
and  
$\zeta_{k}$ and $\xi_{k}$ are the amplitudes of the ice-water interface and water-air interface, respectively.
$(f_{a}, g_{a}, H_{a})$ and $(f_{l}, g_{l}, H_{l}, H_{s})$ are dimensionless functions with respect to $\eta$ and $y_{*}$, respectively. $k$ is the wavenumber and $\sigma=\sigma^{(r)}+i \sigma^{(i)}$, $\sigma^{(r)}$ and $v_{p} \equiv -\sigma^{(i)}/k$ are the amplification rate and phase velocity of the disturbance, respectively. 
Since the undisturbed part of heat conduction in the ice is assumed to be zero, $\bar{T}_{s}=T_{sl}$ is used throughout this paper. 

\subsubsection{Governing equations for undisturbed 
and disturbed parts of flow and temperature in the air}

Substituting 
$\psi_{a}=\bar{\psi}_{a}+\psi'_{a}
=u_{\infty}\delta_{0}\bar{F}_{a}(\eta)+u_{\infty}f_{a}(\eta)\xi_{k}{\rm exp}[\sigma t+ikx]$ and
$T_{a}=\bar{T}_{a}+T'_{a}
=\bar{T}_{a}+H_{a}(\eta)\bar{G}_{a}\xi_{k}{\rm exp}[\sigma t+ikx]$
into the complete equations (\ref{eq:geq-ua}), (\ref{eq:geq-va}) and (\ref{eq:geq-Ta}),
a set of dimensionless differential equations for the undisturbed part 
$\bar{F}_{a}$, 
$\bar{T}_{a*}=(\bar{T}_{a}-T_{\infty})/(T_{la}-T_{\infty})$ and
for the disturbed part
$f_{a}$, 
$H_{a}$ 
are obtained: 
\begin{equation}
\frac{d^{3}\bar{F}_{a}}{d\eta^{3}}
=-\bar{F}_{a}\frac{d^{2}\bar{F}_{a}}{d\eta^{2}},
\label{eq:geq-basicFa}
\end{equation}
\begin{equation}
\frac{d^{2}\bar{T}_{a*}}{d\eta^{2}}
=-Pr_{a}\bar{F}_{a}\frac{d\bar{T}_{a*}}{d\eta},
\label{eq:geq-basicTa}
\end{equation}
\begin{eqnarray}
\frac{d^{4}f_{a}}{d\eta^4}
&=&-\bar{F}_{a}\frac{d^{3}f_{a}}{d\eta^{3}}
+\left\{2k_{a*}^{2}-(2-i k_{a*}\Rey_{a})\frac{d\bar{F}_{a}}{d\eta}\right\}\frac{d^{2}f_{a}}{d\eta^{2}} 
\nonumber \\
&& +\left\{k_{a*}^{2}\left(\bar{F}_{a}+2\eta\frac{d\bar{F}_{a}}{d\eta}\right)
-\frac{d^{2}\bar{F}_{a}}{d\eta^{2}}\right\}\frac{df_{a}}{d\eta}
-\left\{k_{a*}^{4}+i k_{a*}\Rey_{a}\left(k_{a*}^{2}\frac{d\bar{F}_{a}}{d\eta}+\frac{d^{3}\bar{F}_{a}}{d\eta^{3}}\right)\right\}f_{a}, \nonumber \\
\label{eq:geq-fa} 
\end{eqnarray}
\begin{eqnarray}
\frac{d^{2}(\bar{G}_{a*}H_{a})}{d\eta^{2}}
&=&-Pr_{a}\bar{F}_{a}\frac{d(\bar{G}_{a*}H_{a})}{d\eta}
+\left\{k_{a*}^{2}+Pr_{a}(-1+i k_{a*}\Rey_{a})\frac{d\bar{F}_{a}}{d\eta}\right\}(\bar{G}_{a*}H_{a}) \nonumber \\ 
&& -i k_{a*}Pr_{a}\Rey_{a}\frac{d\bar{T}_{a*}}{d\eta}f_{a},
\label{eq:geq-Ha}
\end{eqnarray}
where $\Rey_{a}=u_{\infty}\delta_{0}/\nu_{a}$ and $Pr_{a}=\nu_{a}/\kappa_{a}$ are the Reynolds number and the Prandtl number of air, respectively,  
$k_{a*}=k\delta_{0}$ is the dimensionless wavenumber normalized by the length $\delta_{0}$,
and $\bar{G}_{a*}\equiv -d\bar{T}_{a*}/d\eta|_{\eta=0}$. 

The undisturbed part of (\ref{eq:bc-ul-vl-ua-va-xi}), 
$\bar{u}_{a}|_{y=\infty}=\partial\bar{\psi}_{a}/\partial y|_{y=\infty}=u_{\infty}$,
$\bar{T}_{a}|_{y=\bar{h}_{0}}=T_{la}$ and $\bar{T}_{a}|_{y=\infty}=T_{\infty}$, 
the disturbed part of (\ref{eq:bc-ul-vl-ua-va-xi}),
$u'_{a}|_{y=\infty}=\partial \psi'_{a}/\partial y|_{y=\infty}=0$,
$v'_{a}|_{y=\infty}=-\partial \psi'_{a}/\partial x|_{y=\infty}=0$, 
the disturbed part of (\ref{eq:bc-Tla}) and $T_{a}'|_{y=\infty}=0$ 
yield the following boundary conditions for
$\bar{F}_{a}$, 
$\bar{T}_{a*}$,
$f_{a}$ and 
$H_{a}$, respectively,
\begin{eqnarray}
\frac{d\bar{F}_{a}}{d\eta}\Big|_{\eta=0}&=&\frac{u_{la}}{u_{\infty}}, \qquad
\bar{F}_{a}|_{\eta=0}=0, \qquad
\frac{d\bar{F}_{a}}{d\eta}\Big|_{\eta=\infty}=1, \nonumber \\
&&\bar{T}_{a*}|_{\eta=0}=1, \qquad
\bar{T}_{a*}|_{\eta=\infty}=0,
\label{eq:bc-basic_air}
\end{eqnarray} 
\begin{eqnarray}
\frac{df_{a}}{d\eta}\Big|_{\eta=0}
=\left(-1+\frac{\mu_{a}}{\mu_{l}}\right)\frac{d^{2}\bar{F}_{a}}{d\eta^{2}}\Big|_{\eta=0}
-\frac{\delta_{0}}{\bar{h}_{0}}\frac{u_{la}}{u_{\infty}}
\frac{df_{l}}{dy_{*}}\Big|_{y_{*}=1}\Big/f_{l}|_{y_{*}=1}, 
\nonumber \\
f_{a}|_{\eta=0}=-\frac{u_{la}}{u_{\infty}}f_{l}|_{y_{*}=1}, \qquad
\frac{df_{a}}{d\eta}\Big|_{\eta=\infty}=0, \qquad
f_{a}|_{\eta=\infty}=0,
\label{eq:bc-fa}
\end{eqnarray}
\begin{equation}
H_{a}|_{\eta=0}=1, \qquad
H_{a}|_{\eta=\infty}=0.
\label{eq:bc-Ha}
\end{equation}

\subsubsection{Governing equations for undisturbed 
and disturbed parts of flow and temperature in the water film}

The undisturbed part of (\ref{eq:heatflux-xi}) yields \cite{UF10, UF11}
\begin{equation}
T_{sl}-T_{la}
=-\frac{K_{a}}{K_{l}}\frac{\bar{h}_{0}}{(\delta_{0}/\bar{G}_{a*})}T_{\infty}.
\label{eq:Tla}
\end{equation}
Substituting (\ref{eq:Tla}) into the undisturbed part of (\ref{eq:heatflux-zeta}), 
the undisturbed ice growth rate is obtained: \cite{UF10, UF11}
\begin{equation}
\bar{V}=-\frac{K_{a}T_{\infty}}{L(\delta_{0}/\bar{G}_{a*})}.
\label{eq:V}
\end{equation}
The length $\delta_{0}/\bar{G}_{a*}$ in (\ref{eq:Tla}) and (\ref{eq:V}) is regarded as the air boundary layer thickness $\delta$ in figure~\ref{fig:ice-water-air}. 

Substituting 
$\psi_{l}=\bar{\psi}_{l}+\psi'_{l}
=u_{la}\bar{h}_{0}\bar{F}_{l}(y_{*})+u_{la}f_{l}(y_{*})\zeta_{k}{\rm exp}[\sigma t+ikx]$ and
$T_{l}=\bar{T}_{l}+T'_{l}
=\bar{T}_{l}+H_{l}(y_{*})\bar{G}_{l}\zeta_{k}{\rm exp}[\sigma t+ikx]$
into the complete equations (\ref{eq:geq-ul}), (\ref{eq:geq-vl}) and (\ref{eq:geq-Tl}),
a set of dimensionless differential equations for the undisturbed part 
$\bar{u}_{l*}\equiv\bar{u}_{l}/u_{la}=d\bar{F}_{l}/dy_{*}$, 
$\bar{T}_{l*}=(\bar{T}_{l}-T_{sl})/(T_{sl}-T_{la})$ and
for the disturbed part
$f_{l}$,
$H_{l}$ 
are obtained: 
\begin{equation}
\frac{d^{2}\bar{u}_{l*}}{dy_{*}^{2}}
=-\frac{g\bar{h}_{0}^{2}\sin\theta}{\nu_{l}u_{la}},
\label{eq:geq-basicFl}
\end{equation}
\begin{equation}
\frac{d^{2}\bar{T}_{l*}}{dy_{*}^{2}}
=0,
\label{eq:geq-basicTl}
\end{equation}
\begin{equation}
\frac{d^{4}f_{l}}{dy_{*}^{4}}
=\left(2k_{l*}^{2}+ik_{l*} \Rey_{l}\bar{u}_{l*}\right)\frac{d^{2}f_{l}}{dy_{*}^{2}}
-\left\{k_{l*}^{4}+ik_{l*} \Rey_{l}\left(k_{l*}^{2}\bar{u}_{l*}+\frac{d^{2}\bar{u}_{l*}}{dy_{*}^{2}}\right)\right
\}f_{l},
\label{eq:geq-fl}
\end{equation}
\begin{equation}
\frac{d^{2}(\bar{G}_{l*}H_{l})}{dy_{*}^{2}}
=\left(k_{l*}^{2}+ik_{l*} \Pec_{l}\bar{u}_{l*}\right)(\bar{G}_{l*}H_{l}) 
-ik_{l*} \Pec_{l}\frac{d\bar{T}_{l*}}{d y_{*}}f_{l},
\label{eq:geq-Hl}
\end{equation}
where $\Rey_{l}=u_{la}\bar{h}_{0}/\nu_{l}$ and $\Pec_{l}=u_{la}\bar{h}_{0}/\kappa_{l}$ 
are the Reynolds number and ${\rm P\acute{e}clet}$ number of water, respectively,
$k_{l*}=k\bar{h}_{0}$ is the dimensionless wavenumber normalized by the length $\bar{h}_{0}$,
and $\bar{G}_{l*}\equiv -d\bar{T}_{l*}/dy_{*}|_{y_{*}=0}$.
When deriving (\ref{eq:geq-basicTl}), $\bar{h}_{0}/(T_{sl}-T_{la})d(T_{sl}-T_{la})/dx=d\bar{h}_{0}/dx-\bar{h}_{0}/(2x)$
is used, which is obtained by differentiating (\ref{eq:Tla}) with respect to $x$. 
Equations (\ref{eq:geq-basicFl})--(\ref{eq:geq-Hl}) are finally obtained by neglecting the term with $d\bar{h}_{0}/dx \ll 1$ and $\bar{h}_{0}/x\ll 1$.
This is in agreement with the more usual lubrication approach.
\cite{Oron97, Myers02_1, Myers02_2}
Using the boundary conditions 
$\bar{u}_{l}|_{y=0}=0$, 
$\mu_{l}\partial\bar{u}_{l}/\partial y|_{y=\bar{h}_{0}}
=\mu_{a}\partial\bar{u}_{a}/\partial y|_{y=\bar{h}_{0}}$,
$\bar{T}_{l}|_{y=0}=T_{sl}$
and 
$\bar{T}_{l}|_{y=\bar{h}_{0}}=T_{la}$, 
the solutions of the dimensionless undisturbed velocity and temperature profiles in the water film are given by
\begin{equation}
\bar{u}_{l*}=
-\frac{g\bar{h}_{0}^{2}\sin\theta}{2\nu_{l}u_{la}}y_{*}^{2}
+\left(\frac{g\bar{h}_{0}^{2}\sin\theta}{\nu_{l}u_{la}}
+\frac{\mu_{a}u_{\infty}\bar{h}_{0}}{\mu_{l}u_{la}\delta_{0}}\frac{d^{2}\bar{F_{a}}}{d\eta^{2}}\Big|_{\eta=0}\right)y_{*}, \qquad
\bar{T}_{l*}=-y_{*}.
\label{eq:Ul-Tl}
\end{equation}

Linearization of the disturbed part of (\ref{eq:bc-ul-vl-zeta}) at $y=0$, as well as 
(\ref{eq:bc-shear-stress-xi}), (\ref{eq:bc-normal-stress-xi}), (\ref{eq:bc-Tla}) and 
(\ref{eq:heatflux-xi}) at $y=\bar{h}_{0}$
yield the boundary conditions for $f_{l}$ and $H_{l}$:
\begin{equation}
\frac{df_{l}}{dy_{*}}\Big|_{y_{*}=0}+\frac{d\bar{u}_{l*}}{dy_{*}}\Big|_{y_{*}=0}=0, \qquad
f_{l}|_{y_{*}=0}=0,
\label{eq:bc-fl0-dfl0} 
\end{equation}
\begin{equation}
\frac{d^{2}f_{l}}{dy_{*}^{2}}\Big|_{y_{*}=1}
+\left(k_{l*}^{2}-\frac{d^{2}\bar{u}_{l*}}{dy_{*}^{2}}\Big|_{y_{*}=1}+\Sigma_{a}\right)f_{l}|_{y_{*}=1}=0, 
\label{eq:bc-shearstress}
\end{equation}
\begin{eqnarray}
\frac{d^{3}f_{l}}{dy_{*}^{3}}\Big|_{y_{*}=1}
-\left(3k_{l*}^{2}+ik_{l*} \Rey_{l}\right)\frac{df_{l}}{dy_{*}}\Big|_{y_{*}=1} 
\nonumber \\
+ik_{l*}\Rey_{l}
\left(\frac{d\bar{u}_{l*}}{dy_{*}}\Big|_{y_{*}=1}
+\frac{\cos\theta}{Fr^{2}}+Wek_{l*}^{2}+\Pi_{a}\right)f_{l}|_{y_{*}=1}=0,
\label{eq:bc-normalstress}
\end{eqnarray}
\begin{equation}
H_{l}|_{y_{*}=1}+f_{l}|_{y_{*}=1}=0,
\label{eq:Tla-xi}
\end{equation}
\begin{equation}
\frac{dH_{l}}{dy_{*}}\Big|_{y_{*}=1}
-\frac{\bar{h}_{0}}{\delta_{0}}\left(-\frac{dH_{a}}{d\eta}\Big|_{\eta=0}\right)f_{l}|_{y_{*}=1}=0.
\label{eq:heatflux-xi-h0}
\end{equation}
Here, $\Sigma_{a}$ in (\ref{eq:bc-shearstress}) and $\Pi_{a}$ in (\ref{eq:bc-normalstress})
are defined by
\begin{equation}
\Sigma_{a}=\frac{\mu_{a}}{\mu_{l}}\frac{u_{\infty}}{u_{la}}\left(\frac{\bar{h}_{0}}{\delta_{0}}\right)^{2}
\left(\frac{d^{2}f_{a}}{d\eta^{2}}\Big|_{\eta=0}+k_{a*}^{2}f_{a}|_{\eta=0}\right),
\label{eq:shearstress_a}
\end{equation}
\begin{eqnarray}
\Pi_{a}&=&-\frac{\rho_{a}}{\rho_{l}}\left(\frac{u_{\infty}}{u_{la}}\right)^{2}\frac{\bar{h}_{0}}{\delta_{0}}
\frac{1+ik_{a*}\Rey_{a}}{1+(k_{a*}\Rey_{a})^{2}} \nonumber \\
&& \times
\left[
\frac{d^{3}f_{a}}{d\eta^{3}}\Big|_{\eta=0}
-\left\{3k_{a*}^{2}+(-1+ik_{a*}\Rey_{a})\frac{d\bar{F}_{a}}{d\eta}\Big|_{\eta=0}\right\}
\frac{df_{a}}{d\eta}\Big|_{\eta=0} 
\right. \nonumber \\
&&\left.
+ik_{a*}\Rey_{a}\frac{d^{2}\bar{F}_{a}}{d\eta^{2}}\Big|_{\eta=0}f_{a}|_{\eta=0}
\right], 
\label{eq:normalstress_a}
\end{eqnarray}
where
$Fr=u_{la}/(g\bar{h}_{0})^{1/2}$ is the Froude number, and 
$We=\gamma/(\rho_{l}u_{la}^{2}\bar{h}_{0})$ is the Weber number.
It should be noted that the disturbed part of the water flow is affected by the airflow
through the terms $\Sigma_{a}$ in (\ref{eq:bc-shearstress}) and $\Pi_{a}$ in (\ref{eq:bc-normalstress}), which are hereafter referred to as the tangential and normal air shear stress disturbances, respectively. 

Using (\ref{eq:Ul-Tl}) and 
$u'_{l}=\partial \psi'_{l}/\partial y$ ,  
the volumetric water flow rate per width is given by
\begin{eqnarray}
Q/l_{w}=\int_{\zeta}^{\xi}
(\bar{u}_{l}+u'_{l})dy 
&&=\frac{g\bar{h}_{0}^{3}\sin\theta}{3\nu_{l}}
+\frac{\mu_{a}u_{\infty}\bar{h}_{0}^{2}}{2\mu_{l}\delta_{0}}\frac{d^{2}\bar{F}_{a}}{d\eta^{2}}\Big|_{\eta=0}
\nonumber \\
&&+u_{la}\bar{h}_{0}(\xi_{k}+f_{l}|_{y_{*}=1}\zeta_{k})\exp[\sigma t+i kx].
\label{eq:Qoverl-0}
\end{eqnarray}
From the disturbed part of (\ref{eq:bc-kinematic-xi}), the relation between the amplitude of the water-air interface, $\xi_{k}$, and that of the ice-water interface, $\zeta_{k}$ is obtained:
$\xi_{k}=-f_{l}|_{y_{*}=1}\zeta_{k}$.
Therefore, (\ref{eq:Qoverl-0}) can be written as
\begin{equation}
Q/l_{w}=\frac{g\sin\theta}{3\nu_{l}}\bar{h}_{0}^{3}
+\frac{\mu_{a}u_{\infty}}{2\mu_{l}\delta_{0}}\frac{d^{2}\bar{F}_{a}}{d\eta^{2}}\Big|_{\eta=0}\bar{h}_{0}^{2}.
\label{eq:Qoverl}
\end{equation}
Applying the definition $\bar{u}_{l}|_{y=\bar{h}_{0}}=u_{la}$ (or equivalently, $\bar{u}_{l*}|_{y_{*}=1}=1$) to (\ref{eq:Ul-Tl}), 
the value of $u_{la}$ is determined from
\begin{equation}
u_{la}=\frac{g\sin\theta}{2\nu_{l}}\bar{h}_{0}^{2}
+\frac{\mu_{a}u_{\infty}}{\mu_{l}\delta_{0}}\frac{d^{2}\bar{F_{a}}}{d\eta^{2}}\Big|_{\eta=0}\bar{h}_{0}.
\label{eq:ula}
\end{equation}

Linearizing the temperature at the ice-water interface in (\ref{eq:Tsl}), $T_{i}$ can be written as $T_{i}=T_{sl}+\Delta T_{sl}$, where $T_{sl}$ is the temperature at an undisturbed ice-water interface
and $\Delta T_{sl}$ is a deviation from it when the ice-water interface is disturbed.
Substituting 
$T'_{l}=H_{l}(y_{*})\bar{G}_{l}\zeta_{k}{\rm exp}[\sigma t+ikx]$ and
$T'_{s}=H_{s}(y_{*})\bar{G}_{l}\zeta_{k}{\rm exp}[\sigma t+ikx]$
into the disturbed part of (\ref{eq:Tsl}) and (\ref{eq:heatflux-zeta}),
the dimensionless temperature deviation at the ice-water interface, $\Delta T_{sl*}\equiv\Imag[\Delta T_{sl}/(T_{sl}-T_{la})]$,
the dimensionless amplification rate, $\sigma_{*}^{(r)}\equiv \sigma^{(r)}/(\bar{V}/\bar{h}_{0})$,
the dimensionless phase velocity, $v_{p*}\equiv -\sigma^{(i)}/(k\bar{V})$
are determined as follows: 
\begin{equation}
\Delta T_{sl*}
= \delta_{b}(t_{*})\left\{(H_{l}^{(r)}|_{y_{*}=0}-1)\sin[k_{l*}(x_{*}-v_{p*}t_{*})] 
+H_{l}^{(i)}|_{y_{*}=0}\cos[k_{l*}(x_{*}-v_{p*}t_{*})]\right\},
\label{eq:DeltaTsl}
\end{equation}
\begin{equation}
\sigma_{*}^{(r)}=-\frac{dH_{l}^{(r)}}{dy_{*}}\Big|_{y_{*}=0}+K^{s}_{l}k_{l*}(H_{l}^{(r)}|_{y_{*}=0}-1),
\label{eq:amp}
\end{equation}
\begin{equation}
v_{p*}=-\frac{1}{k_{l*}}\left(-\frac{dH_{l}^{(i)}}{dy_{*}}\Big|_{y_{*}=0}+K^{s}_{l}k_{l*}H_{l}^{(i)}|_{y_{*}=0}\right),
\label{eq:phasevel}
\end{equation}
where $\Imag$ denotes the imaginary part of its argument, $H_{l}^{(r)}$ and $H_{l}^{(i)}$ are the real and imaginary parts of $H_{l}$, and $K^{s}_{l}=K_{s}/K_{l}=3.96$ is the ratio of the thermal conductivity of ice to that of water,  
$x_{*}=x/\bar{h}_{0}$, 
$t_{*}=(\bar{V}/\bar{h}_{0})t$, 
$\delta_{b}(t_{*})=\exp(\sigma_{*}^{(r)}t_{*})\delta_{b}$ and
$\delta_{b}=\zeta_{k}/\bar{h}_{0}$.
It should be noted that $\Delta T_{sl*} \rightarrow 0$ in the limit $k_{l*} \rightarrow 0$, and $\Delta T_{sl*}$ for a finite $k_{l*}$ varies because the disturbed temperature distribution in the water layer, $H_{l}$, is affected by both air and water flows (see Ref. \onlinecite{UF11} for more details).

\subsubsection{Numerical procedure}

Since the airflow was not considered in a previous paper, \cite{UFYT10}
$\bar{h}_{0}$ is determined
by the gravity-driven part in (\ref{eq:Qoverl}), 
$(Q/l_{w})_{g}\equiv g\bar{h}_{0}^{3}\sin\theta/(3\nu_{l})$.
This yields
\begin{equation}
\bar{h}_{0}=\left[\frac{3\nu_{l}}{g\sin\theta}\left(\frac{Q}{l_{w}}\right)_{g}\right]^{1/3},
\label{eq:gravity-h0}
\end{equation}
and $u_{la}=g\bar{h}_{0}^{2}\sin\theta/(2\nu_{l})$ from (\ref{eq:ula}), 
and then $\bar{u}_{l*}=-y_{*}^{2}+2y_{*}$ is the half-parabolic form from (\ref{eq:Ul-Tl}).
Hence, the values 
$d\bar{u}_{l*}/dy_{*}|_{y_{*}=0}=2$, 
$d\bar{u}_{l*}/dy_{*}|_{y_{*}=1}=0$ and 
$d^{2}\bar{u}_{l*}/dy_{*}^{2}|_{y_{*}=1}=-2$ 
are obtained.
Noting that
$\Sigma_{a}=0$ in (\ref{eq:bc-shearstress}) and 
$\Pi_{a}=0$ in (\ref{eq:bc-normalstress}) in the absence of airflow, 
and that the term $k_{l*}\Rey_{l}(\cos\theta/Fr^{2}+Wek_{l*}^{2})$ in (\ref{eq:bc-normalstress}) is equivalent to the parameter $\alpha$ of (22) in Ref. \onlinecite{UFYT10},
the boundary conditions (\ref{eq:bc-fl0-dfl0}) (\ref{eq:bc-shearstress}) and 
(\ref{eq:bc-normalstress})
reduce to (23) in the previous paper. \cite{UFYT10}
$\alpha$ was the parameter characterizing the effect of gravity and surface tension on the water-air surface.
Furthermore, in the absence of airflow (\ref{eq:geq-Ha}) herein can be written as $d^{2}H_{a}/d\eta^{2}=k_{a*}^{2}H_{a}$.
Its solution is $H_{a}={\rm e}^{-k_{a*}\eta}$ with the boundary conditions (\ref{eq:bc-Ha}), which yields
$\bar{h}_{0}/\delta_{0}(-dH_{a}/d\eta|_{\eta=0})=k_{l*}$. Hence, the boundary condition (\ref{eq:heatflux-xi-h0}) reduces to (33) in the previous paper. \cite{UFYT10}

In another previous paper, \cite{UF11} since water flows on a horizontal ice surface, $\bar{h}_{0}$ is determined
by the shear-driven part in (\ref{eq:Qoverl}), 
$(Q/l_{w})_{s}\equiv \mu_{a}u_{\infty}d^{2}\bar{F}_{a}/d\eta^{2}|_{\eta=0}\bar{h}_{0}^{2}/(2\mu_{l}\delta_{0})$. 
$\bar{h}_{0}$ can be expressed as
\begin{equation}
\bar{h}_{0}=\left[\frac{2\mu_{l}\delta_{0}}{\mu_{a}u_{\infty}\frac{d^{2}\bar{F}_{a}}{d\eta^{2}}\Big|_{\eta=0}}\left(\frac{Q}{l_{w}}\right)_{s}\right]^{1/2},
\label{eq:shear-h0}
\end{equation}
and
$u_{la}=\mu_{a}u_{\infty}d^{2}\bar{F}_{a}/d\eta^{2}|_{\eta=0}\bar{h}_{0}/(\mu_{l}\delta_{0})$ 
from (\ref{eq:ula}), and then
$\bar{u}_{l*}=y_{*}$ is the linear form from (\ref{eq:Ul-Tl}).
Hence, the values 
$d\bar{u}_{l*}/dy_{*}|_{y_{*}=0}=1$, 
$d\bar{u}_{l*}/dy_{*}|_{y_{*}=1}=1$ and 
$d^{2}\bar{u}_{l*}/dy_{*}^{2}|_{y_{*}=1}=0$ 
are obtained.

\begin{figure}
\begin{center}
\includegraphics[width=5cm,height=5cm,keepaspectratio,clip]{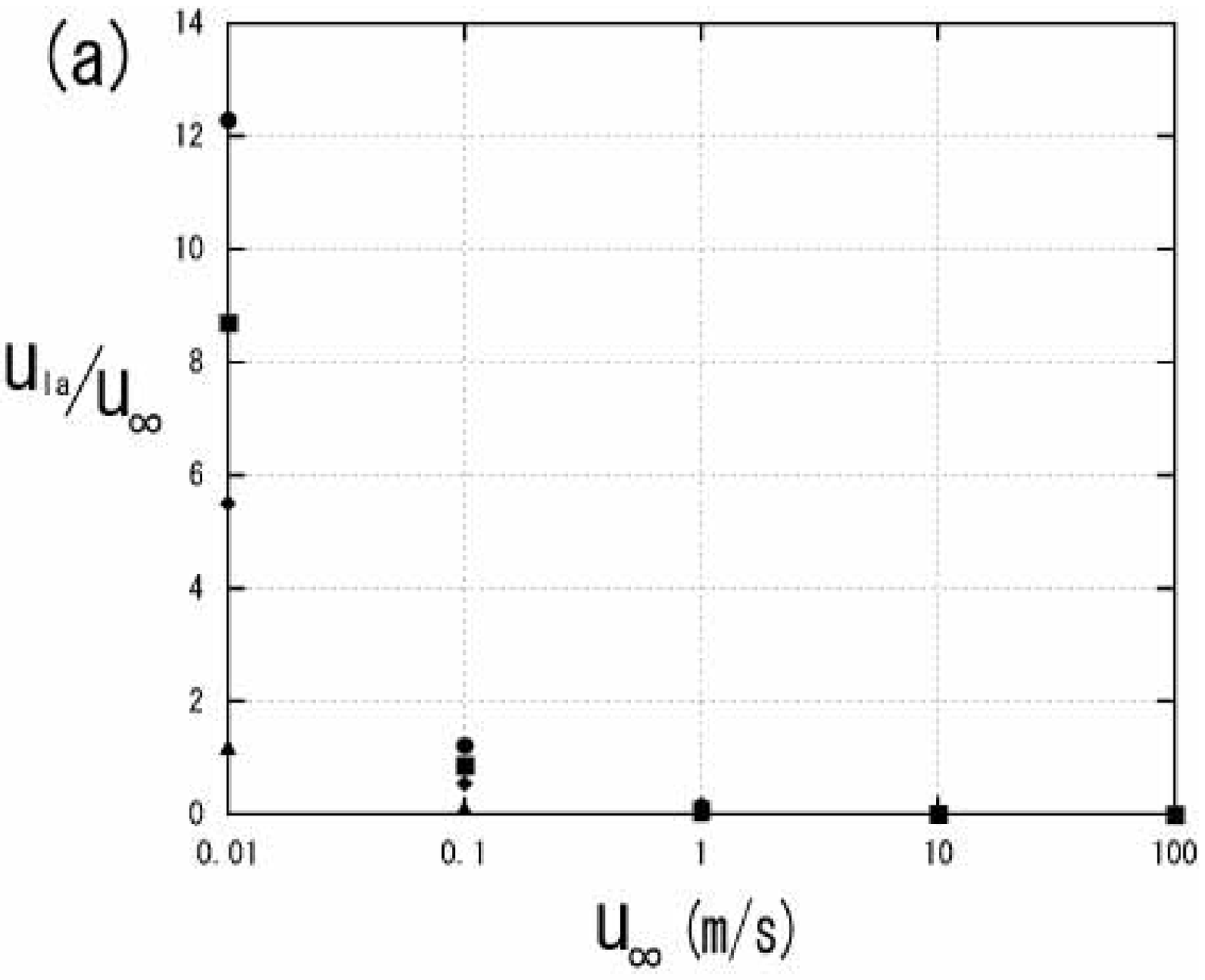}\hspace{3mm}
\includegraphics[width=5cm,height=5cm,keepaspectratio,clip]{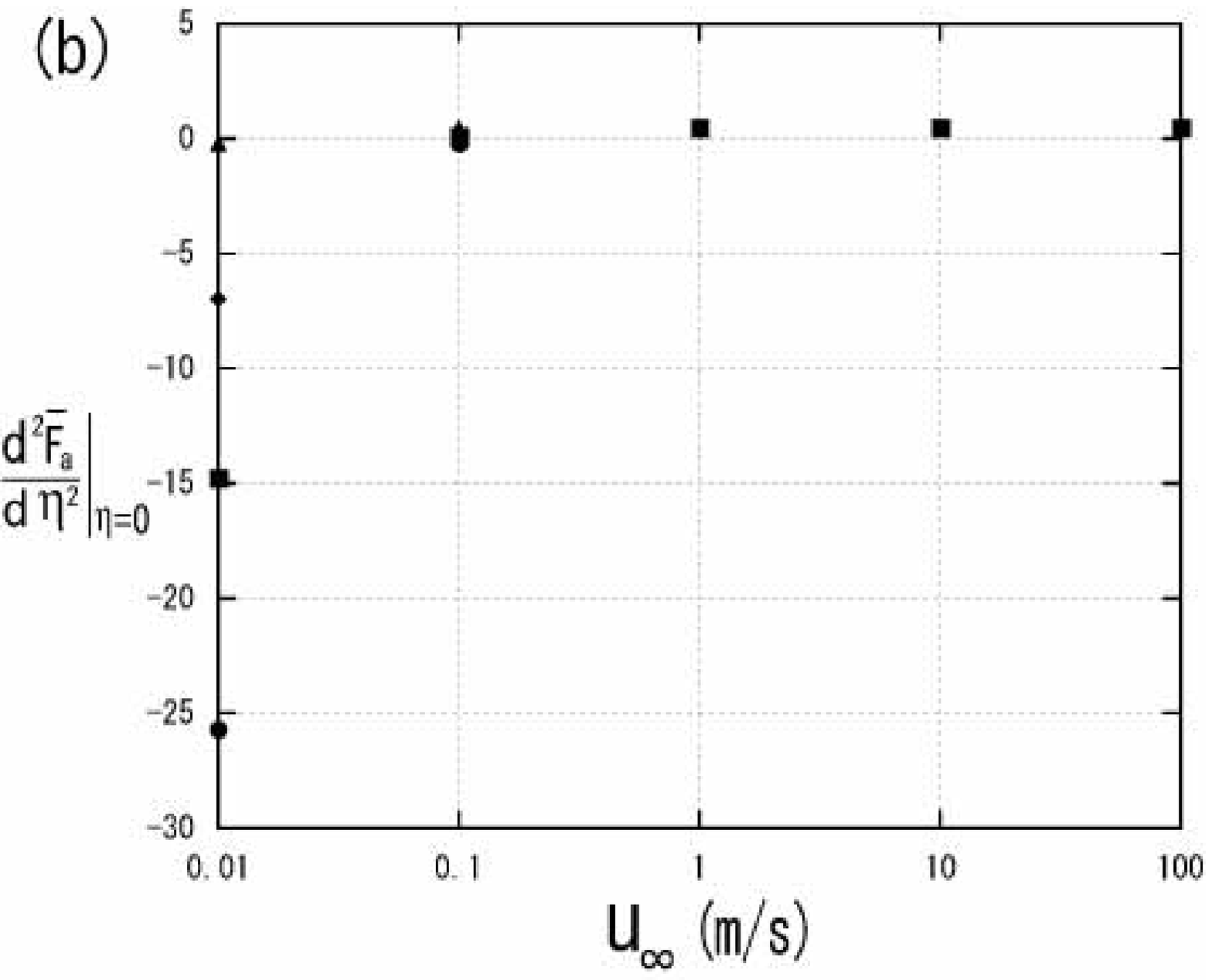}\hspace{3mm}
\includegraphics[width=5cm,height=5cm,keepaspectratio,clip]{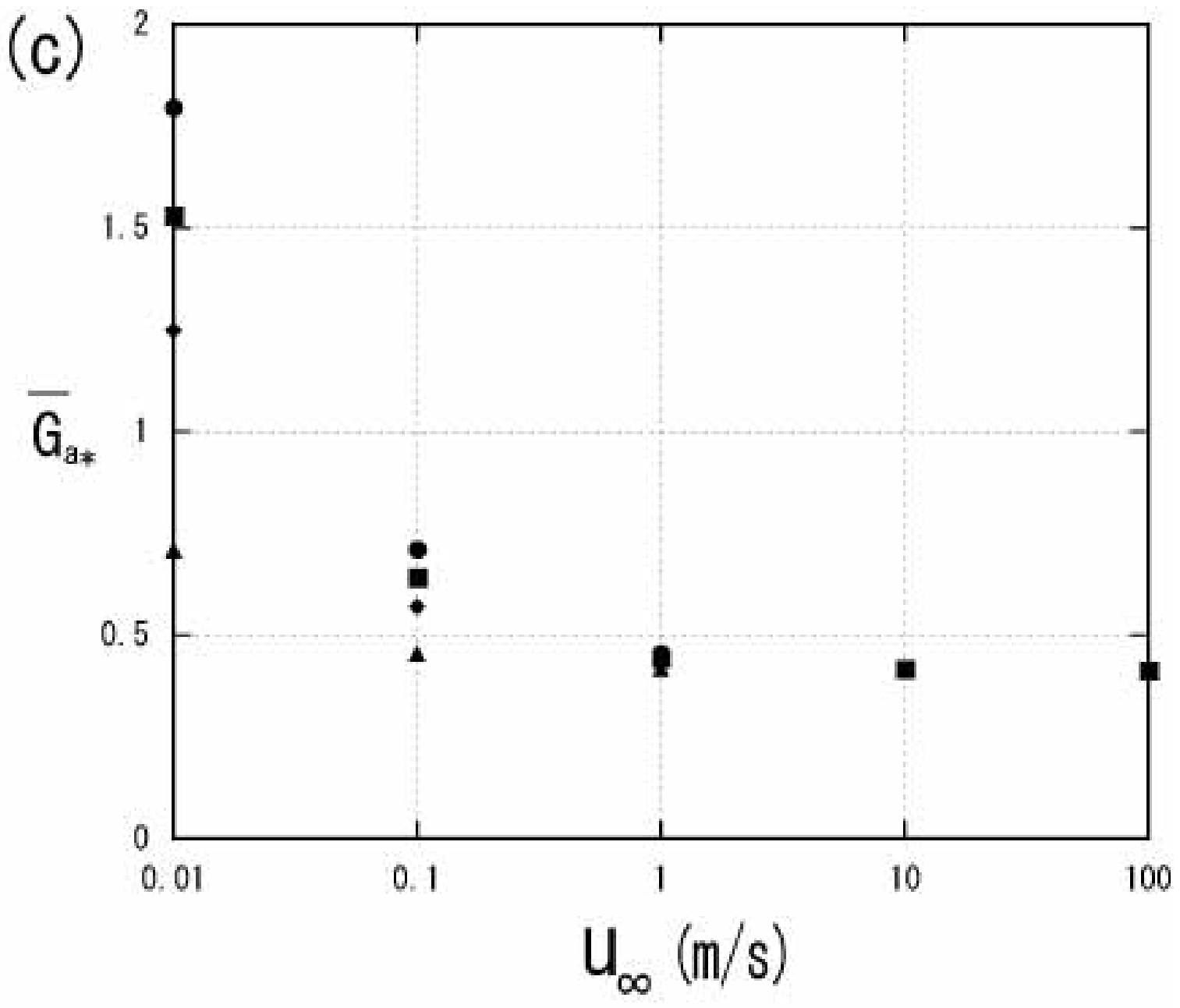}
\end{center}
\caption{Variation of 
(a) $u_{la}/u_{\infty}$,
(b) $d^{2}\bar{F}_{a}/d\eta^{2}|_{\eta=0}$ and
(c) $\bar{G}_{a*}=-d\bar{T}_{a*}/d\eta|_{\eta=0}$ with $u_{\infty}$
for $Q/l_{w}=1692 (\bullet)$, 
1000 $(\blacksquare)$, 
500 $(\blacklozenge)$ and 
160/3 $(\blacktriangle)$ [(ml/h)/cm].}
\label{fig:uinf-dFa-d2Fa-Ga}
\end{figure}

As $u_{\infty}$ increases, the ratio of $u_{la}$ to $u_{\infty}$ approaches 0,
as shown in figure~\ref{fig:uinf-dFa-d2Fa-Ga} (a) for various $Q/l_{w}$.
Then the first equation of (\ref{eq:bc-basic_air}) and the first and second equations of (\ref{eq:bc-fa}) can be approximated as
$d\bar{F}_{a}/d\eta|_{\eta=0}=0$, 
$df_{a}/d\eta|_{\eta=0}=-d^{2}\bar{F}_{a}/d\eta^{2}|_{\eta=0}$ 
and $f_{a}|_{\eta=0}=0$, using the fact that the viscosity ratio of air to water is very small, $\mu_{a}/\mu_{l} \ll 1$.
These conditions are equivalent to  
$\bar{u}_{a}|_{y=\bar{h}_{0}}=0$, $u'_{a}|_{y=\bar{h}_{0}}=0$ and $v'_{a}|_{y=\bar{h}_{0}}=0$, respectively,
which means that the air effectively sees the water as a rigid body.
Accordingly, 
the boundary conditions (\ref{eq:bc-fl0-dfl0}), (\ref{eq:bc-shearstress}) and 
(\ref{eq:bc-normalstress}) with (\ref{eq:shearstress_a}) and (\ref{eq:normalstress_a}) 
reduce to boundary conditions (39), (40) and (41) with (42) and (43) in the previous paper. \cite{UF11}
Furthermore, as $u_{\infty}$ increases, the values of 
$d^{2}\bar{F}_{a}/d\eta^{2}|_{\eta=0}$ and
$\bar{G}_{a*}=-d\bar{T}_{a*}/d\eta|_{\eta=0}$ 
converge to 0.47 and 0.41, respectively, for various $Q/l_{w}$, 
as shown in figures~\ref{fig:uinf-dFa-d2Fa-Ga} (b) and (c).
Then the profiles $\bar{u}_{a*}$ and $\bar{T}_{a*}$ are independent of the parameters $Q/l_{w}$, $\theta$, $u_{\infty}$ and $x$ and become similarity solutions for large $u_{\infty}$. \cite{Schlichting99}
When $u_{la}/u_{\infty} \ll 1$, the solutions in the air are determined independently of the solutions in the water film. Hence, once $\bar{F}_{a}$ is obtained, $\bar{h}_{0}$ is determined from (\ref{eq:shear-h0}).
On the other hand, in order to solve the governing equations for the water flow, the solutions of the airflow are necessary, as indicated in the terms $\Sigma_{a}$ and $\Pi_{a}$. 

By combining the two driving forces, the situation becomes more complex, as explained below.
The air-water-ice multi-phase system considered here consists of (\ref{eq:geq-basicFa})-(\ref{eq:geq-Ha}) 
with boundary conditions (\ref{eq:bc-basic_air})-(\ref{eq:bc-Ha}) in the air,
and (\ref{eq:geq-fl}) and (\ref{eq:geq-Hl}) with (\ref{eq:Ul-Tl}) and the boundary conditions
(\ref{eq:bc-fl0-dfl0})-(\ref{eq:heatflux-xi-h0}) in the water film, as well as a cubic equation (\ref{eq:Qoverl}) for $\bar{h}_{0}$.
For a given $Q/l_{w}$, $\theta$, $u_{\infty}$ and $x$, the value of $\bar{h}_{0}$ 
is numerically determined from (\ref{eq:Qoverl}). 
However, the value $d^{2}\bar{F}_{a}/d\eta^{2}|_{\eta=0}$ is needed. 
When $u_{\infty}$ is less than about 1 m/s, the term $u_{la}/u_{\infty}$ 
in (\ref{eq:bc-basic_air}) and (\ref{eq:bc-fa}) cannot be neglected. 
Furthermore, the first equation in (\ref{eq:bc-fa}) includes $df_{l}/dy_{*}|_{y*=1}/f_{l}|_{y_{*}=1}$, 
whereas $\Sigma_{a}$ in (\ref{eq:bc-shearstress}) and 
$\Pi_{a}$ in (\ref{eq:bc-normalstress}) need the solutions $\bar{F}_{a}$ and $f_{a}$
as indicated in (\ref{eq:shearstress_a}) and (\ref{eq:normalstress_a}).
Since the solutions in the air and in the water film and $\bar{h}_{0}$ are coupled, it is impossible to solve the current system simultaneously. 

To overcome this difficulty,
the system is solved with the following iterative method.
First, the temporal values of 
$\bar{h}_{0}$ and $df_{l}/dy_{*}|_{y*=1}/f_{l}|_{y_{*}=1}$ are set.
Then, the system is solved, except for
(\ref{eq:Qoverl}) and the first equation in (\ref{eq:bc-fa}).
From these solutions and the excluded equations,
new values of $\bar{h}_{0}$ and $df_{l}/dy_{*}|_{y*=1}/f_{l}|_{y_{*}=1}$ 
are obtained. Then, the system is solved again.
After some iterations, $\bar{h}_{0}$ and $df_{l}/dy_{*}|_{y*=1}/f_{l}|_{y_{*}=1}$ settle to constant values,
and the solutions of whole system is finally obtained.
Substituting solution $H_{l}$ into (\ref{eq:amp}) and (\ref{eq:phasevel}) and replacing $k_{l*}$ by $(\bar{h}_{0}/\delta_{0})k_{a*}$, $\sigma_{*}^{(r)}$ and $v_{p*}$ are presented with respect to $k_{a*}$.
 
The variation of the wavelength $\lambda$ of ice ripples formed on an inclined plane for various slope angles $\theta$ of Fig. 8 in the previous paper \cite{UFYT10} was obtained in the absence of airflow. It was confirmed that the current system reproduced the same variation of $\lambda$ with $\theta$ at the limit of low wind speed, for example, $u_{\infty}=0.01$ (m/s), and for the same water supply rate of $Q/l_{w}=160/3$ [(ml/h)/cm] as used in the previous paper. \cite{UFYT10}
The case of $u_{\infty}=0$ is excluded in the current system 
because $\delta_{0}=(2\nu_{a}x/u_{\infty})^{1/2}$ diverges. 

\section{\label{sec:RandD}Results and discussion}

\subsection{\label{sec:h0-parameters}Variation of thickness $\bar{h}_{0}$ and 
surface velocity $u_{la}$ with $\theta$, $u_{\infty}$ and $Q/l_{w}$}

\begin{figure}
\begin{center}
\includegraphics[width=5cm,height=5cm,keepaspectratio,clip]{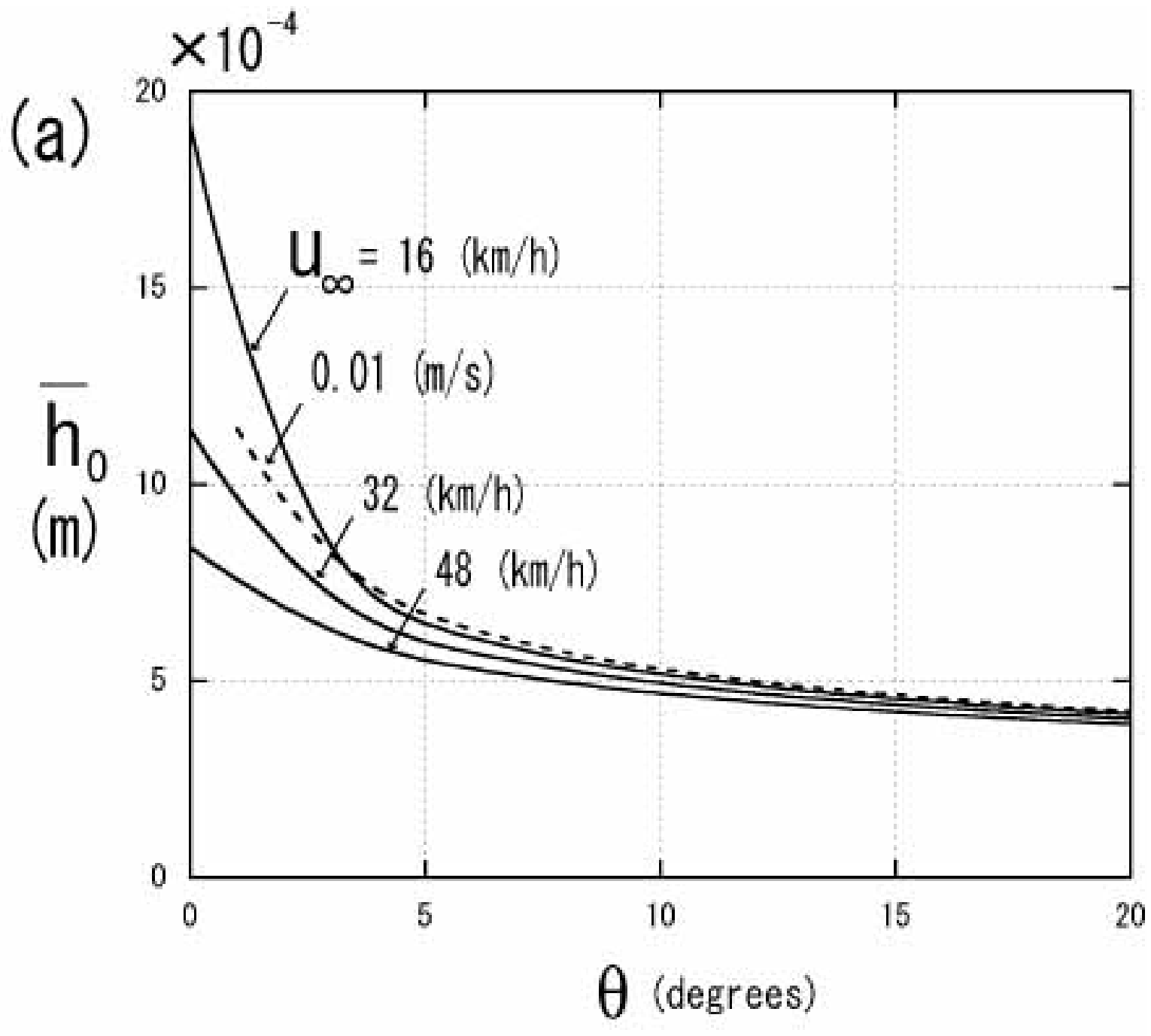}
\includegraphics[width=5cm,height=5cm,keepaspectratio,clip]{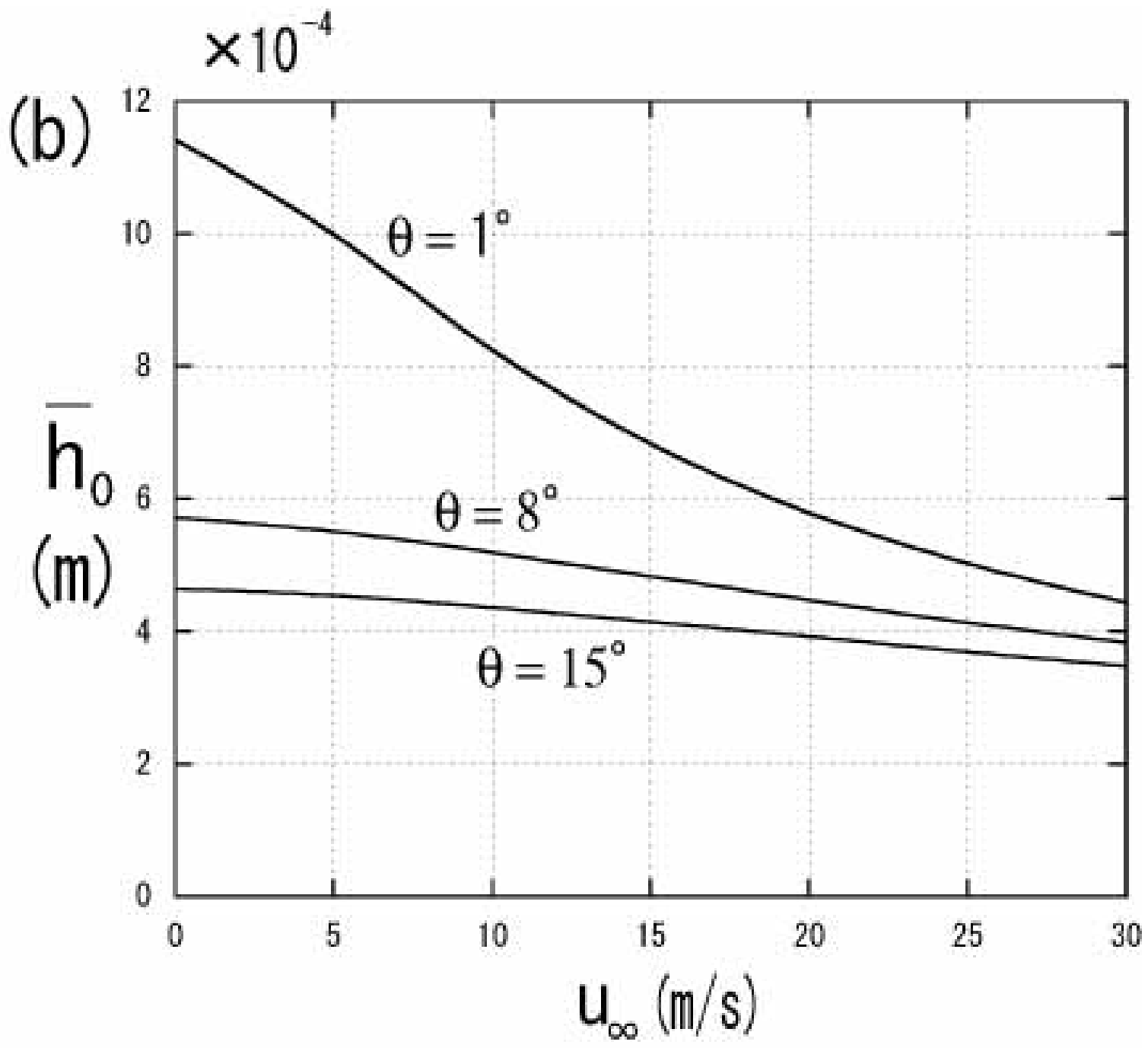}
\includegraphics[width=5cm,height=5cm,keepaspectratio,clip]{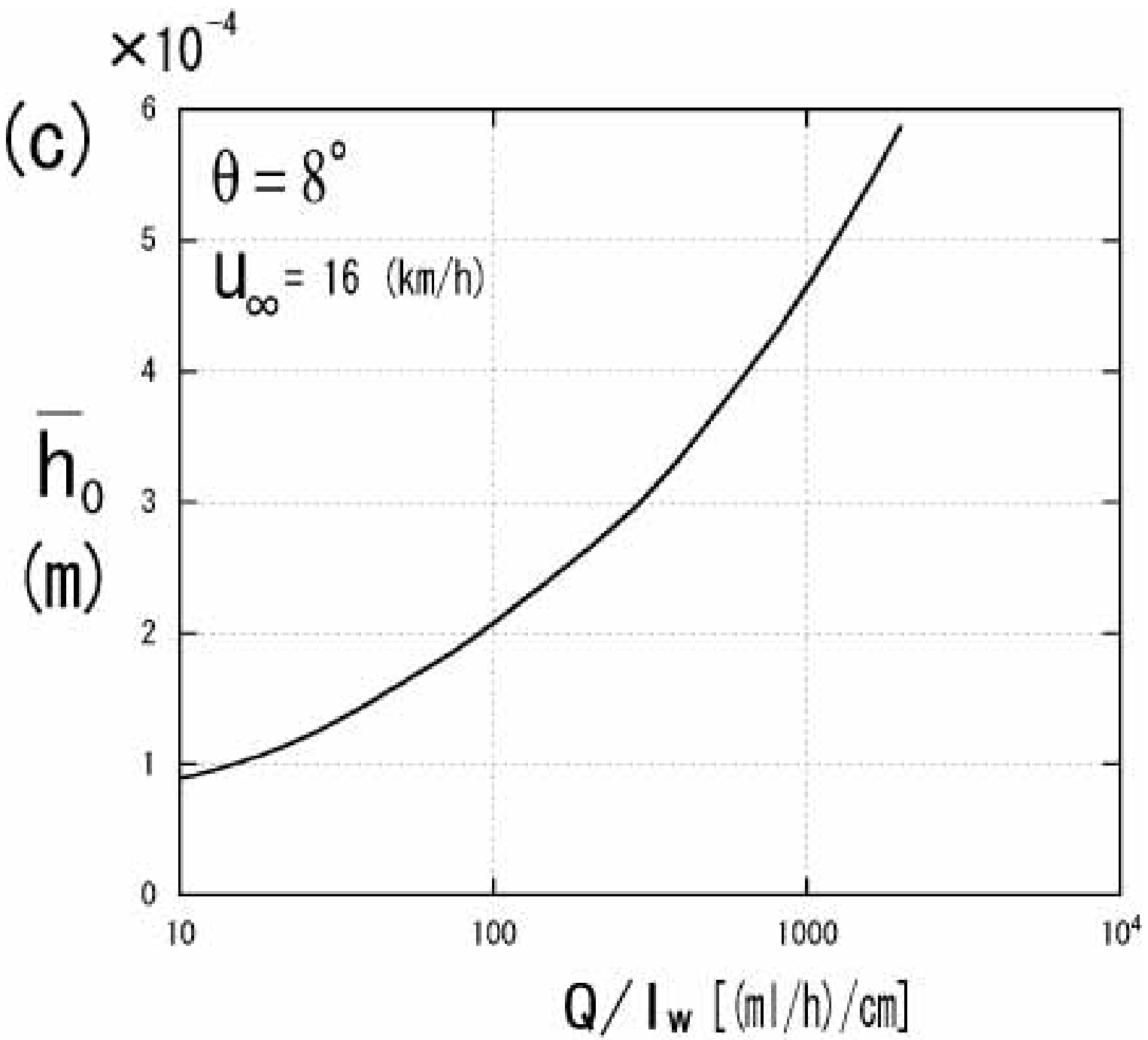}
\includegraphics[width=5cm,height=5cm,keepaspectratio,clip]{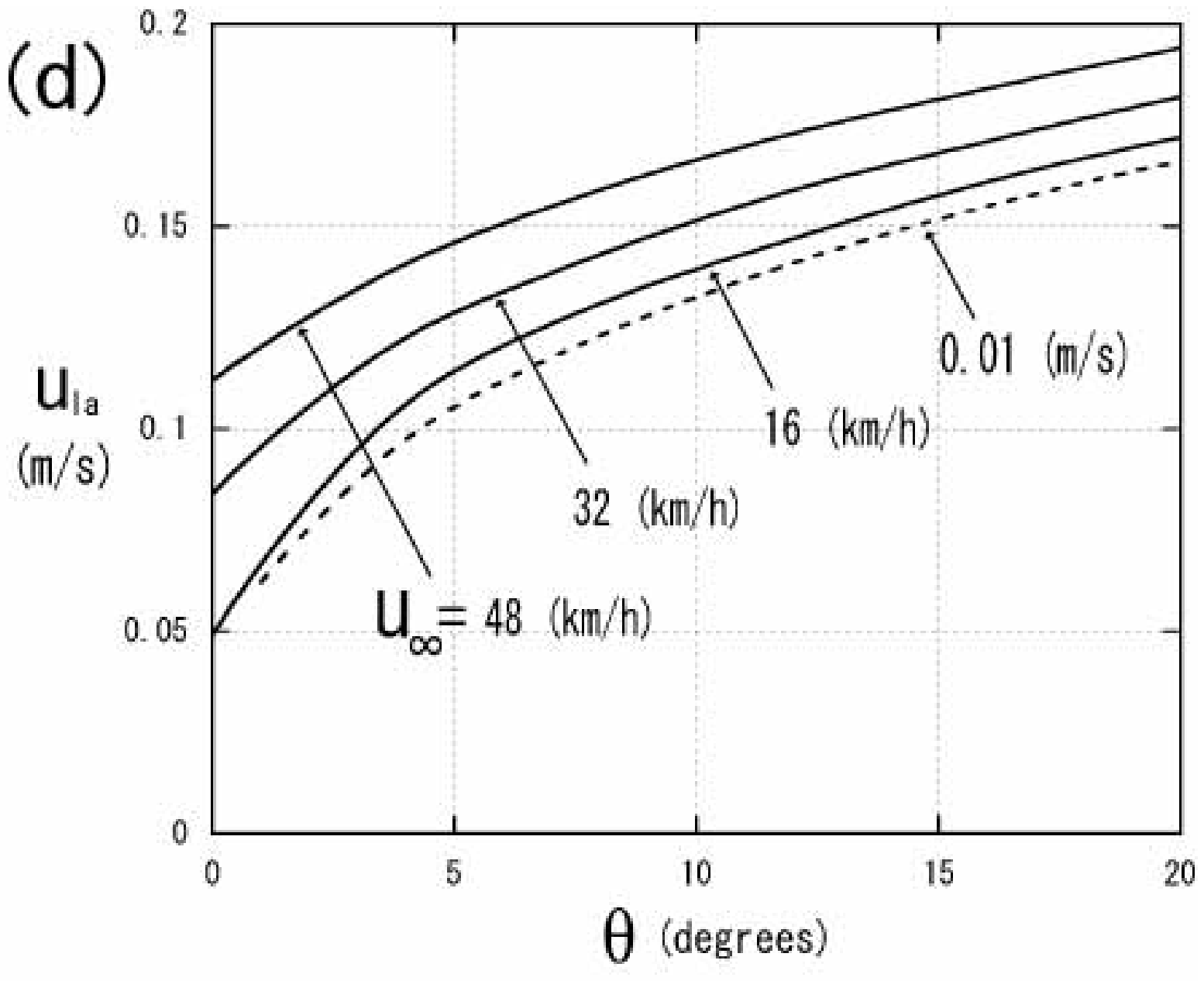}
\includegraphics[width=5cm,height=5cm,keepaspectratio,clip]{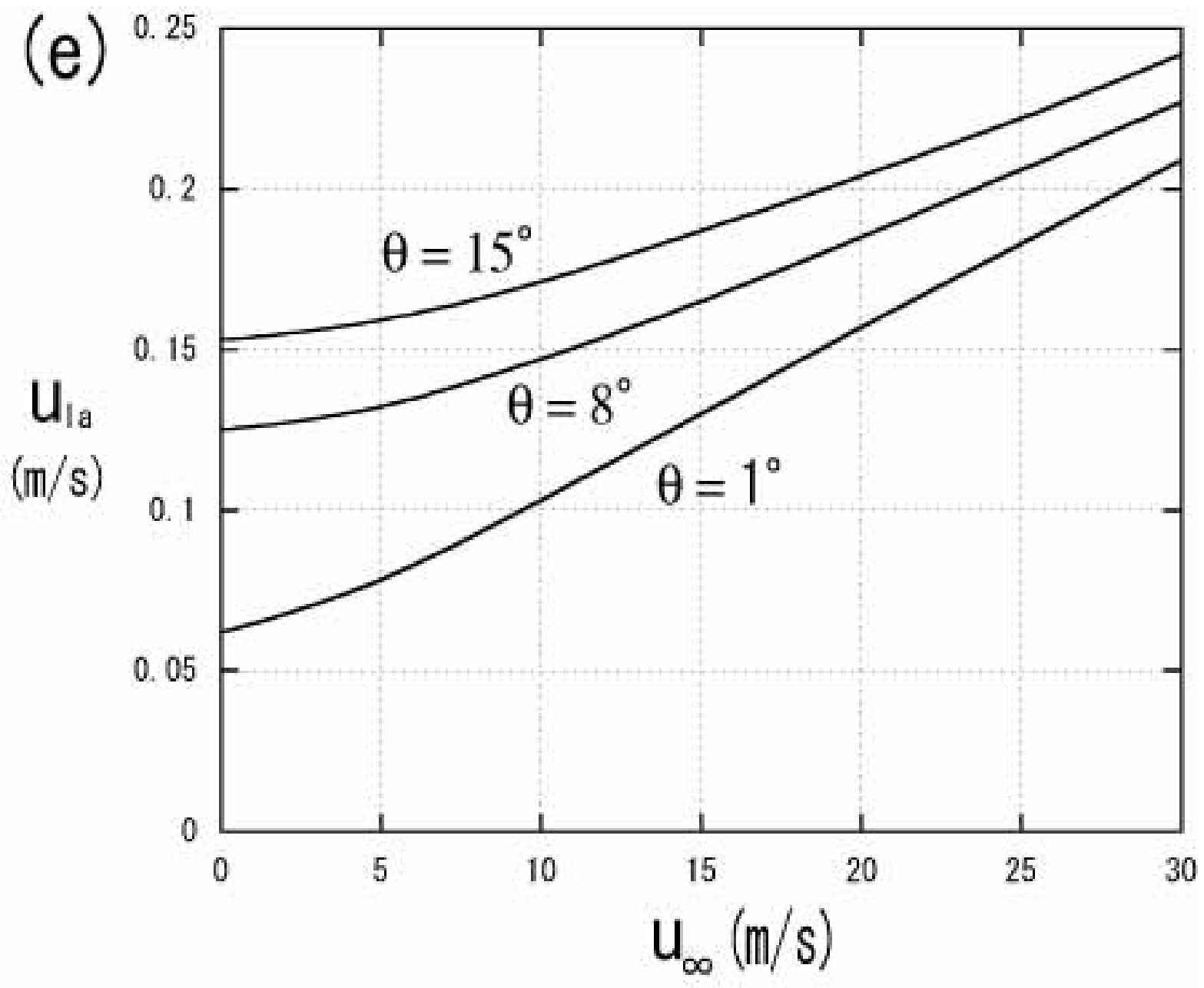}
\includegraphics[width=5cm,height=5cm,keepaspectratio,clip]{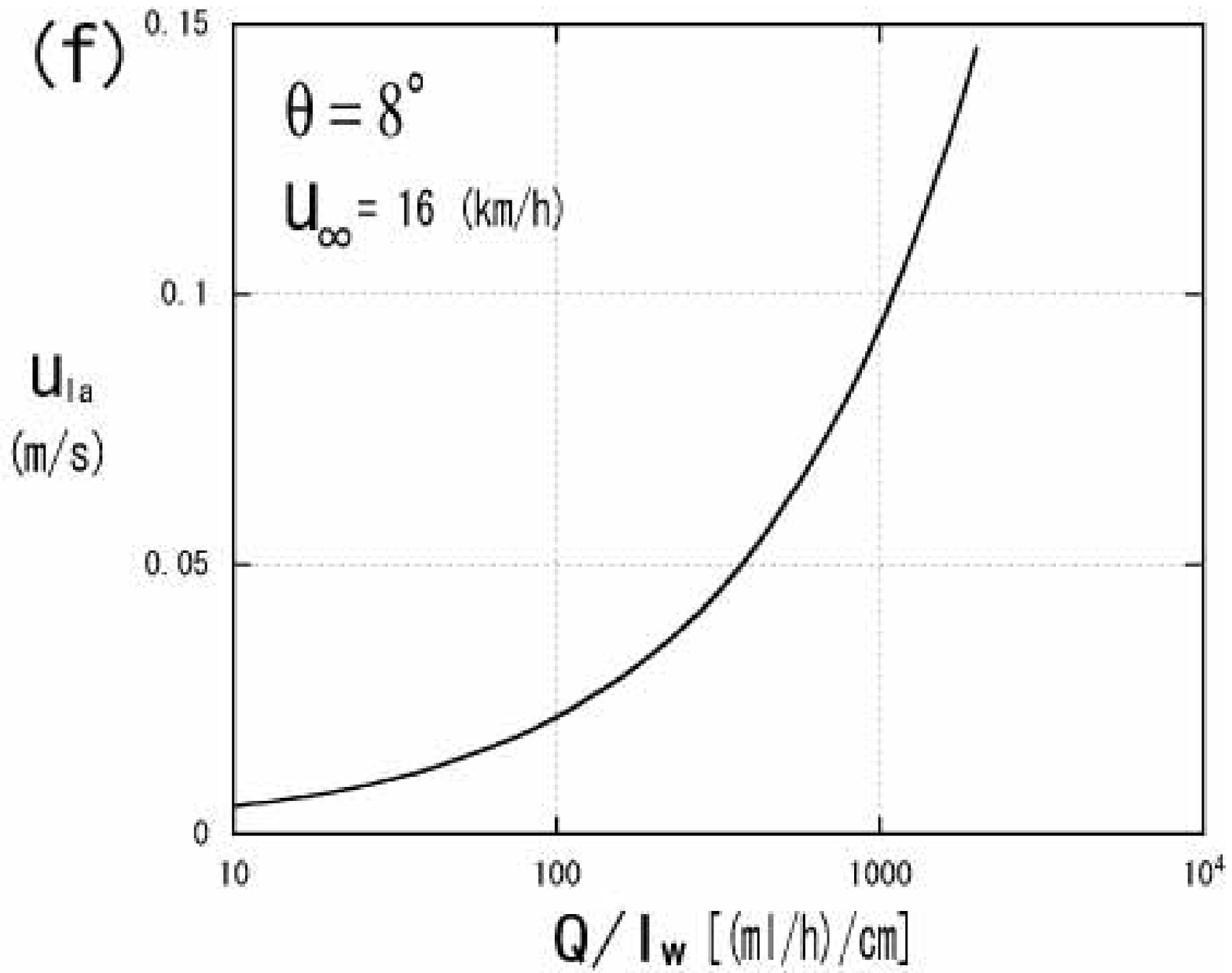}
\includegraphics[width=5cm,height=5cm,keepaspectratio,clip]{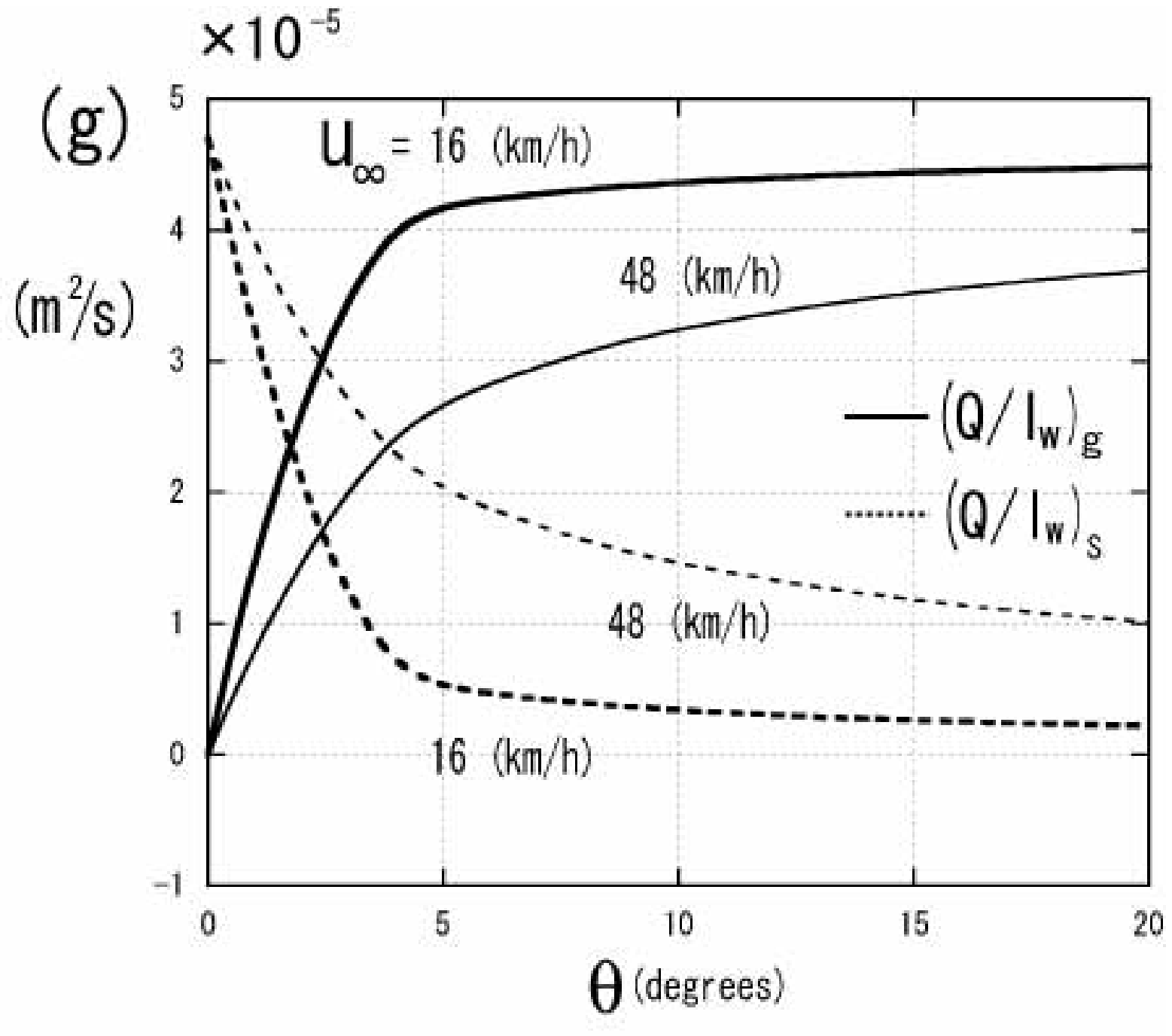}\hspace{3mm}
\includegraphics[width=5cm,height=5cm,keepaspectratio,clip]{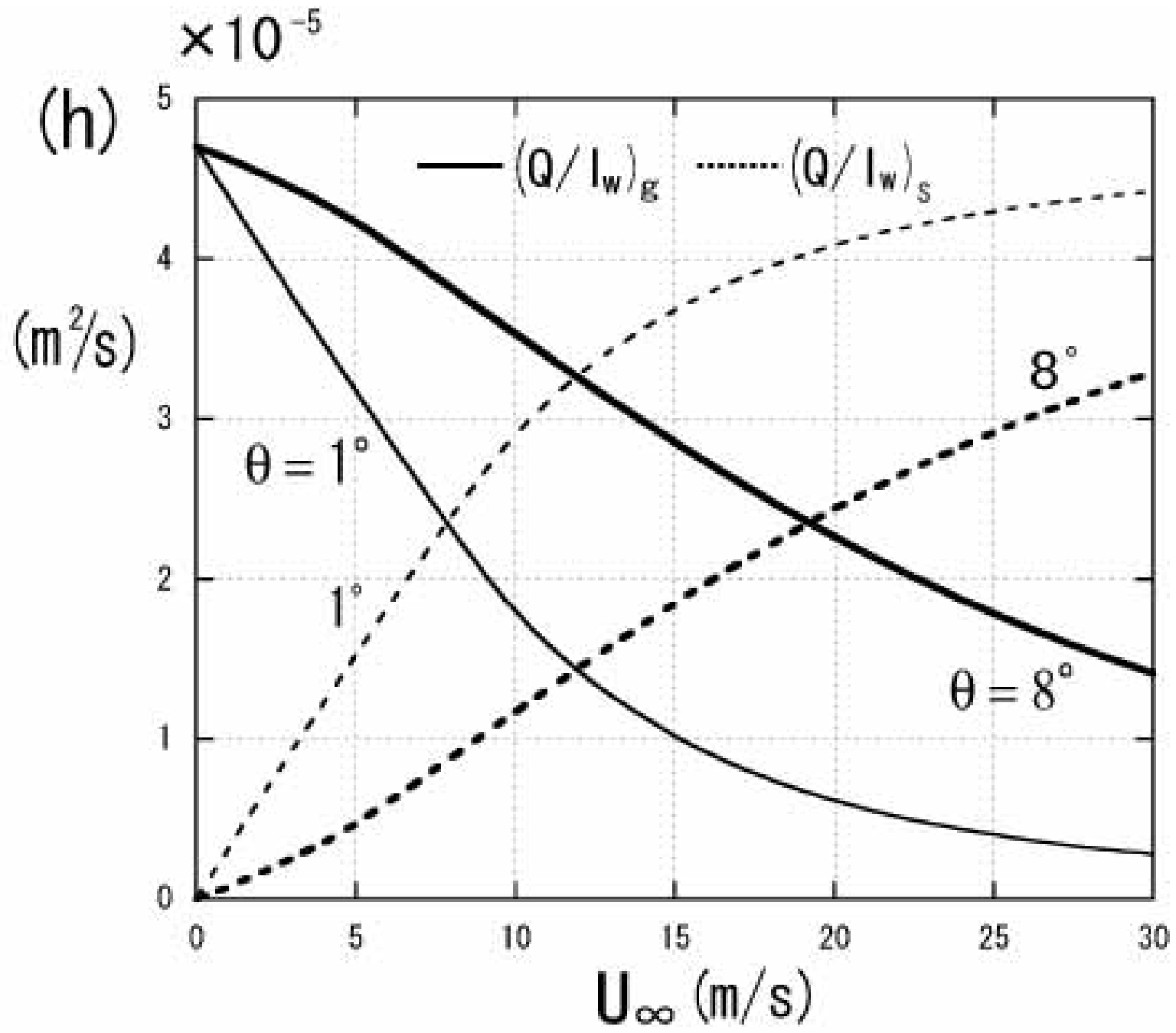}
\end{center}
\caption{For $Q/l_{w}=1692$ [(ml/h)/cm]=$4.7\times 10^{-5}\rm{(m^{2}/s)}$ and $x=0.1$ m, 
variation of $\bar{h}_{0}$ and $u_{la}$
with
(a), (d) $\theta$ for $u_{\infty}=16, 32, 48$ (km/h) (solid curves) and $u_{\infty}=0.01$ (m/s) (dashed curve); 
(b), (e) $u_{\infty}$ for $\theta=1^{\circ}$, $8^{\circ}$, $15^{\circ}$;
(c), (f) $Q/l_{w}$ for $\theta=8^{\circ}$ and $u_{\infty}=16$ (km/h).
Variation of gravity-driven water flow rate $(Q/l_{w})_{g}$ and 
shear-driven water flow rate $(Q/l_{w})_{s}$ with 
(g) $\theta$ for $u_{\infty}=16$, 48 (km/h) and 
(h) $u_{\infty}$ for $\theta=1^{\circ}$, $8^{\circ}$.}
\label{fig:h0-theta-uinf-Qoverl}
\end{figure}

Figures~\ref{fig:h0-theta-uinf-Qoverl} (a), (b) and (c) show the variation of undisturbed water film thickness $\bar{h}_{0}$ with $\theta$, $u_{\infty}$ and $Q/l_{w}$, respectively. $\bar{h}_{0}$ decreases with increasing $\theta$ and $u_{\infty}$, whereas it increases with $Q/l_{w}$. Figures~\ref{fig:h0-theta-uinf-Qoverl} (d), (e) and (f) show the variation of the surface velocity of the water film, $u_{la}$, with $\theta$, $u_{\infty}$ and $Q/l_{w}$, respectively. $u_{la}$ increases with increasing $\theta$, $u_{\infty}$ and $Q/l_{w}$. 
Figure~\ref{fig:h0-theta-uinf-Qoverl} (a) shows that the differences between $\bar{h}_{0}$ for different values of $u_{\infty}$ become smaller with $\theta$, and that the asymptotic form of $\bar{h}_{0}$ can be expressed by (\ref{eq:gravity-h0}).
This means that as $\theta$ increases, the water flow rate is dominated by the gravity-driven part in (\ref{eq:Qoverl}), $(Q/l_{w})_{g}$, as shown by the solid curves in figure~\ref{fig:h0-theta-uinf-Qoverl} (g).

On the other hand, in figure~\ref{fig:h0-theta-uinf-Qoverl} (b), the differences between $\bar{h}_{0}$ for different values of $\theta$ become smaller with $u_{\infty}$, and 
the water flow rate is dominated by the shear-driven part in (\ref{eq:Qoverl}),
$(Q/l_{w})_{s}$, as shown by the dashed curves in figure~\ref{fig:h0-theta-uinf-Qoverl} (h). 
The asymptotic form of $\bar{h}_{0}$ can be expressed by (\ref{eq:shear-h0}).
From $(Q/l_{w})_{g}=(Q/l_{w})_{s}$, 
the shear-driven flow changes to a gravity-driven one at 
\begin{equation}
\theta_{\rm c}=\sin^{-1}\left(\frac{3\mu_a\nu_{l}u_{\infty}^{3/2}d\bar{F}_{a}/d\eta^{2}|_{\eta=0}}
{2g\mu_{l}(2\nu_{a}x)^{1/2}\bar{h}_{0}}\right),
\label{eq:theta_c}
\end{equation}
as $\theta$ increases in figure~\ref{fig:h0-theta-uinf-Qoverl} (g).
On the other hand, 
the gravity-driven flow changes to a shear-driven one at 
\begin{equation}
u_{\infty \rm c}=\left\{\frac{2\mu_{l}(2\nu_{a}x)^{1/2}\bar{h}_{0}^{2}\sin\theta}
{3\mu_a\nu_{l}d\bar{F}_{a}/d\eta^{2}|_{\eta=0}}\right\}^{2/3},
\label{eq:uinf_c}
\end{equation} 
as $u_{\infty}$ increases, as shown in figure~\ref{fig:h0-theta-uinf-Qoverl} (h).
The points $\theta_{\rm c}$ and $u_{\infty \rm c}$ move to the right 
in figures~\ref{fig:h0-theta-uinf-Qoverl} (g) and (h)
as $u_{\infty}$ and $\theta$ increase.  

\subsection{\label{sec:lambda-vp-parameters}Variation of wavelength $\lambda$ and 
phase velocity $v_{p*}$ with $\theta$, $u_{\infty}$ and $Q/l_{w}$}

\begin{figure}
\begin{center}
\includegraphics[width=4cm,height=4cm,keepaspectratio,clip]{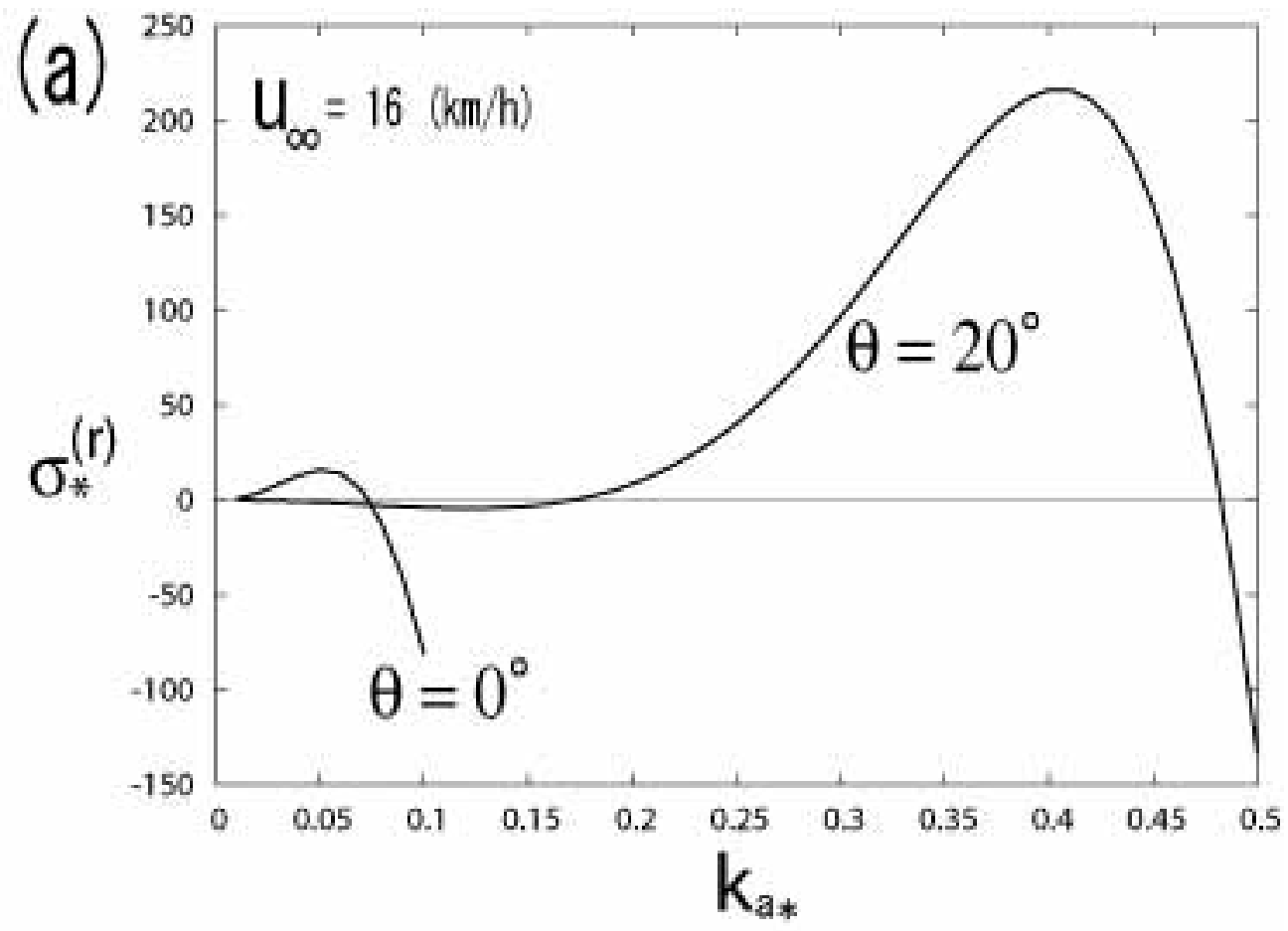}
\includegraphics[width=4cm,height=4cm,keepaspectratio,clip]{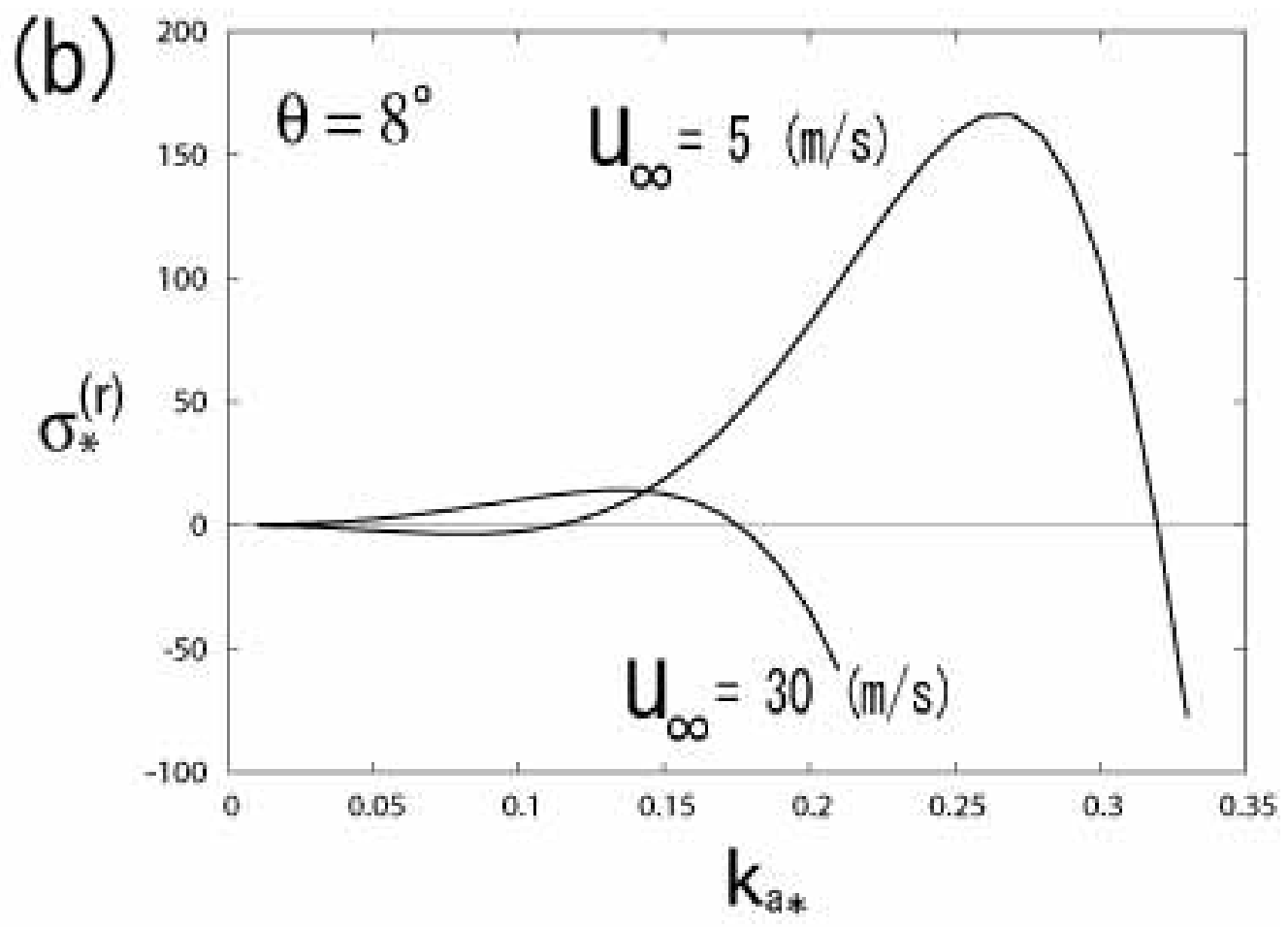}
\includegraphics[width=4cm,height=4cm,keepaspectratio,clip]{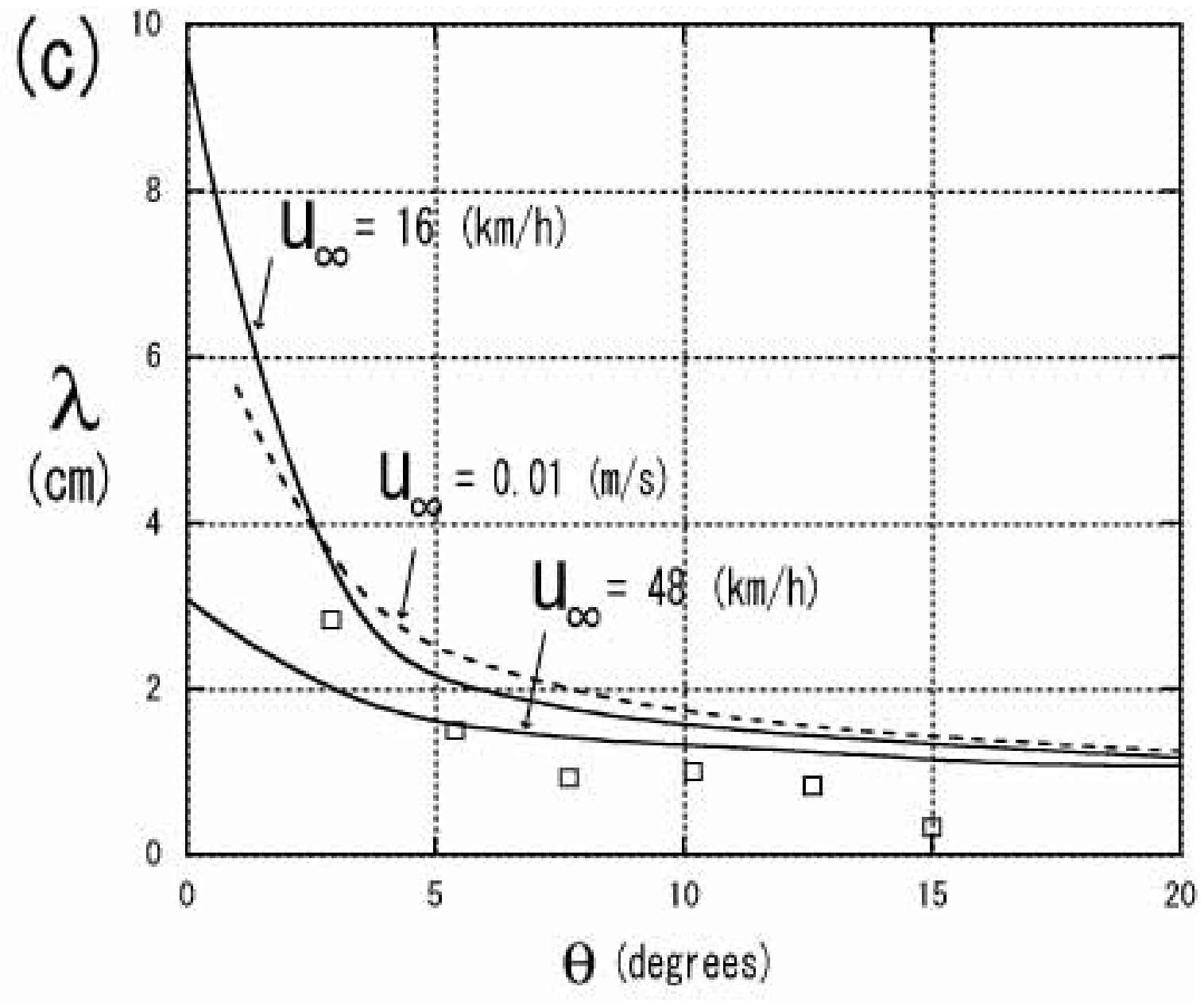}
\includegraphics[width=4cm,height=4cm,keepaspectratio,clip]{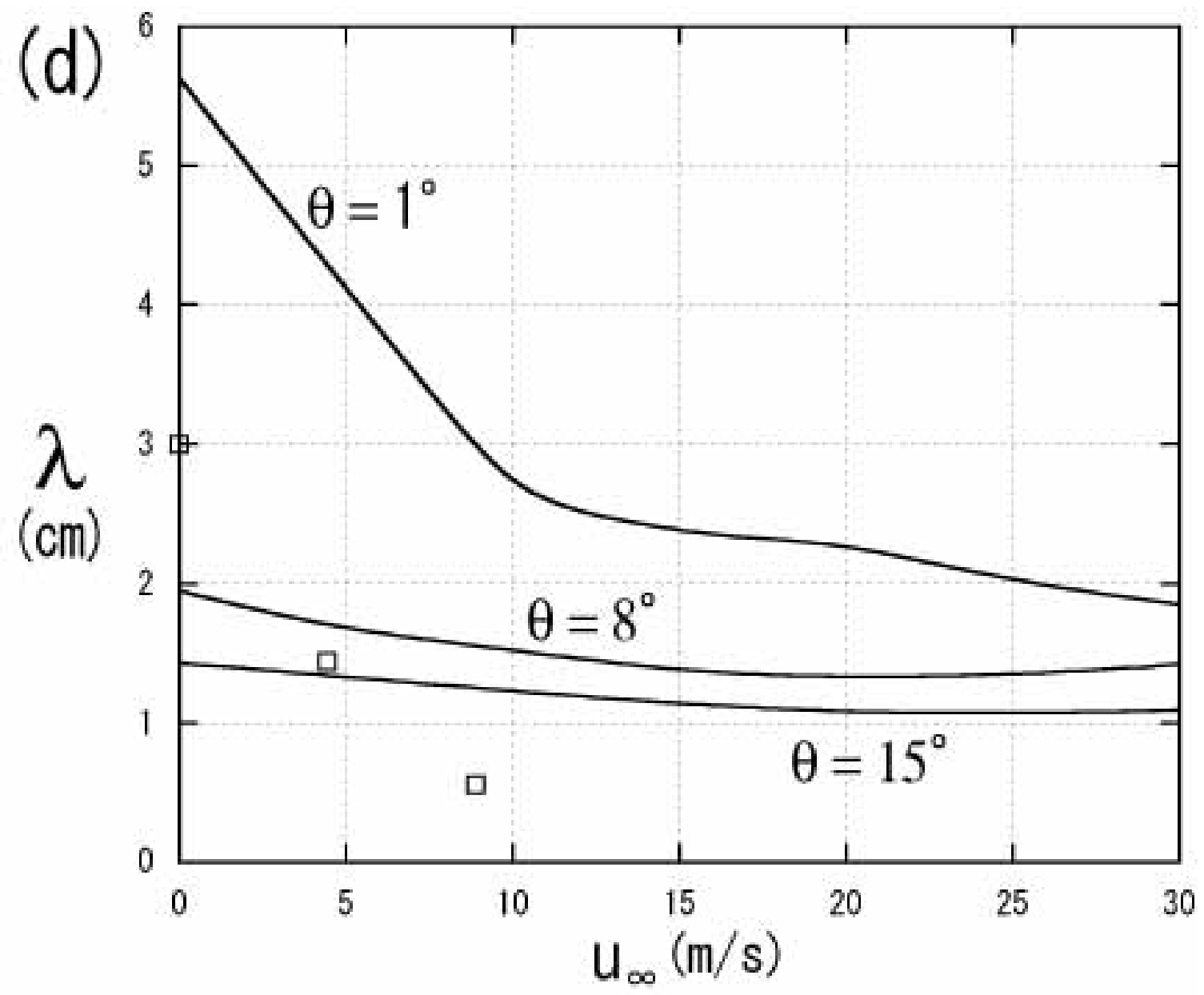}
\includegraphics[width=4cm,height=4cm,keepaspectratio,clip]{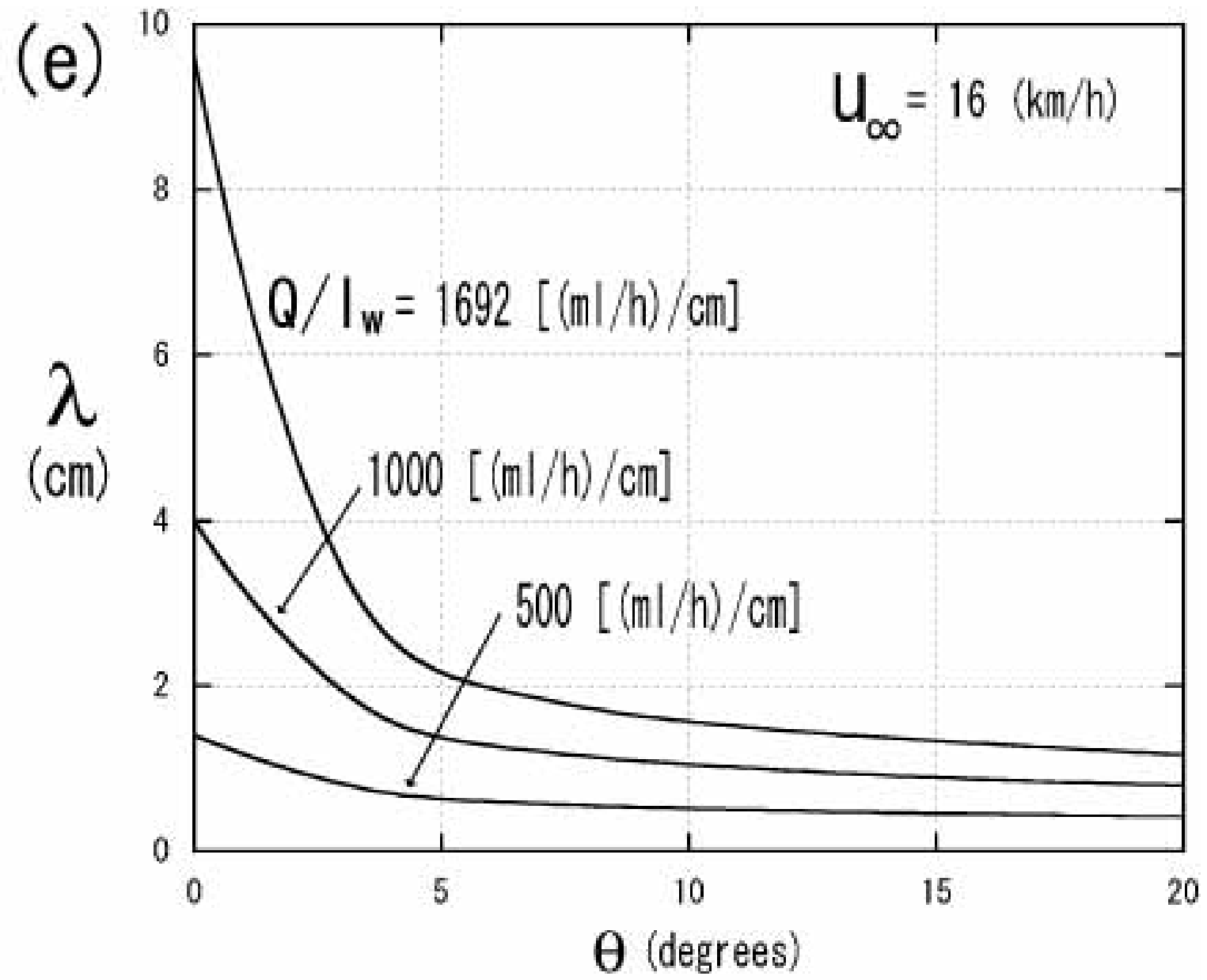}
\includegraphics[width=4cm,height=4cm,keepaspectratio,clip]{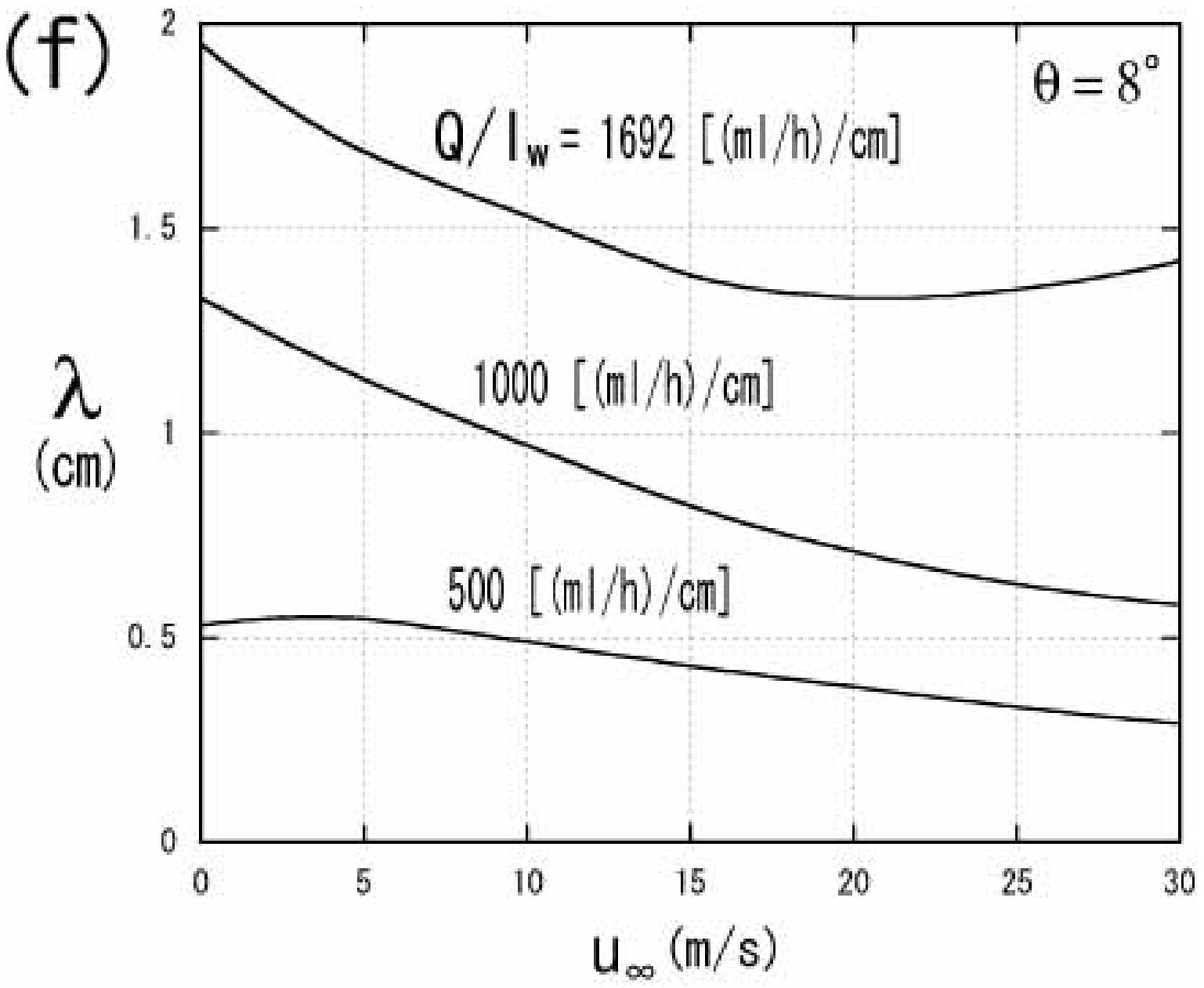}
\includegraphics[width=4cm,height=4cm,keepaspectratio,clip]{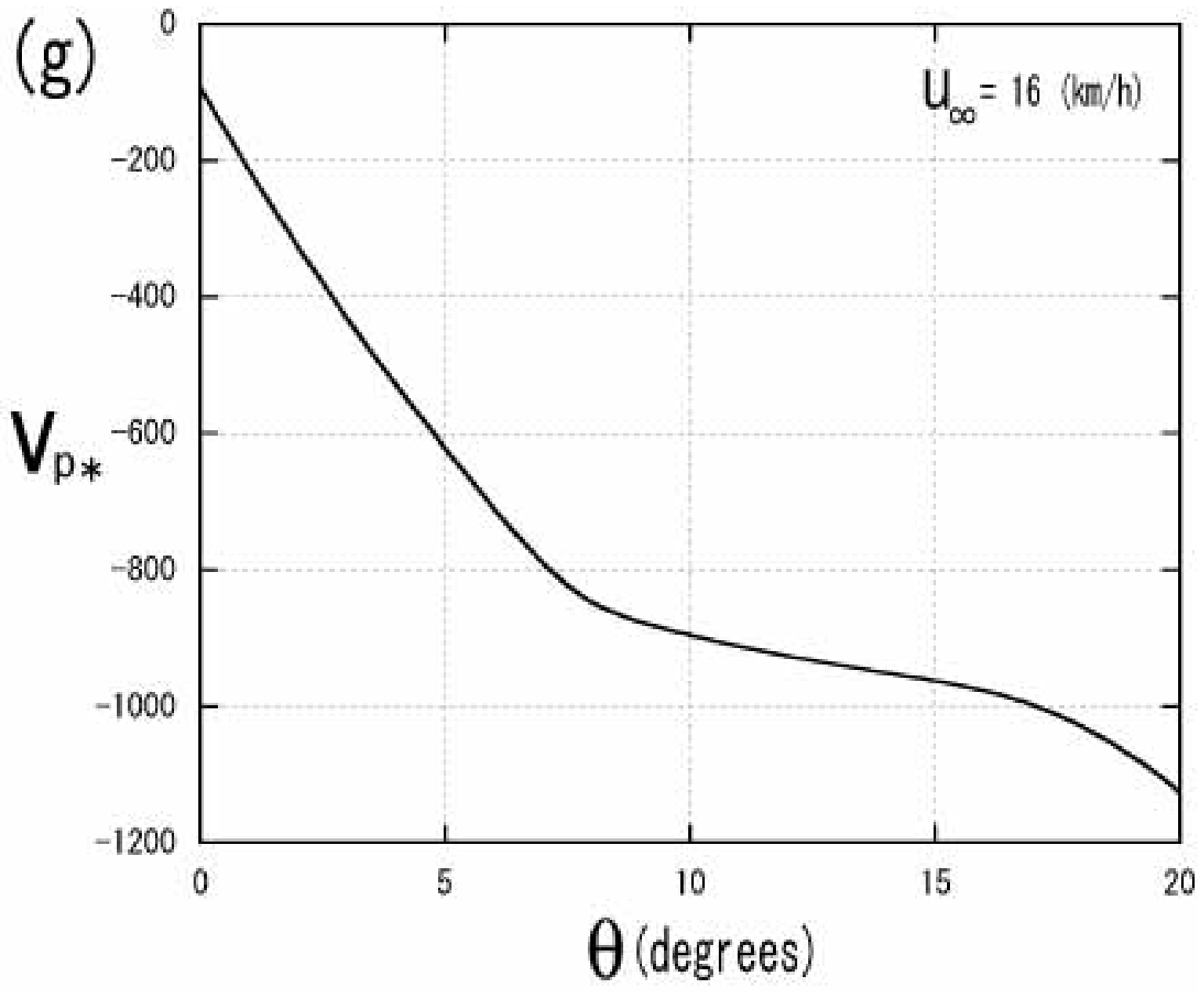}
\includegraphics[width=4cm,height=4cm,keepaspectratio,clip]{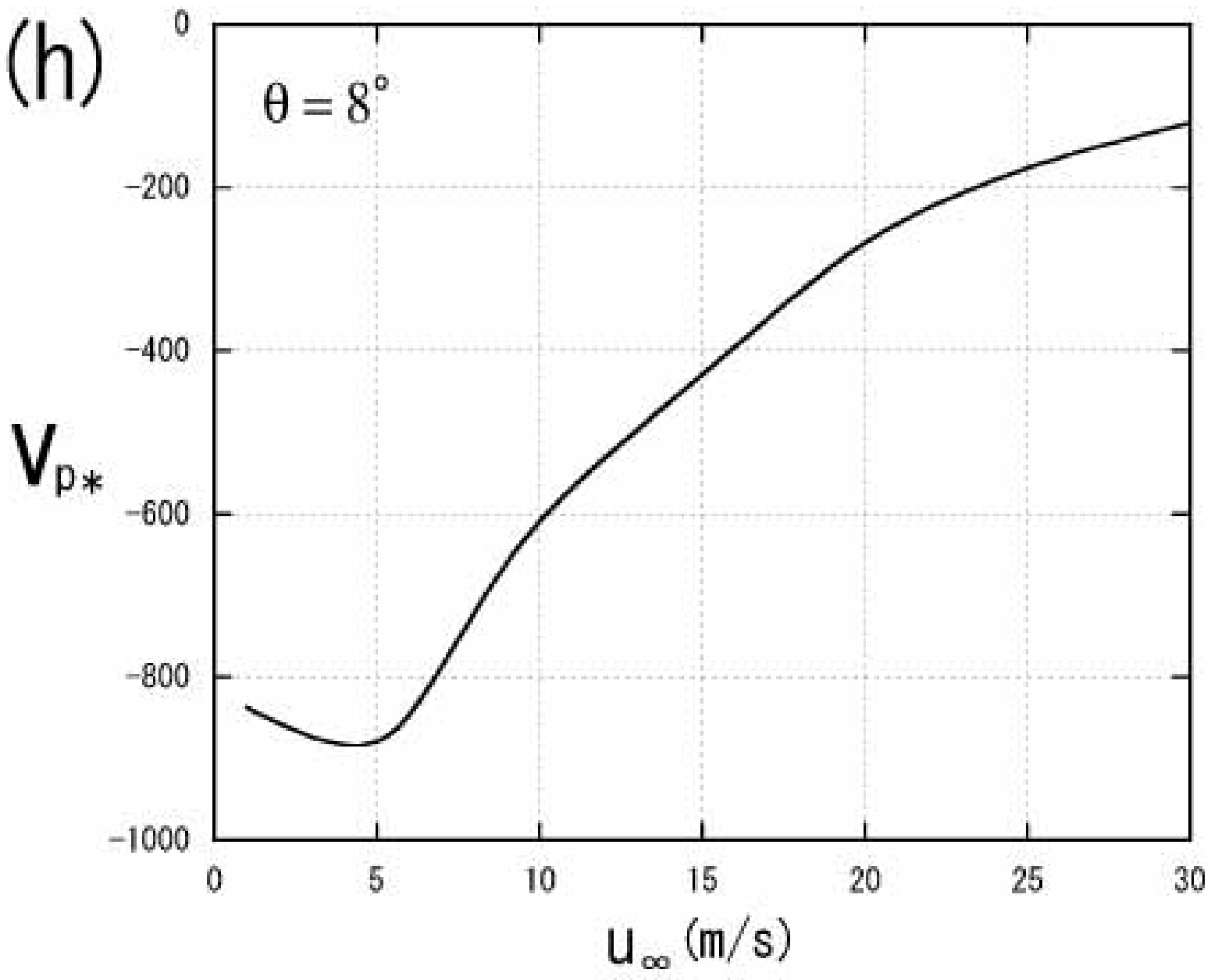}
\end{center}
\caption{
(a) and (b) represent the variation of dimensionless amplification rate $\sigma_{*}^{(r)}=\sigma^{(r)}/(\bar{V}/\bar{h}_{0})$ against dimensionless wave number $k_{a*}=k\delta_{0}$. 
Solid and dashed curves in (c) and (d) represent the variation of theoretically obtained wavelength $\lambda$ with $\theta$ and $u_{\infty}$.
The symbol $\square$ in (c) and (d) represent the experimental results of Ref. \onlinecite{Streitz02} for the variation of roughness spacing with $\theta$ at $u_{\infty}=16$ km/h 
and with $u_{\infty}$ at $\theta=8^{\circ}$, respectively.
(e) and (f) represent
the variation of $\lambda$ with $\theta$ and $u_{\infty}$, respectively, 
for $Q/l_{w}=500, 1000, 1692$ [(ml/h)/cm].
(g) and (h) represent the variation of  
the dimensionless phase velocity $v_{p*}=v_{p}/\bar{V}$ with $\theta$ and $u_{\infty}$, respectively.
The theoretical curves in (a)--(d), (g) and (h) are obtained for the same water supply rate of $Q/l_{w}=1692$ [(ml/h)/cm] as used in the experiments of Ref. \onlinecite{Streitz02}.}
\label{fig:lambda-vp}
\end{figure}

Figures~\ref{fig:lambda-vp} (a) shows the variation of the dimensionless amplification rates $\sigma_{*}^{(r)}=\sigma^{(r)}/(\bar{V}/\bar{h}_{0})$ against the dimensionless wavenumber $k_{a*}=k\delta_{0}$ for $\theta=0^{\circ}$, $20^{\circ}$ at $u_{\infty}=16$ km/h and $Q/l_{w}=1692$ [(ml/h)/cm]. On the other hand, figure~\ref{fig:lambda-vp} (b) shows the variation of $\sigma_{*}^{(r)}$ against $k_{a*}$ for $u_{\infty}=5$, $30$ m/s at $\theta=8^{\circ}$ and $Q/l_{w}=1692$ [(ml/h)/cm].
An ice pattern with a wave number at which the amplification rate acquires a maximum is expected to be observed. 
For example, for $\theta=20^{\circ}$ in figure~\ref{fig:lambda-vp} (a), $\sigma_{*}^{(r)}$ acquires a maximum value of $\sigma^{(r)}_{*\rm max}=216.9$ at $k_{a*}=0.41$. Since the wave number $k$ is normalized by $\delta_{0}$, the corresponding wavelength of the ice pattern is 1.19 cm from $\lambda=2\pi\delta_{0}/k_{a*}$. Here, $\delta_{0}=(2\nu_{a}x/u_{\infty})^{1/2}=7.65\times 10^{-4}$ m estimated from $x=0.1$ m and $u_{\infty}=16$ km/h is used. 

Using this method, the wavelength $\lambda$ was calculated for various $\theta$ and $u_{\infty}$. The variation of $\lambda$ with $\theta$ at $u_{\infty}=16$, 48 km/h and $u_{\infty}=0.01$ m/s is presented in figure~\ref{fig:lambda-vp} (c), whereas figure~\ref{fig:lambda-vp} (d) shows the variation of $\lambda$ with $u_{\infty}$ at $\theta=1^{\circ}$, $8^{\circ}$ and $15^{\circ}$. Here the theoretical results are shown by the solid and dashed curves. The symbol $\square$ in Figs. \ref{fig:lambda-vp} (c) and (d) represents the measured roughness spacing in the experiments \cite{Streitz02} for $u_{\infty}=16$ km/h and $\theta=8^{\circ}$, respectively. Both theoretical and experimental results at $u_{\infty}=16$ km/h in Figs. \ref{fig:lambda-vp} (c) show that $\lambda$ decreases with increasing $\theta$. The variation of $\lambda$ for $u_{\infty}=0.01$ m/s also shows the same trend as that for $u_{\infty}=16$ and 48 km/h, despite large difference of $u_{\infty}$. 
The experimental results ($\square$) for $\theta=8^{\circ}$ in figure~\ref{fig:lambda-vp} (d) shows that $\lambda$ rapidly decreases with increasing $u_{\infty}$. 
On the other hand, theoretically obtained $\lambda$ for the same $\theta=8^{\circ}$ gradually decreases for small values of $u_{\infty}$, and increases very slightly with $u_{\infty}$. For much larger values of $u_{\infty}$, which is not shown in figure~\ref{fig:lambda-vp} (d), $\lambda$ decreases again. The variation of $\lambda$ for $\theta=15^{\circ}$ also shows the same trend as that for $\theta=8^{\circ}$. $\lambda$ at $\theta=1^{\circ}$ rapidly decreases for small values of $u_{\infty}$ and gradually decreases for larger $u_{\infty}$. 
Figures~\ref{fig:lambda-vp} (e) and (f) show the variation of $\lambda$ with $\theta$ at $u_{\infty}=16$ km/h and with $u_{\infty}$ at $\theta=8^{\circ}$, respectively, for $Q/l_{w}=500, 1000, 1692$ [(ml/h)/cm]. These figures show that $\lambda$ increases with $Q/l_{w}$. 
It was found by comparing
figures~\ref{fig:h0-theta-uinf-Qoverl} (a) and \ref{fig:lambda-vp} (c), as well as
figures~\ref{fig:h0-theta-uinf-Qoverl} (b) and \ref{fig:lambda-vp} (d)
that the variations of $\lambda$ with $\theta$ and $u_{\infty}$ show almost the same trends as for $\bar{h}_{0}$. This indicates that $\bar{h}_{0}$ is the most important parameter to determine ice roughness spacing. Since the experimental data ($\square$) for $\theta=8^{\circ}$ in figure~\ref{fig:lambda-vp} (d) are scarce, it is difficult to conclude that there is large disagreement between the experimental and theoretical results. In addition, based on theoretical considerations, 
it seems impossible that the experimental results ($\square$) shown in figure~\ref{fig:lambda-vp} (d)
decrease rapidly with $u_{\infty}$.
In order for that to be true, $\bar{h}_{0}$ at $\theta=8^{\circ}$ in figure~\ref{fig:h0-theta-uinf-Qoverl} (b) must decrease rapidly with $u_{\infty}$, like $\bar{h}_{0}$ at $\theta=1^{\circ}$. 

Finally, figures~\ref{fig:lambda-vp} (g) and (h) show the variation of the dimensionless phase velocity $v_{p*}=v_{p}/\bar{V}$ with $\theta$ at $u_{\infty}=16$ km/h and with $u_{\infty}$ at $\theta=8^{\circ}$. The magnitude of $v_{p*}$ was defined from the wavenumber at which $\sigma_{*}^{(r)}$ acquires a maximum value. It was found that $v_{p*}$ is negative for all $\theta$ and $u_{\infty}$, which indicates that ice pattern moves in the direction opposite to the water flow since ice grows faster just upstream of any protrusion and slower downstream (see FIG. 5 in Ref. \onlinecite{UF11}). 

It is well known that a solid surface under a supercooled liquid film is morphological unstable, resulting in dendritic growth. The effect of the water flow on the isotherms on such a microscopic length scale is negligible, and the fundamental building block of the morphological instability of a solidification front is the Mullins-Sekerka theory. \cite{Langer80} In this case, the amplification rate is given by 
$\sigma^{(r)}_{*}=k_{l*}\{1-(d_{0}/\bar{h}_{0})(l_{\rm th}/\bar{h}_{0})k_{l*}^{2}(1+K^{s}_{l})\}$, 
and the characteristic wavelength is $\lambda_{\rm micro}=2\pi\{3l_{\rm th}d_{0}(1+K^{s}_{l})\}^{1/2}$, where 
$l_{\rm th}=\kappa_{l}/\bar{V}$ and $d_{0}=T_{sl}\Gamma C_{pl}/L^{2}$ are a macroscopic and microscopic characteristic length, respectively, $\Gamma$ is the ice-water interface tension and $C_{pl}$ is the specific heat at constant pressure of the water. \cite{UF11} It should be noted that $\lambda_{\rm micro}$ depends on only $u_{\infty}$ because $l_{\rm th} \sim \delta_{0} \sim u_{\infty}^{-1/2}$, while $\lambda$ based on the macro-scale morphological instability under a supercooled liquid film herein depends on $\theta$, $u_{\infty}$ and $Q/l_{w}$, as shown in figures~\ref{fig:lambda-vp} (c), (d), (e) and (f).  

\subsection{\label{sec:sigma-parameters}Variation of amplification rate 
$\sigma_{*\rm max}^{(r)}$ with $\theta$, $u_{\infty}$ and $Q/l_{w}$}

\begin{figure}
\begin{center}
\includegraphics[width=6cm,height=6cm,keepaspectratio,clip]{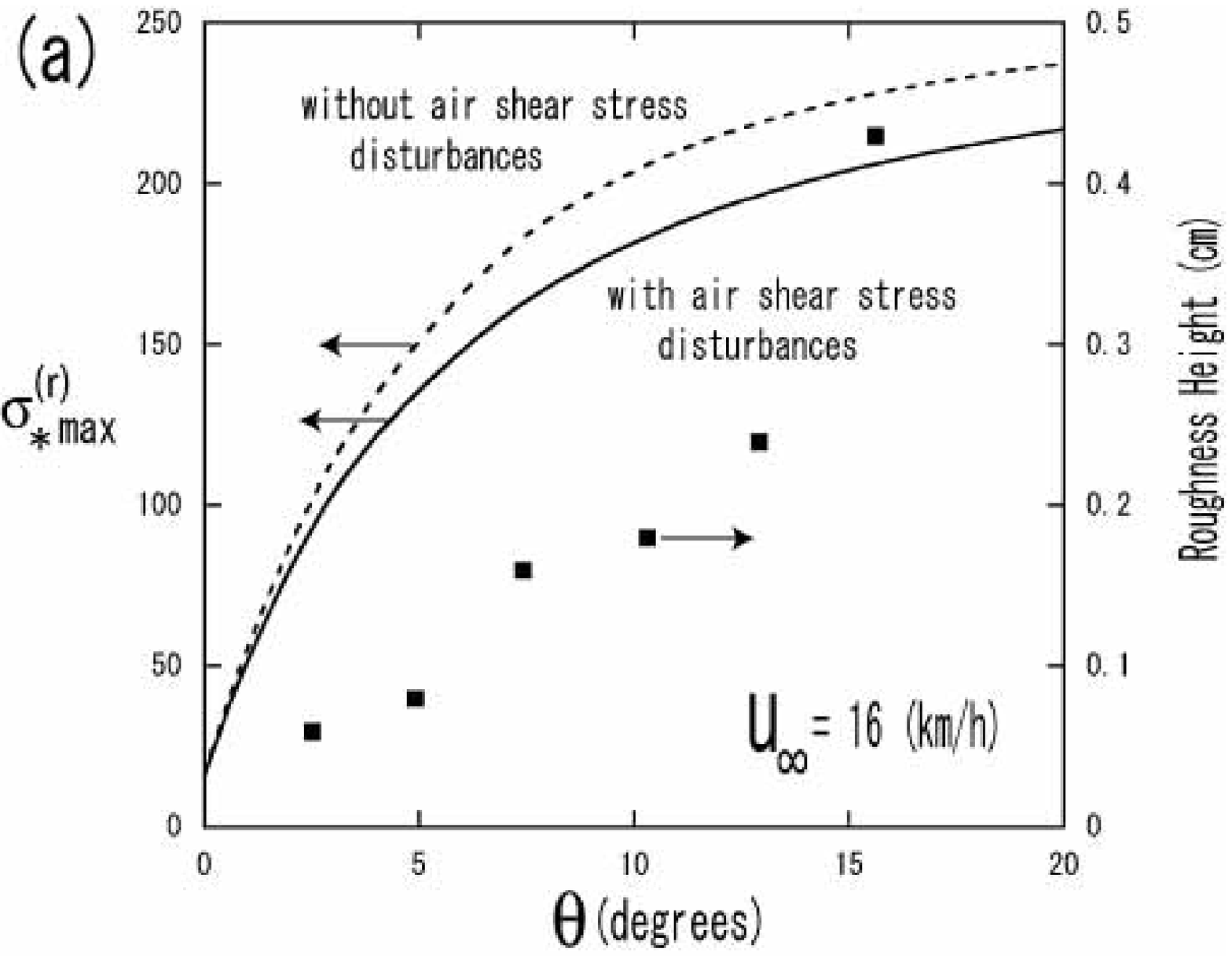}
\includegraphics[width=6cm,height=6cm,keepaspectratio,clip]{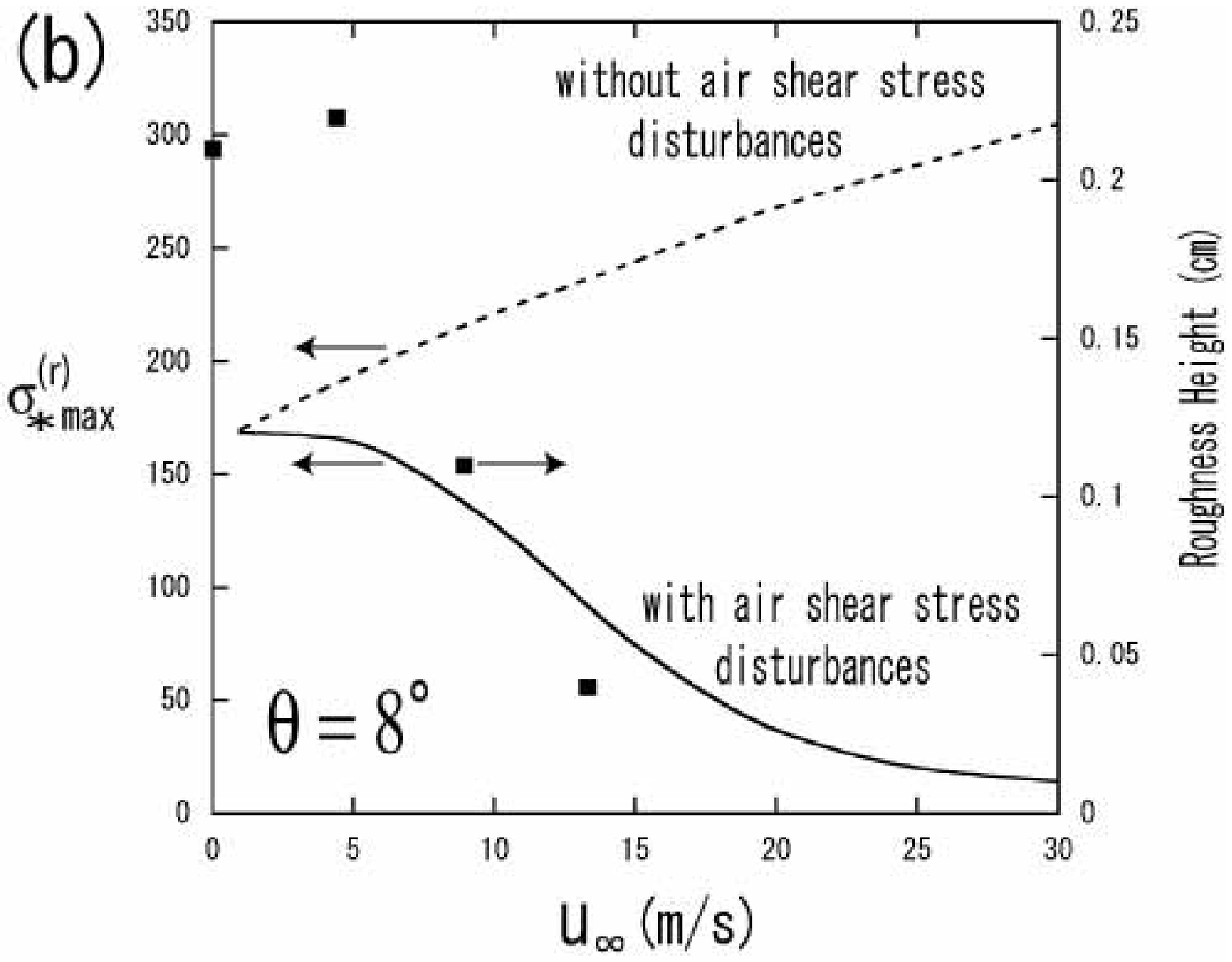}\\[2mm]
\includegraphics[width=6cm,height=6cm,keepaspectratio,clip]{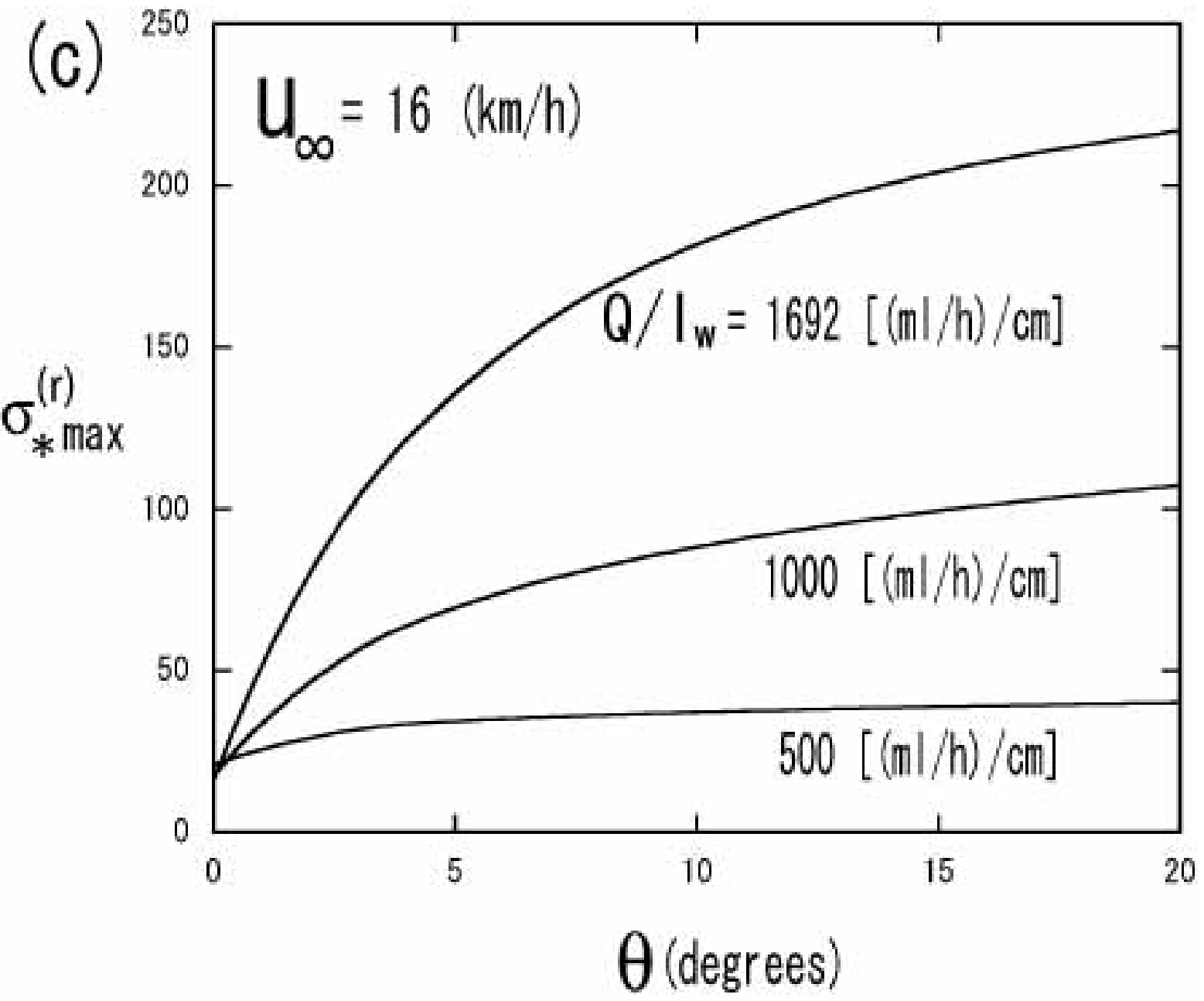}
\includegraphics[width=6cm,height=6cm,keepaspectratio,clip]{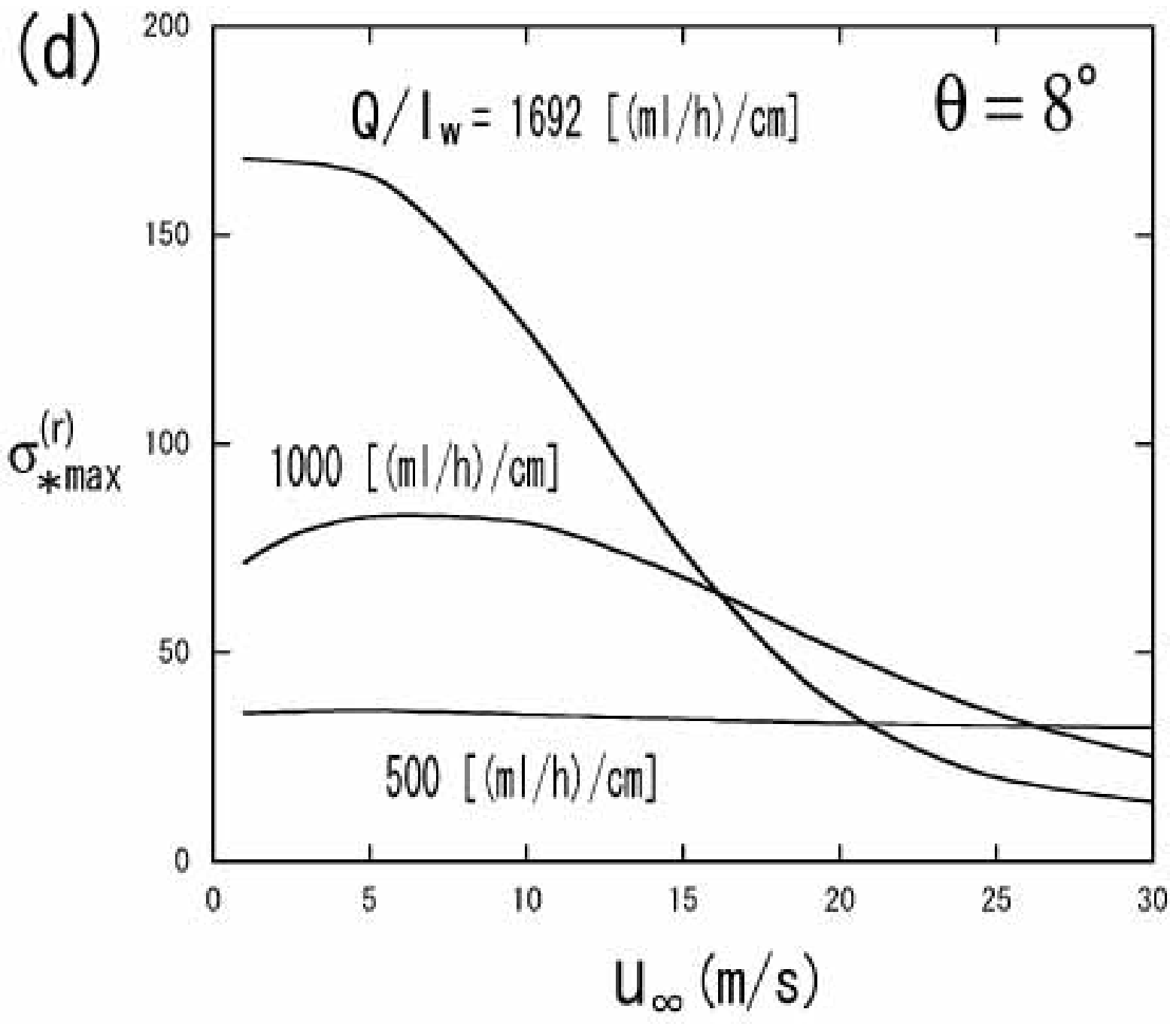}\\[2mm]
\end{center}
\caption{For $Q/l_{w}=1692$ [(ml/h)/cm] and $x=0.1$ m, 
variation of dimensionless amplification rate $\sigma_{*\rm max}^{(r)}$ with 
(a) $\theta$ at $u_{\infty}=16$ (km/h) and
(b) $u_{\infty}$ at $\theta=8^{\circ}$. 
The symbol $\blacksquare$ represents the measured roughness height in the experiments of Ref. \onlinecite{Streitz02}.
Variation of $\sigma_{*\rm max}^{(r)}$ with 
(c) $\theta$ at $u_{\infty}=16$ (km/h) and (d) $u_{\infty}$ at $\theta=8^{\circ}$, 
for $Q/l_{w}=500, 1000, 1692$ [(ml/h)/cm].}
\label{fig:amp-parameters}
\end{figure}

The symbol $\blacksquare$ in figures~\ref{fig:amp-parameters} (a) and (b) shows the variation of measured roughness height by Ref. \onlinecite{Streitz02} with $\theta$ for a light wind of $u_{\infty}=16$ km/h and with $u_{\infty}$ for a mild slope of $\theta=8^{\circ}$, respectively, for $Q/l_{w}=1692$ [(ml/h)/cm].
These experimental results indicate that roughness height increases with increasing slope and decreases with increasing wind speed.
The variation of $\sigma_{*\rm max}^{(r)}$ with $\theta$ and $u_{\infty}$ are shown by the solid and dashed curves with the same parameters as used in the experiments.
Small disturbances of the ice-water interface are assumed to be sinusoidal, expressed as 
\begin{equation} 
y_{*}=\zeta_{*}=\delta_{b}\Imag[{\rm exp}(\sigma_{*}t_{*}+ik_{l*}x_{*})]
=\delta_{b}(t_{*})\sin[k_{l*}(x_{*}-v_{p*}t_{*})].
\label{eq:zeta}
\end{equation}
In linear stability analysis, the amplitude of the ice-water interface disturbance of the most unstable mode increases with time as follows:
$\delta_{b}(t_{*})=\exp(\sigma_{* \rm max}^{(r)}t_{*})\delta_{b}$,
where $\delta_{b}=\zeta_{k}/\bar{h}_{0}$ is a dimensionless initial infinitesimal amplitude.
However, the linear theory is unable to clarify further features related to the development of disturbances.
Hence, only comparison of trends between the solid, dashed curves and the experimental results ($\blacksquare$) in figures~\ref{fig:amp-parameters} (a) and (b) is meaningful since $\delta_{b}(t_{*})$ do not provide an actual roughness height of the ice-water interface.

The solid curves consider the effect of the tangential and normal air shear stress disturbances on the water-air interface, which are represented by $\Sigma_{a}$ and $\Pi_{a}$ in (\ref{eq:shearstress_a}) and (\ref{eq:normalstress_a}), respectively. On the other hand, the dashed curves are obtained by neglecting $\Sigma_{a}$ and $\Pi_{a}$ in (\ref{eq:bc-shearstress}) and (\ref{eq:bc-normalstress}). It was found that the roughness height to be expected from $\sigma_{*\rm max}^{(r)}$ increases with $\theta$ and decreases with $u_{\infty}$, as shown by the solid curves in figures~\ref{fig:amp-parameters} (a) and (b), which show the same trends as for the experimental results ($\blacksquare$). If the air shear stress disturbances are neglected, $\sigma_{*\rm max}^{(r)}$ is slightly overestimated for $\theta$, and drastically so for $u_{\infty}$, as shown by the dashed curves in figures~\ref{fig:amp-parameters} (a) and (b), respectively. The model which takes into account the effect of the air shear stress disturbances on the water-air interface is supported by the experimental results. Streitz and Ettema state that "it is probable that roughness height would attain a maximum with slope, and become smaller with steeper slopes". \cite{Streitz02} However, for much larger values of $\theta$, which is not shown in figure~\ref{fig:amp-parameters} (a), $\sigma_{*\rm max}^{(r)}$ increases only gradually with $\theta$ and never attain a maximum value. Since wind flow increases heat loss to air, it should be expected that the roughness height increases with increasing wind speed. However, the roughness height shown in figure~\ref{fig:amp-parameters} (b) is contrary to this expectation. The physical explanation of that will be given in the next section.
Finally, figures~\ref{fig:amp-parameters} (c) and (d) show the variation of $\sigma_{*\rm max}^{(r)}$ with $\theta$ at $u_{\infty}=16$ km/h and with $u_{\infty}$ at $\theta=8^{\circ}$,
 respectively, for $Q/l_{w}=500, 1000, 1692$ [(ml/h)/cm]. It was found that $\sigma_{*\rm max}^{(r)}$ increases with $Q/l_{w}$ for any $\theta$, whereas $\sigma_{*\rm max}^{(r)}$ increases with $Q/l_{w}$ for low $u_{\infty}$, but decreases with $Q/l_{w}$ for higher $u_{\infty}$. 
 
\subsection{\label{sec:qa-Ga'-xi-ka}Variation of the disturbed part of convective 
heat transfer rate at the water-interface with $\theta$ and $u_{\infty}$}

Since the dimensionless amplitude of the water-air interface, $\delta_{t}=\xi_{k}/\bar{h}_{0}$, and that of the ice-water interface, $\delta_{b}=\zeta_{k}/\bar{h}_{0}$
are related as $\delta_{t}=-f_{l}|_{{y_{*}}=1}\delta_{b}$, 
the disturbance of the water-air interface due to change in ice shape can be expressed as follows:
\begin{eqnarray} 
y_{*}=\xi_{*}&=&1+\Imag[\delta_{t}{\rm exp}(\sigma_{*}t_{*}+ik_{l*}x_{*})] \nonumber \\
&=&1+\delta_{b}(t_{*})|f_{l}|_{y_{*}=1}|\sin[k_{l*}(x_{*}-v_{p*}t_{*})-\Theta_{\xi_{*}}],
\label{eq:xi}
\end{eqnarray}
where 
$|f_{l}|_{y_{*}=1}|=[(f_{l}^{(r)}|_{y_{*}=1})^{2}+(f_{l}^{(i)}|_{y_{*}=1})^2]^{1/2}$ is the amplitude,
and 
$\Theta_{\xi_{*}}$ is the phase difference between the water-air and ice-water interfaces. 
Here $f_{l}^{(r)}$ and $f_{l}^{(i)}$ are the real and imaginary parts of $f_{l}$.
The water-air interface disturbance 
causes disturbance of convective heat transfer rate from the water-air interface to the air. 
Its dimensionless form can be written as  
$q'_{a*}\equiv \Imag[-K_{a}\partial T'_{a}/\partial y|_{y=\xi}]/K_{a}\bar{G}_{a}
=\Imag[G'_{a}\xi_{k}/\bar{h}_{0}{\rm exp}(\sigma_{*}t_{*}+ik_{l*}x_{*})]$,
where 
$T'_{a}=H_{a}(\eta)\bar{G}_{a}\xi_{k}{\rm exp}[\sigma t+ikx]$ is used and
$G'_{a}\equiv (\bar{h}_{0}/\delta_{0})(-dH_{a}/d\eta)|_{\eta=0}$ 
represents the disturbed part of the air temperature gradient at the water-air interface.
It should be noted that $q'_{a*}$ includes $G'_{a}$ and $\xi_{k}$. 
Using $\xi_{k}=-f_{l}|_{{y_{*}}=1}\zeta_{k}$, $q'_{a*}$ can be expressed as
\begin{equation}
q'_{a*}
=\delta_{b}(t_{*})|q'_{a*}|\sin[k_{l*}(x_{*}-v_{p*}t_{*})-\Theta_{q'_{a*}}],
\label{eq:qa}
\end{equation}
where
$|q'_{a*}|
=[(G'^{(r)}_{a}f_{l}^{(r)}|_{y_{*}=1}-G'^{(i)}_{a}f_{l}^{(i)}|_{y_{*}=1})^{2}
 +(G'^{(r)}_{a}f_{l}^{(i)}|_{y_{*}=1}+G'^{(i)}_{a}f_{l}^{(r)}|_{y_{*}=1})^{2}]^{1/2}$
is the amplitude and $\Theta_{q'_{a*}}$ is the phase difference between the disturbed heat flux at the water-air interface and ice-water interface. 
Here
$G'^{(r)}_{a}\equiv (\bar{h}_{0}/\delta_{0})(-dH_{a}^{(r)}/d\eta)|_{\eta=0}$ and
$G'^{(i)}_{a}\equiv (\bar{h}_{0}/\delta_{0})(-dH_{a}^{(i)}/d\eta)|_{\eta=0}$
represent the real and imaginary parts of $G'_{a}$.
Defining the undisturbed part of local convective heat transfer coefficient at the water-air interface and
the disturbed part of it as 
$\bar{h}_{x}=-K_{a}\partial\bar{T}_{a}/\partial y|_{y=\bar{h}_{0}}/(T_{la}-T_{\infty})$ and 
$h'_{x}=\Imag[-K_{a}\partial T'_{a}/\partial y|_{y=\bar{h}_{0}}/(T_{la}-T_{\infty})]$, respectively,
$h'_{x}/\bar{h}_{x}$ is equivalent to $q'_{a*}$. \cite[]{UF11}

\begin{figure}
\begin{center}
\includegraphics[width=5cm,height=5cm,keepaspectratio,clip]{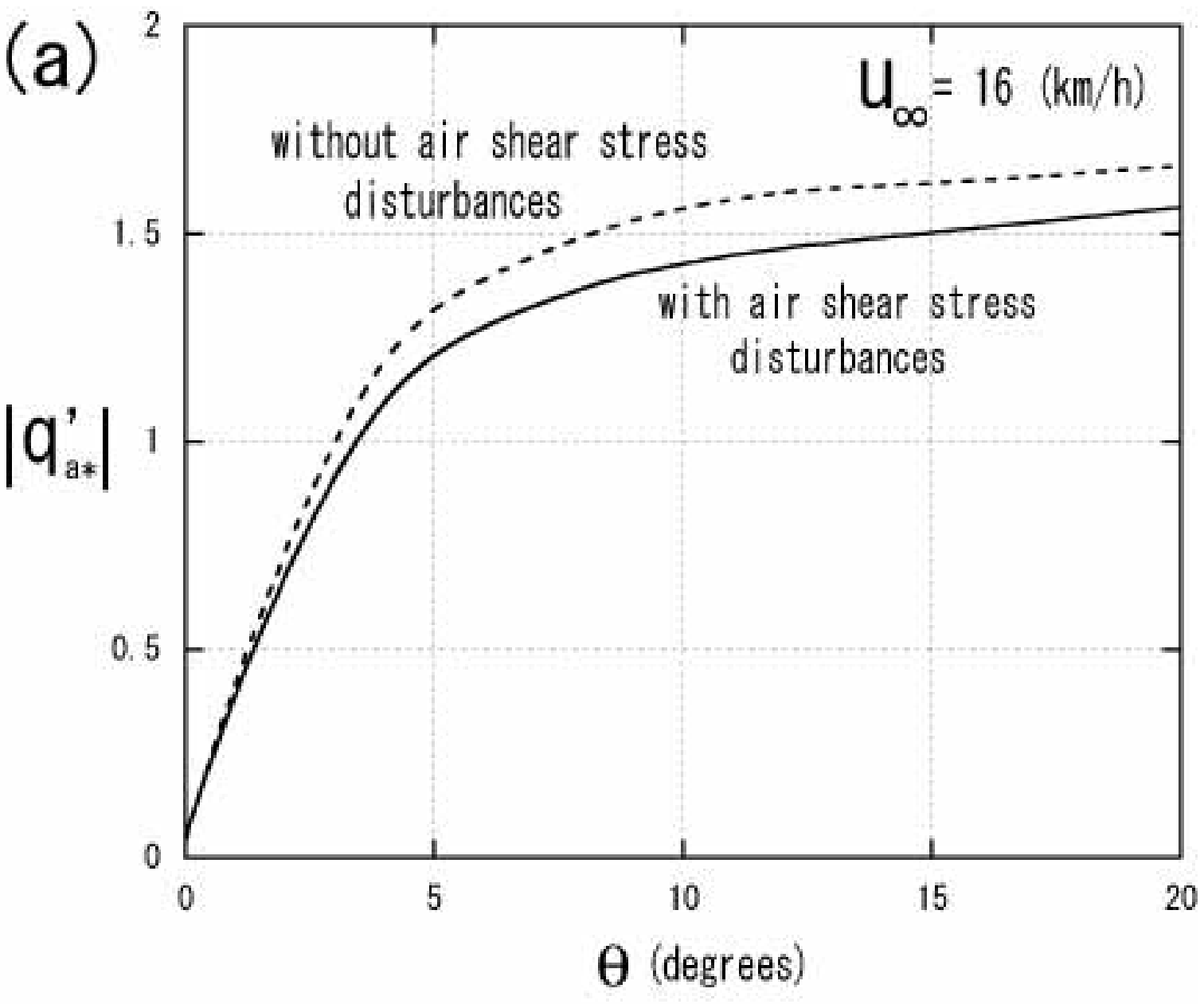}
\includegraphics[width=5cm,height=5cm,keepaspectratio,clip]{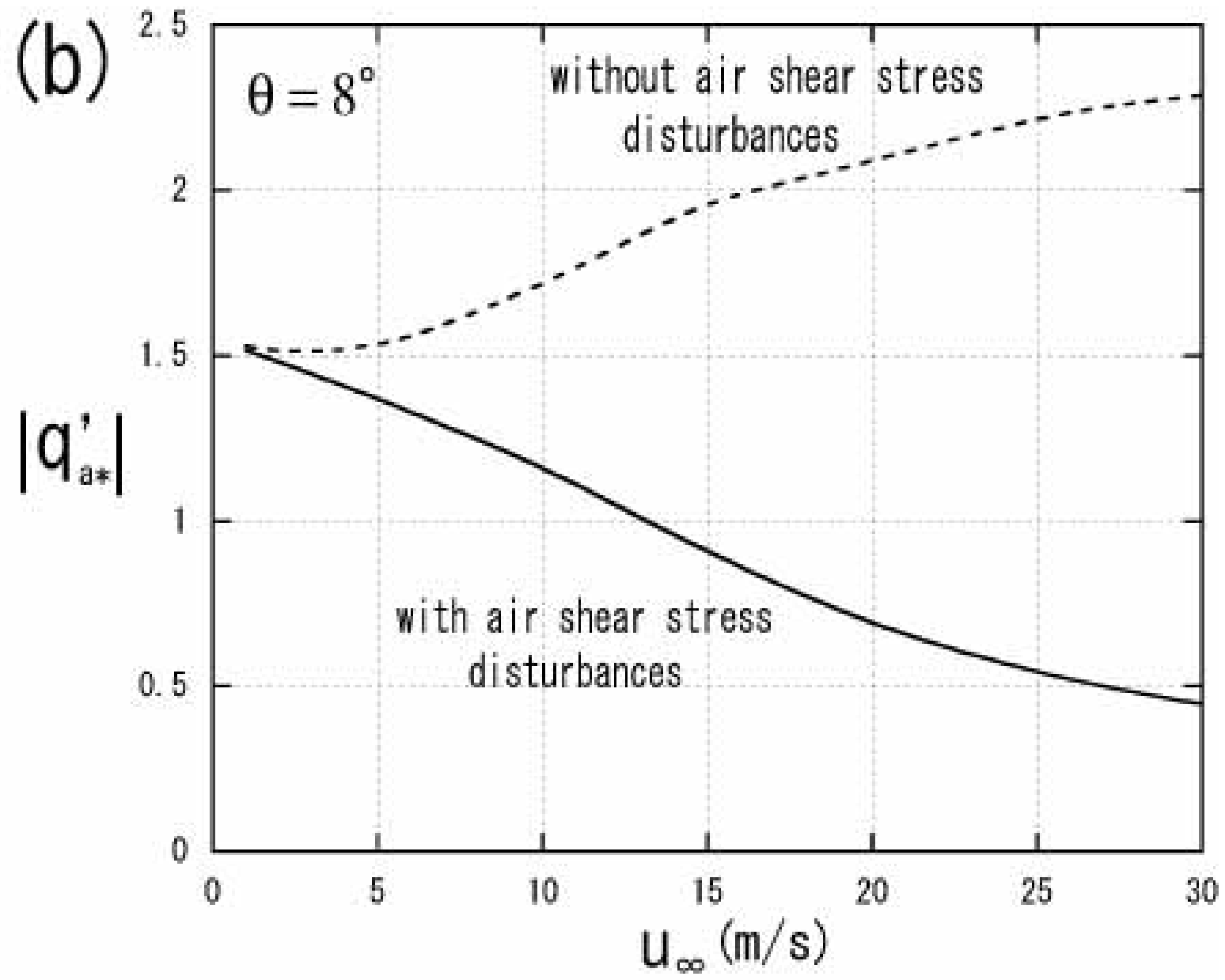}
\includegraphics[width=5cm,height=5cm,keepaspectratio,clip]{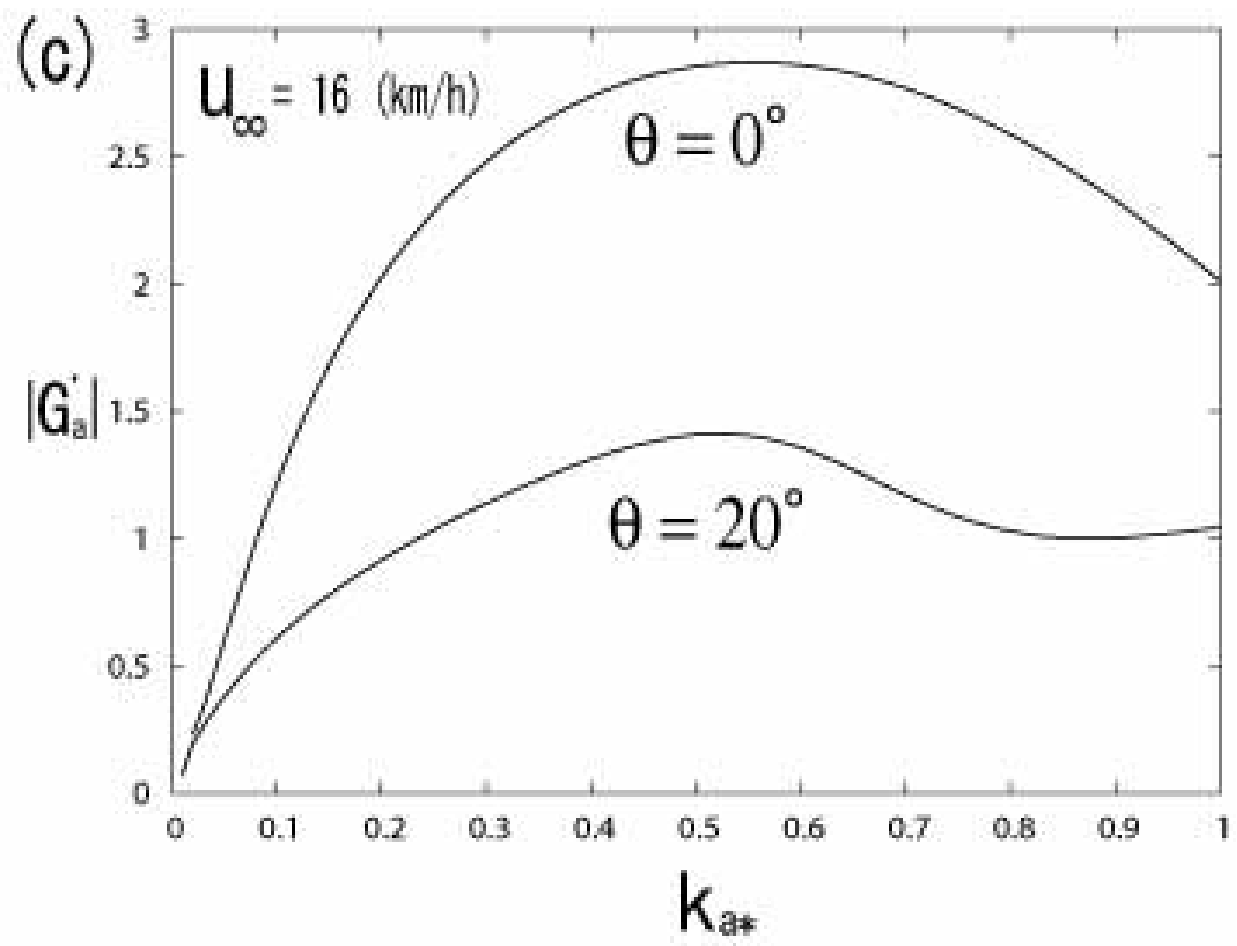}
\includegraphics[width=5cm,height=5cm,keepaspectratio,clip]{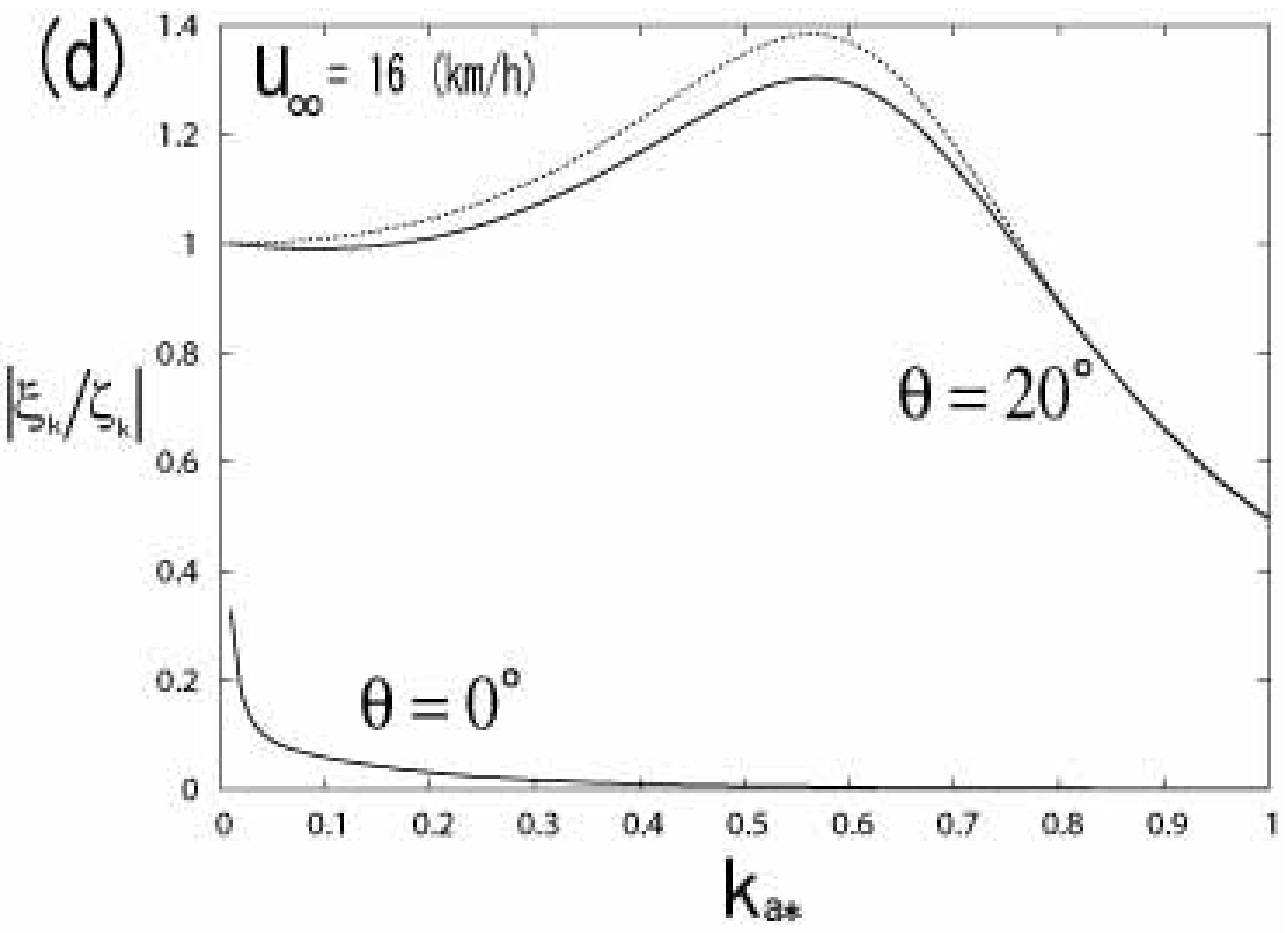}
\includegraphics[width=5cm,height=5cm,keepaspectratio,clip]{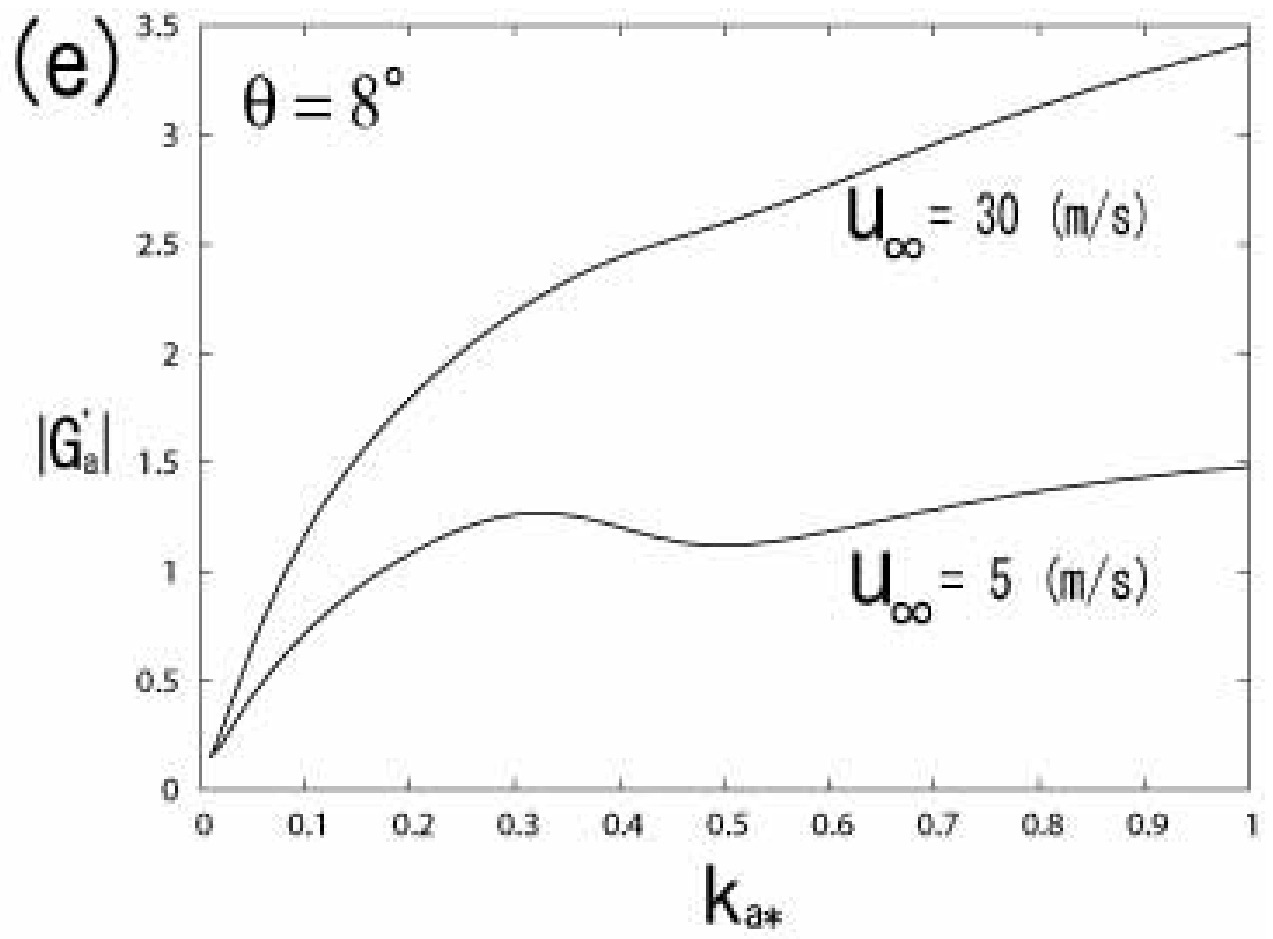}
\includegraphics[width=5cm,height=5cm,keepaspectratio,clip]{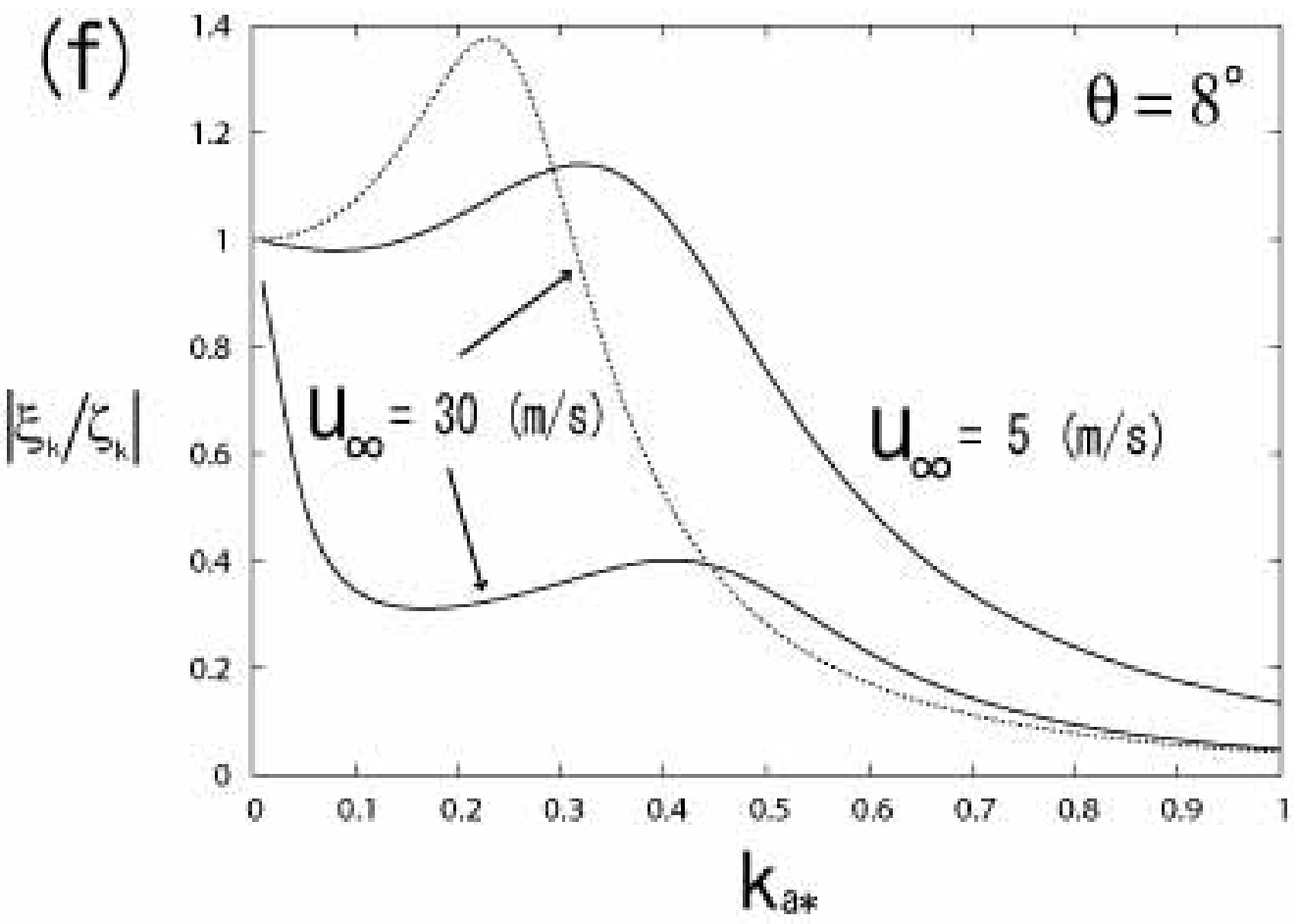}
\end{center}
\caption{For $Q/l_{w}=1692$ [(ml/h)/cm] and $x=0.1$ m,
variation of 
amplitude of dimensionless disturbed heat flux at the water-air interface, $|q'_{a*}|$, with 
(a) $\theta$ at $u_{\infty}=16$ (km/h) and 
(b) $u_{\infty}$ at $\theta=8^{\circ}$.
Variation of magnitude of disturbed temperature gradient at the water-air interface, $|G'_{a}|$, and 
the ratio of amplitude of the water-air interface to that of the ice-water interface, 
$|\xi_{k}/\zeta_{k}|$, with $k_{a*}$ 
for $\theta=0^{\circ}$ and $20^{\circ}$ ((c) and (d))
and for $u_{\infty}=5$ and 30 (m/s) ((e) and (f)).
The solid curves consider the effect of the tangential and normal air shear stress disturbances on the water-air interface, and the dashed curves do not consider this effect.}
\label{fig:qa-Ga'-xi-ka}
\end{figure}

For $Q/l_{w}=1692$ [(ml/h)/cm], figures~\ref{fig:qa-Ga'-xi-ka} (a) and (b) show the variation of $|q'_{a*}|$ with $\theta$ at $u_{\infty}=16$ km/h and with $u_{\infty}$ at $\theta=8^{\circ}$, respectively. Here $|q'_{a*}|$ is evaluated from the wavenumber at which $\sigma_{*}^{(r)}$ acquires a maximum value.  
In the results represented by the solid curves, the effect of the tangential and normal air shear stress disturbances on the water-air interface is considered, which is not the case for the dashed curves.
In response to the temperature distribution in the neighborhood of the ice-water interface, the amplification rate or roughness height of the ice-water interface is determined.
Since the boundary condition (\ref{eq:heatflux-xi-h0}) can be written as $dH_{l}/dy_{*}|_{y_{*}=1}-G'_{a}f_{l}|_{y_{*}=1}=0$, 
the disturbed temperature distribution, $H_{l}$, in the water film is affected by $q'_{a*}$. 
Therefore, figures~\ref{fig:amp-parameters} (a) and \ref{fig:qa-Ga'-xi-ka} (a), as well as
figures~\ref{fig:amp-parameters} (b) and \ref{fig:qa-Ga'-xi-ka}(b) show the same trends with respect to $\theta$ and $u_{\infty}$, respectively.

Figures~\ref{fig:qa-Ga'-xi-ka} (c), (d) and figures~\ref{fig:qa-Ga'-xi-ka} (e), (f) show the variation of the amplitude of the disturbed temperature gradient at the water-air interface, $|G'_{a}|=[(G'^{(r)}_{a})^{2}+(G'^{(i)}_{a})^{2}]^{1/2}$, and the ratio of amplitude of the water-air interface to that of the ice-water interface, $|\xi_{k}/\zeta_{k}|$, against $k_{a*}$ for $\theta=0^{\circ}$, $20^{\circ}$ 
and for $u_{\infty}=5$, 30 (m/s), respectively.
For extremely small values of $u_{\infty}$, the influence of airflow on the air temperature distribution can be neglected,
and (\ref{eq:geq-Ha}) can be approximated as $d^{2}H_{a}/d\eta^{2}=k_{a*}^{2}H_{a}$.
Its solution is $H_{a}={\rm e}^{-k_{a*}\eta}$ with the boundary conditions (\ref{eq:bc-Ha}), which yields
$|G'_{a}|=(\bar{h}_{0}/\delta_{0})k_{a*}$. 
For example, for $u_{\infty}=0.01$ m/s and $Q/l_{w}=1692$ [(ml/h)/cm], the following values are obtained: $\delta_{0}=16.1$ mm, and $\bar{h}_{0}=0.42$ mm for $\theta=20^{\circ}$, $\bar{h}_{0}=0.56$ mm for $\theta=8^{\circ}$.  
Therefore, each $|G'_{a}|$ is represented by $0.026k_{a*}$ and $0.035 k_{a*}$, which are very much below the solid curves in figures~\ref{fig:qa-Ga'-xi-ka}(c) and (e). This indicates that the disturbed temperature gradient at the water-air interface in the presence of airflow becomes extremely large.

In figure~\ref{fig:qa-Ga'-xi-ka}(c), $|G'_{a}|$ for $\theta=0^{\circ}$ is greater than that for $\theta=20^{\circ}$. 
In figure~\ref{fig:qa-Ga'-xi-ka}(e), $|G'_{a}|$ for $u_{\infty}=30$ m/s is greater than that for $u_{\infty}=5$ m/s.
Therefore, it might be expected that the heat transfer rate at the water-air interface
for $\theta=0^{\circ}$ and $u_{\infty}=30$ m/s is larger than that for $\theta=20^{\circ}$ and $u_{\infty}=5$ m/s. 
However, it should be stressed that the convective heat transfer rate $q'_{a*}$ depends not only on the disturbed temperature gradient at the water-air interface, $G'_{a}$, but also on the amplitude of the water-air interface, $\xi_{k}=-f_{l}|_{{y_{*}}=1}\zeta_{k}$. 
Figure~\ref{fig:qa-Ga'-xi-ka} (d) shows that $|\xi_{k}/\zeta_{k}|$ at $\theta=0^{\circ}$ decreases more rapidly with $k_{a*}$ than that at $\theta=20^{\circ}$. For $Q/l_{w}=1692$ [(ml/h)/cm] and $u_{\infty}=16$ km/h, $\sigma_{*}^{(r)}$ acquires a maximum value at $k_{a*}=0.05$ for $\theta=0^{\circ}$ and $k_{a*}=0.41$ for $\theta=20^{\circ}$.
Estimating $|G'_{a}|$ and $|\xi_{k}/\zeta_{k}|$ at these $k_{a*}$, $|q'_{a*}|$ at $\theta=20^{\circ}$ is greater than at $\theta=0^{\circ}$, as shown in figure~\ref{fig:qa-Ga'-xi-ka} (a). 
Hence, the roughness height in figure~\ref{fig:amp-parameters} (a) increases with $\theta$. 
On the other hand, 
figure~\ref{fig:qa-Ga'-xi-ka} (f) shows that $|\xi_{k}/\zeta_{k}|$ decreases with $u_{\infty}$. 
For $Q/l_{w}=1692$ [(ml/h)/cm] and $\theta=8^{\circ}$, $\sigma_{*}^{(r)}$ acquires a maximum value at $k_{a*}=0.27$ for $u_{\infty}=5$ m/s and $k_{a*}=0.13$ for $u_{\infty}=30$ m/s. Estimating $|G'_{a}|$ and $|\xi_{k}/\zeta_{k}|$ at these $k_{a*}$, $|q'_{a*}|$ at $u_{\infty}=30$ m/s is less than at $u_{\infty}=5$ m/s, as shown in figure~\ref{fig:qa-Ga'-xi-ka} (b). Hence, the roughness height in figure~\ref{fig:amp-parameters} (b) decreases with $u_{\infty}$. 
The undisturbed part of the local convective heat transfer coefficient can be written as $\bar{h}_{x}=0.292K_{a}\sqrt{u_{\infty}/(\nu_{a}x})$ for larger $u_{\infty}$, \cite{Gent00,UF11} which indicates that as wind speed increases, the undisturbed part of heat transfer from the water-air interface to the air increases and the undisturbed ice growth rate, $\bar{V}=-\bar{h}_{x}T_{\infty}/L$ is enhanced by the airflow. \cite{UF11} However, the disturbed part $q'_{a*}=h'_{x}/\bar{h}_{x}$ does not necessarily increase with $u_{\infty}$. 

When the air shear stress disturbances are considered, the wavelength is $\lambda=1.42$ cm from $k_{a*}=0.13$ for $u_{\infty}=30$ m/s, $Q/l_{w}=1692$ [(ml/h)/cm] and $\theta=8^{\circ}$. If we neglect the air shear stress disturbances, $\sigma_{*}^{(r)}$ acquires a maximum value at $k_{a*}=0.18$ for the same parameters, and $|\xi_{k}/\zeta_{k}|$ is overestimated for small region of $k_{a*}$, as shown by the dashed curve in figure~\ref{fig:qa-Ga'-xi-ka} (f). This leads to an overestimation of $|q'_{a*}|$, as shown by the dashed curve in figure~\ref{fig:qa-Ga'-xi-ka} (b), and $\lambda=1.03$ cm from $k_{a*}=0.18$.
These results indicate that if we neglect the effect of the air shear stress disturbances on the water-air interface, the roughness spacing and height are erroneously estimated. In particular, the results show that the variation of the roughness height with $u_{\infty}$ cannot predicted even qualitatively. 
 
\subsection{\label{sec:parameters-extraction}The most dominant term contributing to 
the behavior of wavelength $\lambda$ and amplification rate $\sigma_{*}^{(r)}$}

Since the governing equations 
(\ref{eq:geq-fa}), (\ref{eq:geq-Ha}), (\ref{eq:geq-fl}), (\ref{eq:geq-Hl}),
(\ref{eq:bc-fl0-dfl0}), (\ref{eq:bc-shearstress}), (\ref{eq:bc-normalstress}), 
(\ref{eq:shearstress_a}) and (\ref{eq:normalstress_a}) include many values:
$\Rey_{a}$, $\Rey_{l}$, $\Pec_{l}$, 
$d\bar{u}_{l*}/dy_{*}|_{y_{*}=0}$, 
$d\bar{u}_{l*}/dy_{*}|_{y_{*}=1}$, 
$d^{2}\bar{u}_{l*}/dy_{*}^{2}|_{y_{*}=1}$,
$\cos\theta/Fr^{2}$,
$Wek_{l*}^{2}$,
$\Sigma_{a}^{(r)}$, $\Sigma_{a}^{(i)}$, $\Pi_{a}^{(r)}$ and $\Pi_{a}^{(i)}$,
it is necessary to extract the most essential ones contributing to the behavior of $\lambda$ in figures~\ref{fig:lambda-vp} (c), (d) and $\sigma_{* \rm max}^{(r)}$ in figures~\ref{fig:amp-parameters} (a), (b).
Here,
$\Sigma_{a}^{(r)}$, $\Sigma_{a}^{(i)}$ and $\Pi_{a}^{(r)}$, $\Pi_{a}^{(i)}$ are the real and imaginary parts of tangential and normal air shear stress disturbances, $\Sigma_{a}$ and $\Pi_{a}$, respectively.
In the following, the values $Wek_{l*}^{2}$, $\Sigma_{a}^{(r)}$, $\Sigma_{a}^{(i)}$, $\Pi_{a}^{(r)}$ and $\Pi_{a}^{(i)}$ are estimated from the wavenumber at which $\sigma_{*}^{(r)}$ acquires a maximum value.   

\begin{figure}
\begin{center}
\includegraphics[width=5cm,height=5cm,keepaspectratio,clip]{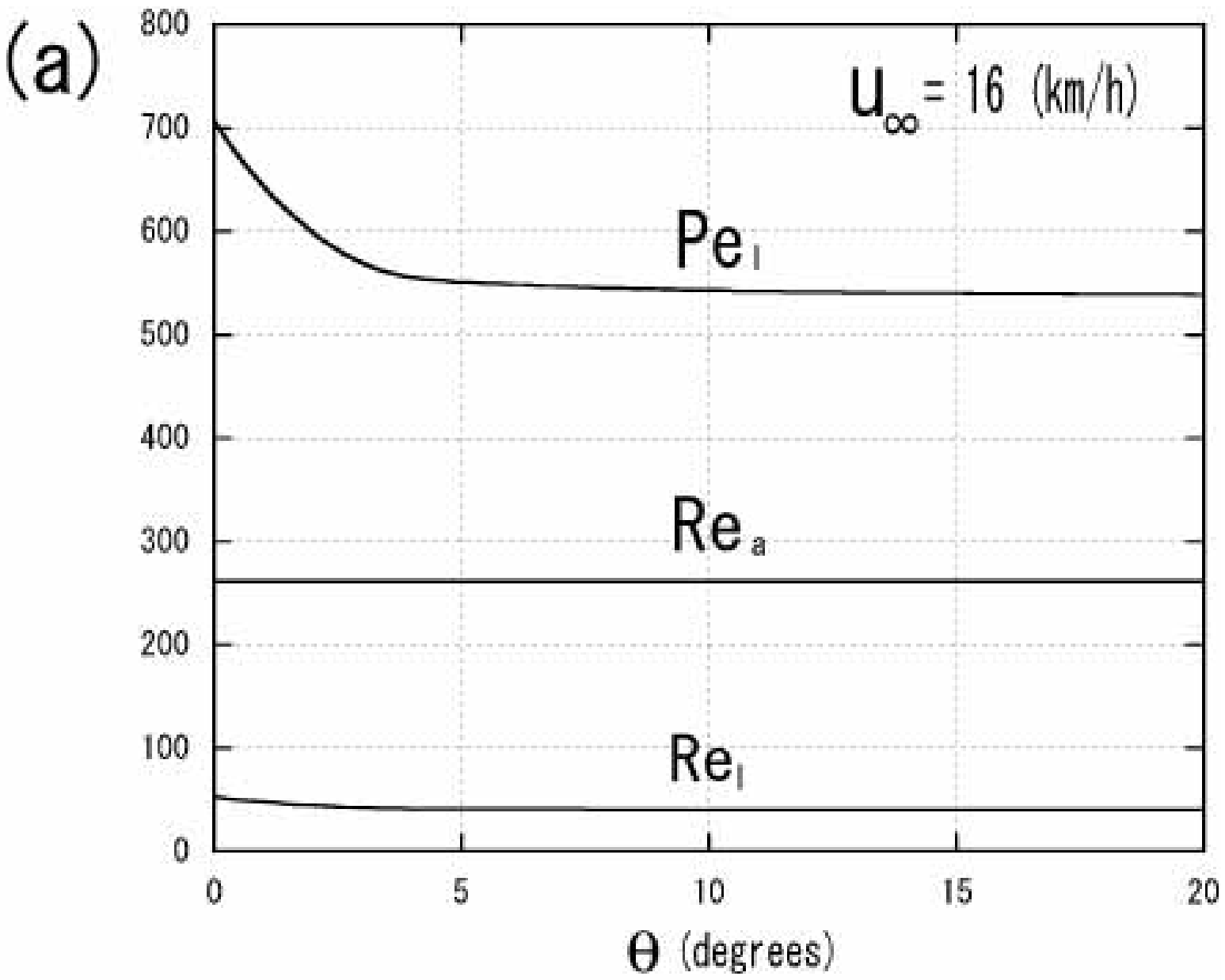}
\includegraphics[width=5cm,height=5cm,keepaspectratio,clip]{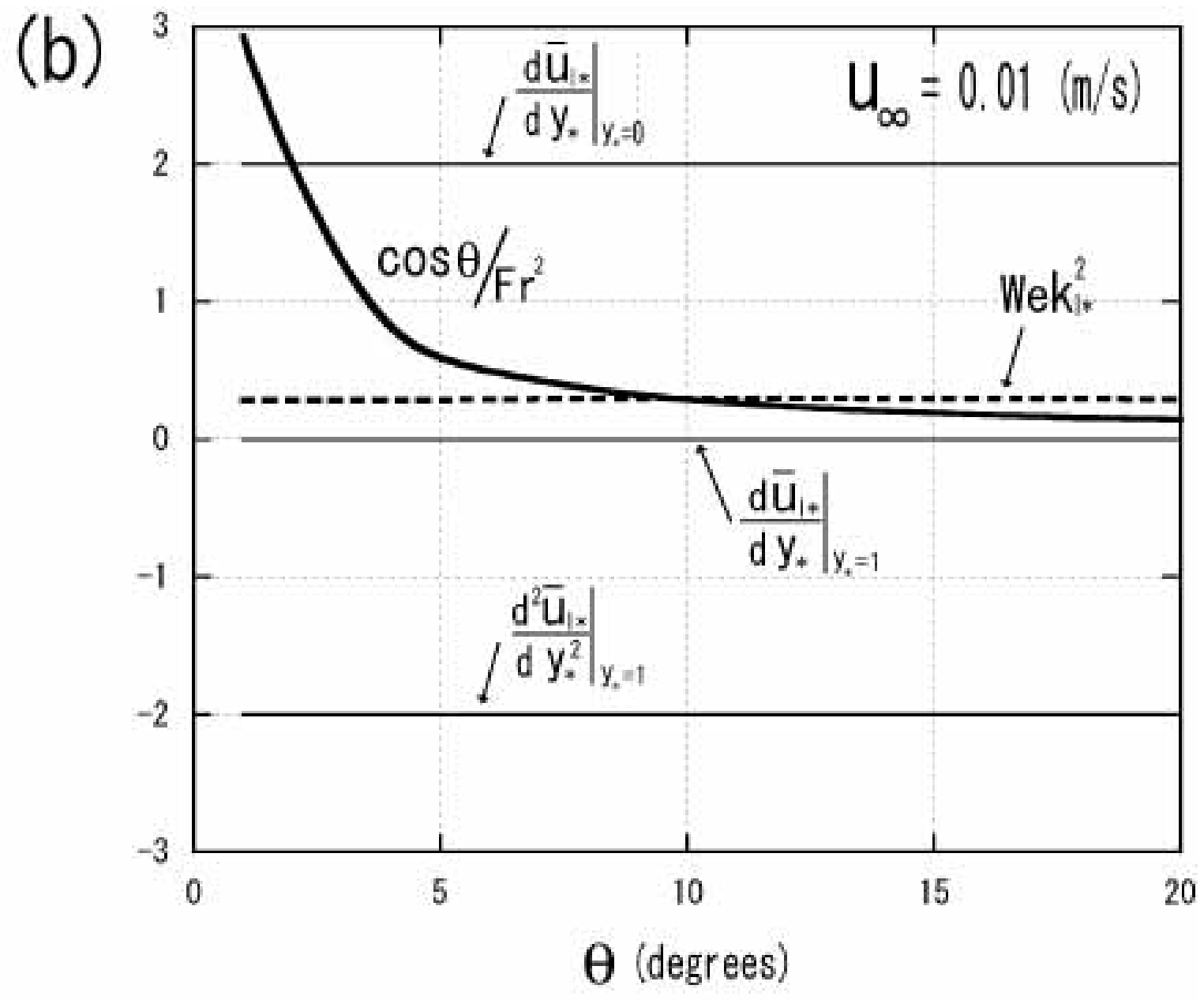}
\includegraphics[width=5cm,height=5cm,keepaspectratio,clip]{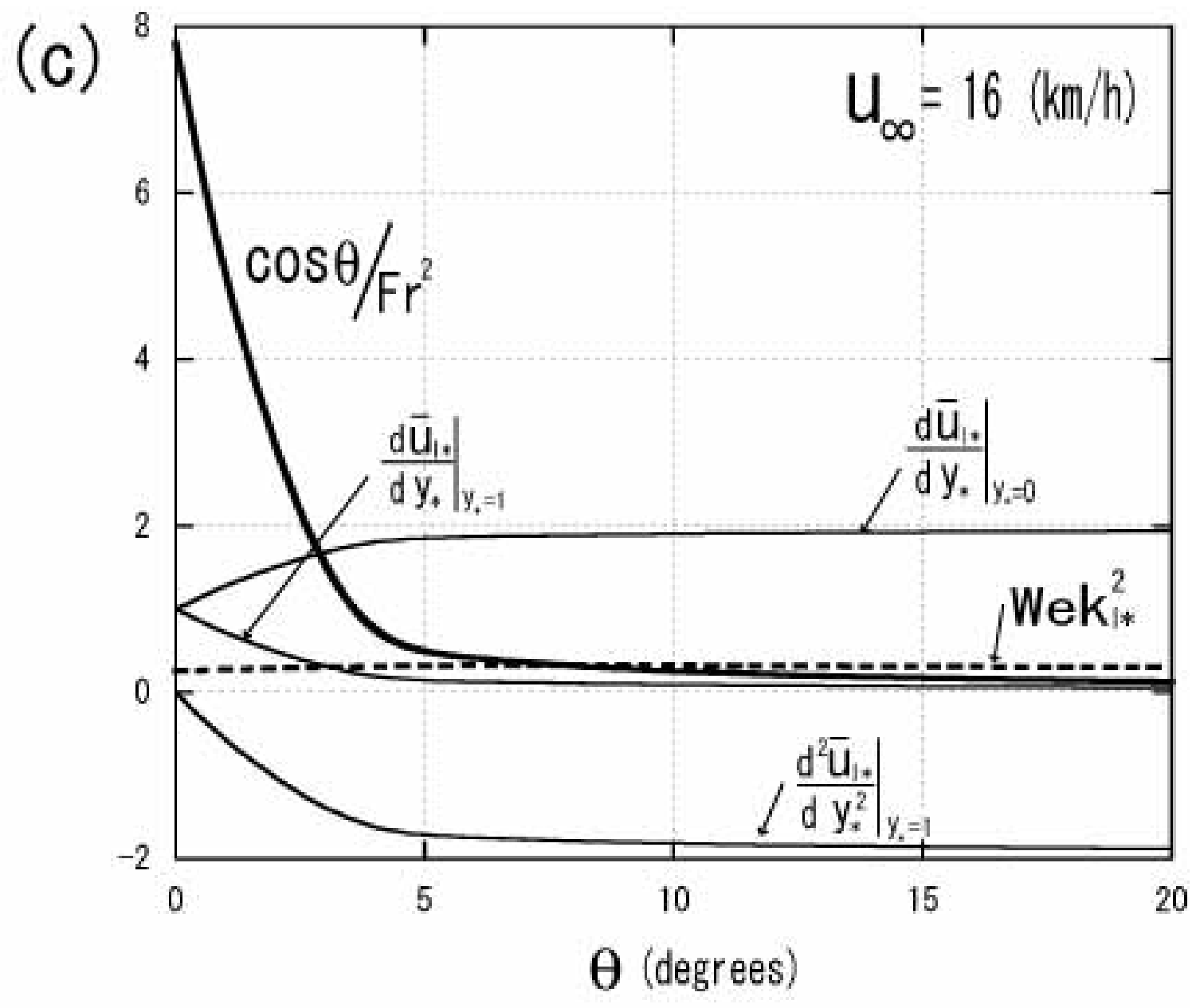}
\includegraphics[width=5cm,height=5cm,keepaspectratio,clip]{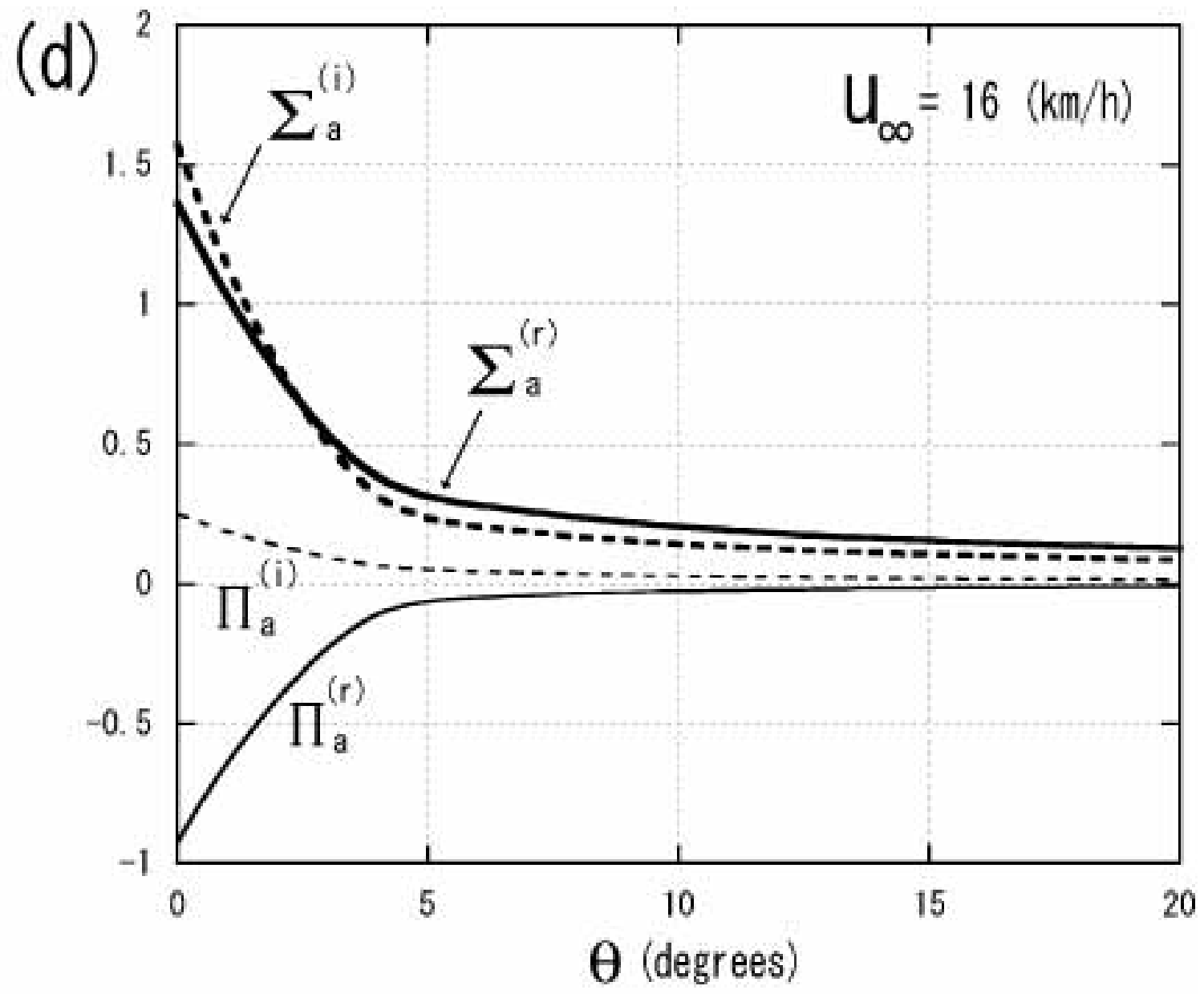}
\includegraphics[width=5cm,height=5cm,keepaspectratio,clip]{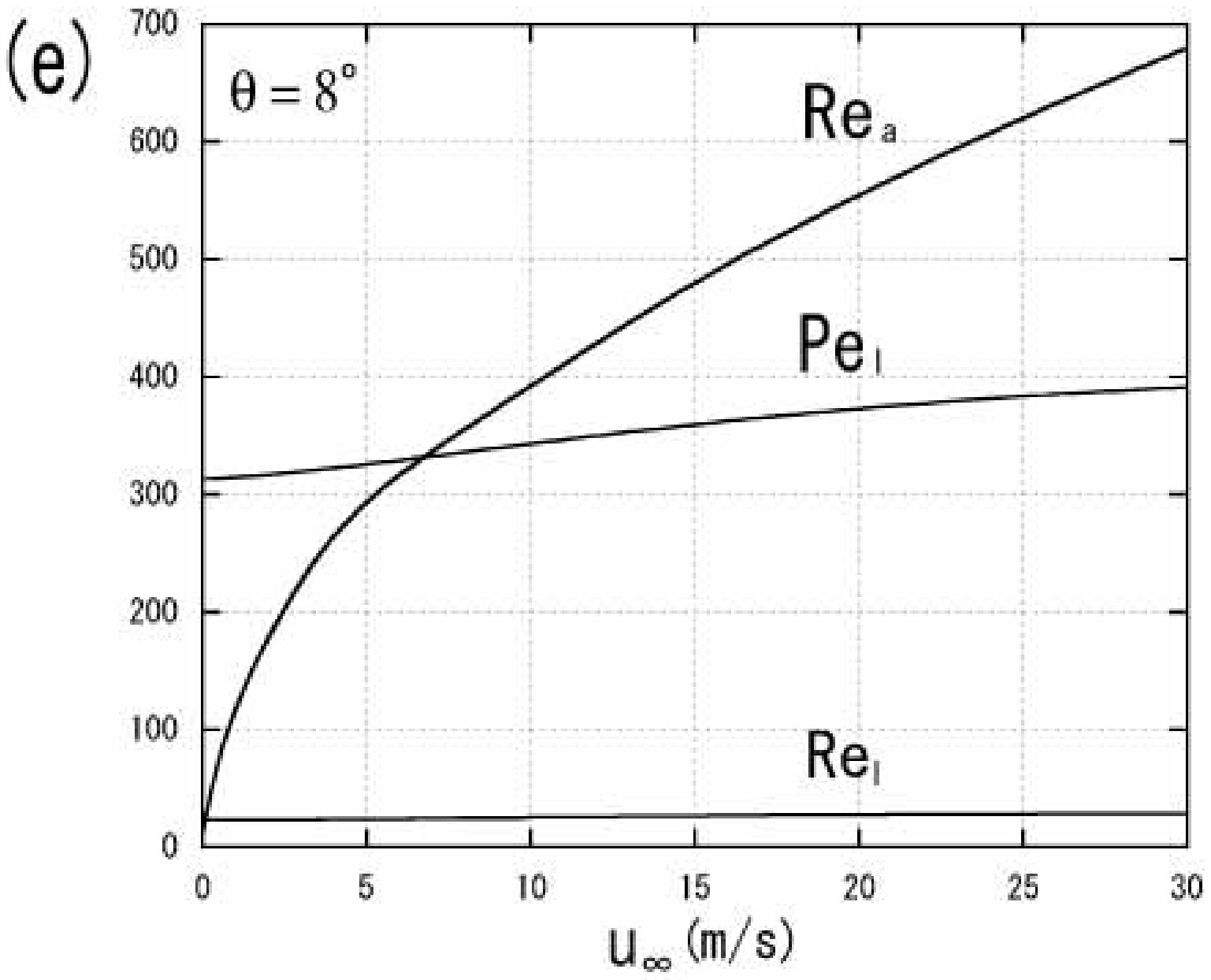}
\includegraphics[width=5cm,height=5cm,keepaspectratio,clip]{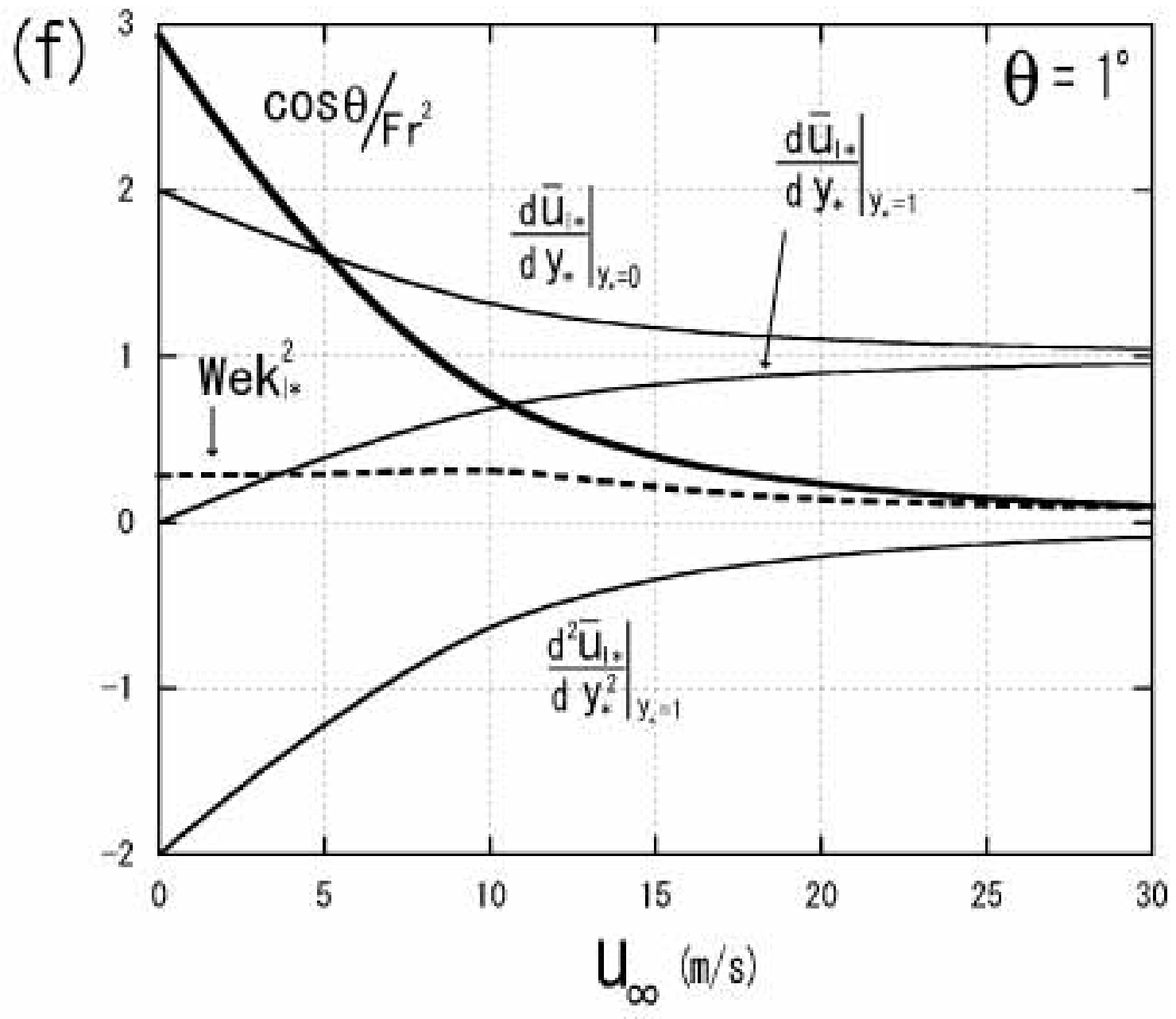}
\includegraphics[width=5cm,height=5cm,keepaspectratio,clip]{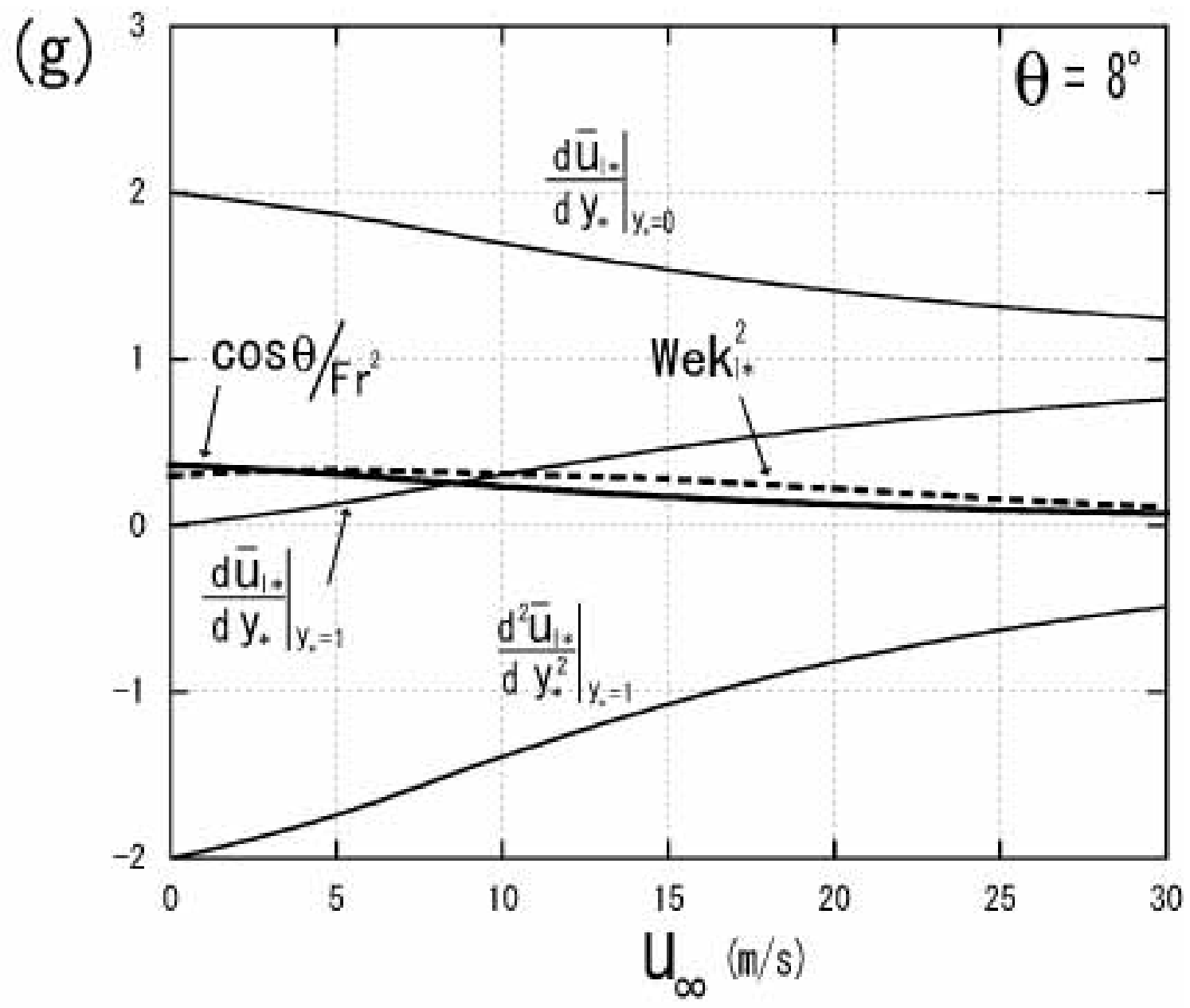}
\includegraphics[width=5cm,height=5cm,keepaspectratio,clip]{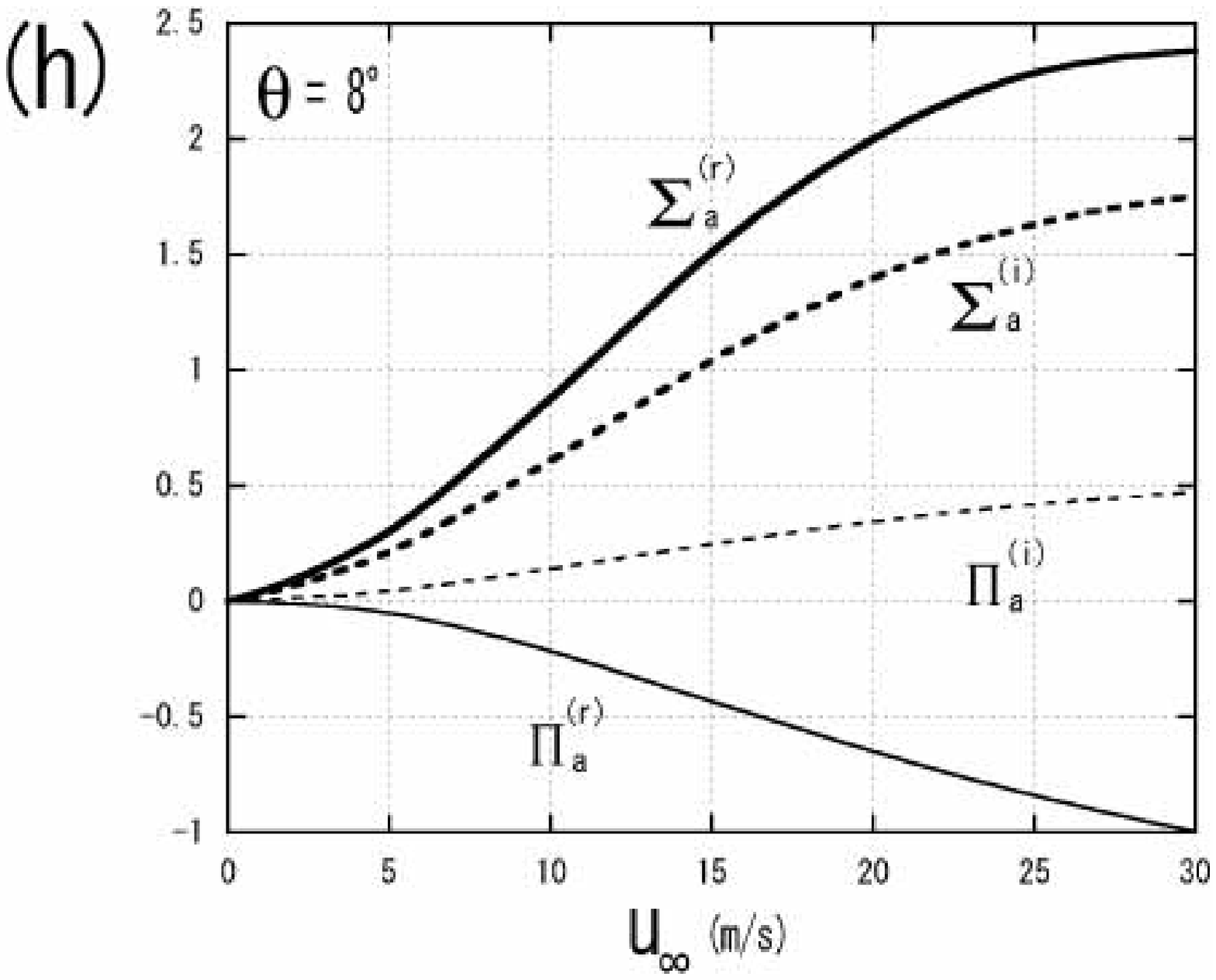}
\end{center}
\caption{For $Q/l_{w}=1692$ [(ml/h)/cm] and $x=0.1$ m,
variation of values
(a) $\Rey_{a}$, $\Rey_{l}$, $\Pec_{l}$;
(b), (c) $d\bar{u}_{l*}/dy_{*}|_{y_{*}=0}$, 
    $d\bar{u}_{l*}/dy_{*}|_{y_{*}=1}$, 
    $d^{2}\bar{u}_{l*}/dy_{*}^{2}|_{y_{*}=1}$,
    $\cos\theta/Fr^{2}$,
    $Wek_{l*}^{2}$;
(d) $\Sigma_{a}^{(r)}$, $\Sigma_{a}^{(i)}$, $\Pi_{a}^{(r)}$, $\Pi_{a}^{(i)}$, with $\theta$.
(e), (f), (g) and (h) represent the variation of these parameters with $u_{\infty}$.}
\label{fig:parameters}
\end{figure}

First, let us consider the dependence of these values on $\theta$.
Figure~\ref{fig:parameters} (a) shows that the variation of $\Rey_{a}$, $\Rey_{l}$, $\Pec_{l}$ with $\theta$
is small except for small $\theta$ values. 
For $u_{\infty}=0.01$ m/s in figure~\ref{fig:parameters} (b), gravity-driven flow is dominant and so, the
velocity profile in the water film is $\bar{u}_{l*}=-y_{*}^{2}+2y_{*}$, which yields values
$d\bar{u}_{l*}/dy_{*}|_{y_{*}=0}=2$, 
$d\bar{u}_{l*}/dy_{*}|_{y_{*}=1}=0$, 
$d^{2}\bar{u}_{l*}/dy_{*}^{2}|_{y_{*}=1}=-2$ for any $\theta$.
On the other hand, for $u_{\infty}=16$ km/h in figure~\ref{fig:parameters} (c), shear stress-driven flow is dominant at $\theta=0^{\circ}$ and so, the velocity profile in the water film is $\bar{u}_{l*}=y_{*}$,
which yields values
$d\bar{u}_{l*}/dy_{*}|_{y_{*}=0}=1$, 
$d\bar{u}_{l*}/dy_{*}|_{y_{*}=1}=1$, 
$d^{2}\bar{u}_{l*}/dy_{*}^{2}|_{y_{*}=1}=0$. 
As $\theta$ increases, the profile changes from $\bar{u}_{l*}=y_{*}$ to $\bar{u}_{l*}=-y_{*}^{2}+2y_{*}$. 
Therefore, 
$d\bar{u}_{l*}/dy_{*}|_{y_{*}=0}$, 
$d\bar{u}_{l*}/dy_{*}|_{y_{*}=1}$, 
$d^{2}\bar{u}_{l*}/dy_{*}^{2}|_{y_{*}=1}$ approach the values (2, 0, -2) for larger $\theta$ values in figure~\ref{fig:parameters} (c).
The effect of gravity and surface tension on the water-air interface is due to the terms
$\cos\theta/Fr^{2}=g\bar{h}_{0}\cos\theta/u_{la}^{2}$ and $Wek_{l*}^{2}=\gamma/(\rho_{l}u_{la}^{2}\bar{h}_{0})k_{l*}^{2}$ in (\ref{eq:bc-normalstress}), respectively.
The term $\cos\theta/Fr^{2}$ for $u_{\infty}=16$ km/h in figure~\ref{fig:parameters} (c) is greater than for $u_{\infty}=0.01$ m/s in figure~\ref{fig:parameters} (b), for small $\theta$ values. However, the difference decreases with $\theta$, and finally the term $Wek_{l*}^{2}$ is more dominant than the term $\cos\theta/Fr^{2}$ for higher wavenumber.
Figure~\ref{fig:parameters} (d) shows that the values $\Sigma_{a}^{(r)}$, $\Sigma_{a}^{(i)}$, $|\Pi_{a}^{(r)}|$ and $\Pi_{a}^{(i)}$ for $u_{\infty}=16$ km/h are large for small $\theta$ values. From (\ref{eq:shearstress_a}) and (\ref{eq:normalstress_a}), the values $\Sigma_{a}^{(r)}$, $\Sigma_{a}^{(i)}$, $\Pi_{a}^{(r)}$ and $\Pi_{a}^{(i)}$ decrease with $\theta$ because $\bar{h}_{0}$ decreases and $u_{la}$ increases with $\theta$, as shown in figures~\ref{fig:h0-theta-uinf-Qoverl} (a) and (d). 
It is found from the above estimates that the most dominant term in the third term of (\ref{eq:bc-normalstress}) for small $\theta$ values is $\cos\theta/Fr^{2}$. A slight difference between wavelength $\lambda$ for $u_{\infty}=16$ km/h and that for $u_{\infty}=0.01$ m/s in figure~\ref{fig:lambda-vp} (c) appears only for very small values of $\theta$, but the difference between them decreases with $\theta$. This trend is almost the same as the variation of $\cos\theta/Fr^{2}$ with $\theta$.
Since the values 
$d\bar{u}_{l*}/dy_{*}|_{y_{*}=1}$, $\cos\theta/Fr^{2}$, $We$ and $|\Pi_{a}^{(r)}|$ 
decrease with $\theta$,
the third term in (\ref{eq:bc-normalstress}) becomes more effective for higher wavenumber (or equivalently, small wavelength) with $\theta$, which results in a decrease of $\lambda$ with $\theta$, as shown in figure~\ref{fig:lambda-vp} (c).
Moreover, the amplitude of the water-air interface becomes smaller by the restoring force mainly due to gravity for small $\theta$ values, as shown by the solid curve of $\theta=0^{\circ}$ in figure~\ref{fig:qa-Ga'-xi-ka} (d). As $\theta$ increases, the action of gravity on the water-air interface becomes smaller since the value $\cos\theta/Fr^{2}$ decreases. Hence, the amplitude of the water-air interface becomes larger, as shown by the solid curve of $\theta=20^{\circ}$ in figure~\ref{fig:qa-Ga'-xi-ka} (d). 
This large disturbance of the water-air interface causes the increase of $\sigma_{* \rm max}^{(r)}$ or roughness height with $\theta$ in figure~\ref{fig:amp-parameters} (a).

Second, let us consider the dependence of the above values on $u_{\infty}$.
Figure~\ref{fig:parameters} (e) shows that the value of $\Rey_{a}$ increases rapidly with $u_{\infty}$, while the values of $\Rey_{l}$ and $\Pec_{l}$ increase only gradually. 
For $\theta=1^{\circ}$ in figure~\ref{fig:parameters} (f), the gravity-driven flow is dominant for small $u_{\infty}$ values, whereas the shear stress-driven flow is dominant for large $u_{\infty}$ values. 
Hence, the values of
$d\bar{u}_{l*}/dy_{*}|_{y_{*}=0}$, 
$d\bar{u}_{l*}/dy_{*}|_{y_{*}=1}$, 
$d^{2}\bar{u}_{l*}/dy_{*}^{2}|_{y_{*}=1}$ in figure~\ref{fig:parameters} (f)
start at (2, 0, -2) and approach (1, 1, 0).
That trend is also the case for $\theta=8^{\circ}$ in figure~\ref{fig:parameters} (g), but the shear-driven effect appears later than for $\theta=1^{\circ}$.
The values $\cos\theta/Fr^{2}$ and $Wek_{l*}^{2}$ in figures~\ref{fig:parameters} (f) and (g) decrease with $u_{\infty}$ because $\bar{h}_{0}$ decreases and $u_{la}$ increases as $u_{\infty}$ increases, as shown in figures~\ref{fig:h0-theta-uinf-Qoverl} (b) and (e).
On the other hand, the values $\Sigma_{a}^{(r)}$, $\Sigma_{a}^{(i)}$, $\Pi_{a}^{(r)}$ and $\Pi_{a}^{(i)}$ in figure~\ref{fig:parameters} (h) increase with $u_{\infty}$ because (\ref{eq:shearstress_a}) and (\ref{eq:normalstress_a}) increase with $u_{\infty}$. 
For $\theta=1^{\circ}$, the term $\cos\theta/Fr^{2}$ is the most dominant in the third term of (\ref{eq:bc-normalstress}) for small $u_{\infty}$ values, but $\cos\theta/Fr^{2}$ is comparable to $Wek_{l*}^{2}$ with $u_{\infty}$.
On the other hand, for $\theta=8^{\circ}$, $\cos\theta/Fr^{2}$ is comparable to $Wek_{l*}^{2}$ for any $u_{\infty}$, as shown in figure~\ref{fig:parameters} (g).
The value $\cos\theta/Fr^{2}$ for $\theta=1^{\circ}$ is much greater than that for $\theta=8^{\circ}$, for small $u_{\infty}$ values.  
Hence, the variation of $\lambda$ with $u_{\infty}$ for $\theta=1^{\circ}$ is large compared to that for $\theta=8^{\circ}$, as shown in figure~\ref{fig:lambda-vp} (d).
For $u_{\infty}=5$ m/s in figures~\ref{fig:parameters} (g) and (h), since the values $\cos\theta/Fr^{2}$, 
$Wek_{l*}^{2}$, $\Sigma_{a}$ and $\Pi_{a}$ are comparable and small, the amplitude of the water-air interface becomes large, as shown by the solid curve of $u_{\infty}=5$ m/s in figure~\ref{fig:qa-Ga'-xi-ka} (f). 
On the other hand,
for $u_{\infty}=30$ m/s in figures~\ref{fig:parameters} (g) and (h), the values $\Sigma_{a}$ and $\Pi_{a}$ are much greater than the values $\cos\theta/Fr^{2}$ and $Wek_{l*}^{2}$. 
Hence, the effect of the shear stress disturbances on the water-air interface is more dominant than that of gravity and surface tension. Moreover, the amplitude of the water-air interface for small $k_{a*}$ values becomes small compared to that for $u_{\infty}=5$ m/s, as shown by the solid curve of $u_{\infty}=30$ m/s in figure~\ref{fig:qa-Ga'-xi-ka} (f). 
This causes the decrease of $\sigma_{* \rm max}^{(r)}$ or roughness height with $u_{\infty}$ in figure~\ref{fig:amp-parameters} (b).  
As the wavenumber increases,
the term $Wek_{l*}^{2}$ becomes the most dominant one, and the water-air interface tends to be flat because of the restoring force due to the surface tension, as shown in figures~\ref{fig:qa-Ga'-xi-ka} (d) and (f). 

\section{\label{sec:concl}Conclusion}

A theoretical model to explain the roughness characteristics in an initial aufeis (icings) formation
observed in the experiments of Ref. \onlinecite{Streitz02} was proposed, 
from a new morphological instability of an ice surface during growth under a supercooled water film driven by gravity and air drag.  
A numerical method to solve complex air-water-ice multi-phase system, where air and water flows and temperature fields are highly coupled with water film thickness, was also proposed.
Using linear stability analysis, roughness characteristics such as roughness spacing and height of the ice-water interface were derived for various water supply rates, plane slopes and airspeeds. 
Major findings are as follows:
(1) The roughness spacing decreases with increasing slope and airspeed, whose trends are in qualitative agreement with the experimental results of Ref. \onlinecite{Streitz02}. Moreover, the roughness spacing was found to increase with the water supply rate. In particular, roughness spacing depends mainly on water layer thickness. 
(2) The upstream propagation of the ice-water interface disturbance was predicted.
(3) The amplification rate of the ice-water interface disturbance increases with slope, but decreases with airspeed. 
In the linear stability analysis, the amplification rate of the ice-water interface disturbance is expected to be relevant with the initial roughness height.
In the experiments, \cite{Streitz02} the roughness height increased with slope, but decreased with airspeed. Therefore, the theoretical results herein are consistent with the experimental ones. Also, the amplification rate was found to increase with the water supply rate for low airspeeds, which suggests that roughness height increases with water supply rate.
The most important finding of this study is that in order to predict the roughness height at higher airspeeds, the influence of air shear stress disturbances on the water-air interface must be taken into account. Otherwise, the disturbed part of the convective heat transfer rate at the water-air interface is not correctly predicted, and the roughness height is erroneously estimated. 

The roughness characteristics shown in the initial aufeis formation have common features with the roughness features observed in glaze icing and geological pattern formations. In order to extend the present model to practical aircraft and structural icing problems, we have to consider air, water and ice interactions near the stagnation point of objects, as shown in figure~\ref{fig:ripples} (c). In that case, water film on the object is formed by unfrozen impinging water droplets, the free stream velocity is not constant around the object and the angle changes locally along the object. Therefore, the water supply rate $Q/l_{w}$ used in the current model must be replaced by liquid water content (LWC), and the values LWC, $u_{\infty}$ and $\theta$ must change locally along the position of the object. Furthermore, gravity impedes the supercooled water flow due to air shear stress on the upper side of the object, while it magnifies the effect of air shear stress on the lower side of the object. In addition to this asymmetry, it will be necessary to consider heat conduction into the object beneath the ice sheet. Although the method and fundamental ideas developed here can be applied to other phenomena by extending the current model, further research and laboratory experiments are necessary to validate our model proposed in this paper.

\begin{acknowledgements}
This study was carried out within the framework of the NSERC/Hydro-Qu$\acute{\rm e}$bec/UQAC Industrial Chair on Atmospheric Icing of Power Network Equipment (CIGELE) and the Canada Research Chair on Engineering of Power Network Atmospheric Icing (INGIVRE) at the Universit$\acute{\rm e}$ du Qu$\acute{\rm e}$bec $\grave{\rm a}$ Chicoutimi. 
The authors would like to thank all CIGELE partners (Hydro-Qu$\acute{\rm e}$bec, Hydro One, R$\acute{\rm e}$seau Transport d'$\acute{\rm E}$lectricit$\acute{\rm e}$ (RTE) and $\acute{\rm E}$lectricit$\acute{\rm e}$ de France (EDF), Alcan Cable, K-Line Insulators, Tyco Electronics, Dual-ADE, and FUQAC) whose financial support made this research possible. The authors would also like to thank H. Tsuji for his useful comments. 
\end{acknowledgements}


\begin{thebibliography}{99}

\bibitem{Oron97}
A. Oron, S. H. Davis, and S. G. Bankoff,
``Long-scale evolution of thin liquid films,"
Rev. Mod. Phys. \textbf{69}, 931--980 (1997). 

\bibitem{Meakin10}
P. Meakin and B. Jamtveit,  
``Geological pattern formation by growth and dissolution in aqueous systems,"
Proc. R. Soc. A \textbf{466}, 659--694 (2000).

\bibitem{Maeno94}
N. Maeno, L. Makkonen, K. Nishimura, K. Kosugi, and T. Takahashi, 
``Growth rates of icicles," 
J. Glaciol \textbf{40}, 319--326 (1994).

\bibitem{Matsuda97}
S. Matsuda, 
``Experimental study on the wavy pattern of icicle surface,"
M.Sc. thesis, Hokkaido University, 1997. 

\bibitem{UFYT10}
K. Ueno, M. Farzaneh, S. Yamaguchi, and H. Tsuji, 
``Numerical and experimental verification of a theoretical model of ripple formation 
in ice growth under supercooled water film flow,"
Fluid Dyn. Res. \textbf{42}, 025508 (2010).

\bibitem{Makkonen00}
L. Makkonen, 
``Models for the growth of rime, glaze, icicles and wet snow on structures,"
Philos. Trans. R. Soc. London, Ser. A \textbf{358}, 2913--2939 (2000). 

\bibitem{Chen11}
A. S-H. Chen and S. M. Morris,  
``Experiments on the morphology of icicles,"
Phys. Rev. E \textbf{83}, 026307 (2011).

\bibitem{Shin96}
J. Shin, 
``Characteristics of surface roughness associated with leading edge ice accretion,"
J. Aircr. \textbf{33}, 316--321 (1996). 

\bibitem{Gent00}
R. W. Gent, N. P. Dart,and J. T. Cansdale, 
``Aircraft icing,"
Philos. Trans. R. Soc. London, Ser. A \textbf{358}, 2873--2911 (2000). 

\bibitem{Streitz02}
J. T. Streitz and R. Ettema, 
``Observations from an aufeis windtunnel,"
Cold Reg. Sci. Technol. \textbf{34}, 85--96 (2002).

\bibitem{Hammer10}
$\O$. Hammer, D. K. Dysthe, B. Jamtveit,
``Travertine terracing: patterns and mechanisms,"
In \textit{Tufas and Speleothems: Unraveling the Microbial and Physical Controls},
(ed. H. M. Pedley \& M. Rogerson) \textbf{336}, pp.~345--355, the Geological Society of London, 2010.

\bibitem{Pentecost05}
A. Pentecost, 
\textit{Travertine} (Springer, 2005).

\bibitem{Ueno03}
K. Ueno, 
``Pattern formation in crystal growth under parabolic shear flow,"
Phys. Rev. E \textbf{68}, 021603 (2003).

\bibitem{Ueno04}
K. Ueno, 
``Pattern formation in crystal growth under parabolic shear flow II,"
Phys. Rev. E \textbf{69}, 051604 (2004).

\bibitem{Ueno07}
K. Ueno, 
``Characteristic of the wavelength of ripples on icicles,"
Phys. Fluids \textbf{19}, 093602 (2007).

\bibitem{UF10}
K. Ueno and M. Farzaneh, 
``Morphological instability of the solid-liquid interface in crystal growth under supercooled liquid film flow and natural convection airflow,"
Phys Fluids \textbf{22}, 017102 (2010).

\bibitem{Schlichting99}
H. Schlichting and K. Gersten, 
\textit{Boundary Layer Theory} (Springer, 1999).

\bibitem{Tsao98}
J. -C. Tsao and A. P. Rothmayer, 
``A mechanism for ice roughness formation on an airfoil leading edge:
Contributing to glaze ice accretion,"
AIAA 98-0485, Jan. 1--21 (1998).

\bibitem{Tsao00}
J. -C. Tsao and A. P. Rothmayer, 
``Triple-deck simulation of surface glaze ice accretion,"
AIAA 00-0234, Jan. 1--17 (2000).

\bibitem{UF11}
K. Ueno and M. Farzaneh, 
``Linear stability analysis of ice growth under supercooled water film driven by a laminar airflow,"
Phys Fluids in press, arXiv:1103.3007 (2011).

\bibitem{Langer80}
J. S. Langer, 
``Instability and pattern formation in crystal growth,"
Rev. Mod. Phys. \textbf{52}, 1--28 (1980). 

\bibitem{Myers02_1}
T. G. Myers, J. P. F. Charpin, and C. P. Thompson, 
``Slowly accreting ice due to supercooled water impacting on a cold surface,"
Phys Fluids \textbf{14}, 240--256 (2002).

\bibitem{Myers02_2}
T. G. Myers, J. P. F. Charpin, and S. J. Chapman, 
``The flow and solidification of a thin fluid film on an arbitrary three-dimensional surface,"
Phys Fluids \textbf{14}, 2788--2803 (2002). 

\end{thebibliography}
\end{document}